%% file: Dust_MC_20juillet2011.tex
\documentclass[twocolumn,traditabstract,longauth]{aa}
\usepackage{graphicx}
\usepackage{natbib}
\usepackage{txfonts}

\input Planck.tex

\begin{document}
\input{Proj_Ref_7_13_authors_and_institutes.tex}
   \title{\textit{Planck} Early Results. XXV. Thermal dust in nearby molecular clouds}

\date{Received 9 January 2011 /
Accepted 23 June 2011}

\abstract{\Planck\ allows unbiased mapping of Galactic sub-millimetre and millimetre emission from the most diffuse regions to the densest parts of molecular clouds.  We present an early analysis of the Taurus molecular complex, on line-of-sight-averaged data and without component separation.  The emission spectrum measured by \Planck\ and {\it IRAS\/} can be fitted pixel by pixel using a single modified blackbody.  Some systematic residuals are detected at 353\,GHz and 143\,GHz, with amplitudes around $-7$\% and +13\%, respectively, indicating that the measured spectra are likely more complex than a simple modified blackbody. 
Significant positive residuals are also detected in the molecular regions and in the 217\,GHz and 100\,GHz bands, mainly caused by to the contribution of the $J=2\rightarrow1$ and $J=1\rightarrow0$ $^{12}$CO and $^{13}$CO emission lines.  
We derive maps of the dust temperature $T$, the dust spectral emissivity index $\beta$, and the dust optical depth at 250\microns\, $\tau_{250}$. The temperature map illustrates the cooling of the dust particles in thermal equilibrium with the incident radiation field, from 16--17\,K in the diffuse regions to 13--14\,K in the dense parts. The distribution of spectral indices is centred at 1.78, with a standard deviation of 0.08 and a systematic error of 0.07. We detect a significant $T-\beta$ anti-correlation. The dust optical depth map reveals the spatial distribution of the column density of the molecular complex from the densest molecular regions to the faint diffuse regions.  We use near-infrared extinction and \ion{H}{i} data at 21-cm to perform a quantitative analysis of the spatial variations of the measured dust optical depth at 250\microns\, per hydrogen atom $\tau_{250}/N_{\rm H}$. We report an increase of $\tau_{250}/N_{\rm H}$ by a factor of about 2 between the atomic phase and the molecular phase, which has a strong impact on the equilibrium temperature of the dust particles.}


\authorrunning{Planck Collaboration}
\titlerunning{Thermal dust in nearby molecular clouds}

   \maketitle
   \allearlypapers
%
%________________________________________________________________

\section{Introduction}
\Planck\ 
\footnote{\Planck\ (http://www.esa.int/\Planck ) is a project of the European Space Agency (ESA) with instruments provided by two scientific consortia funded by ESA member states (in particular the lead countries France and Italy), with contributions from NASA (USA) and telescope reflectors provided by a collaboration between ESA and a scientific consortium led and funded by Denmark.}
\citep{tauber2010a, planck2011-1.1} is the third-generation space mission to measure the anisotropy of the cosmic microwave background (CMB).  It observes the sky in nine frequency bands covering 30--857\,GHz with high sensitivity and angular resolution from 3$\arcmin$\ to 5$\arcmin$.  The Low Frequency Instrument (LFI; \citealt{Mandolesi2010, Bersanelli2010, planck2011-1.4}) covers the 30, 44, and 70\,GHz bands with amplifiers cooled to 20\,\hbox{K}.  The High Frequency Instrument (HFI; \citealt{Lamarre2010, planck2011-1.5}) covers the 100, 143, 217, 353, 545, and 857\,GHz bands with bolometers cooled to 0.1\,\hbox{K}. Polarisation is measured in all but the highest two bands \citep{Leahy2010, Rosset2010}.  A combination of radiative cooling and three mechanical coolers produces the temperatures needed for the detectors and optics \citep{planck2011-1.3}.  Two data processing centres (DPCs) check and calibrate the data and make maps of the sky \citep{planck2011-1.7, planck2011-1.6}.  \Planck's sensitivity, angular resolution, and frequency coverage make it a powerful instrument for Galactic and extragalactic astrophysics as well as cosmology.  Early astrophysics results are given in Planck Collaboration, 2011h--u. This paper presents the early results of the analysis of Planck/HFI observations of the Taurus molecular cloud. 

{\it IRAS\/} provided a complete view of the interstellar matter in our Galaxy in four photometric bands from 12 to 100\microns\ with angular resolution of about 4$\arcmin$. The 12, 25, and 60\micron\ bands are mainly sensitive to the emission of the smallest interstellar grains.  They are aromatic hydrocarbons (large molecules) and amorphous hydrocarbon grains that undergo stochastic heating upon photon absorption, and tend to emit most of their energy at wavelengths shortward of 100\micron. 
The large particles have dimensions of the order of 100\,nm and make up the bulk of the dust mass. They are in equilibrium between thermal emission and absorption of UV and 
visible photons from incident radiation. Only one {\it IRAS\/} band, at 100\microns, is dominated by the emission of this dust component. DIRBE and FIRAS on board {\it COBE\/} produced all-sky maps at longer wavelengths, with angular resolution lower than {\it IRAS\/} (40$\arcmin$\ and 7\deg, respectively), which allowed the measurement of dust temperatures and spectral indices. The dust temperature was found to be on average $\sim$17.5\,K (with a spectral emissivity index $\beta=2$) in the diffuse atomic medium \citep{Boulanger1996} and to be lower in molecular clouds with no embedded bright stars \citep{Lagache1998}. Small patches of molecular clouds have been observed in more detail from the ground by the JCMT \citep{Johnstone1999}, by balloon-borne experiments PRONAOS \citep{Ristorcelli1998}, Archeops \citep{Desert2008} and BLAST \citep{Netterfield2009}, and from space by \textit{Spitzer} \citep[e.g., ][]{Flagey2009} and {\it Herschel\/} \citep[e.g.,][]{Andre2010, Motte2010, Abergel2010, Juvela2011}. %\citep[e.g.,][]{Andre2010, Motte2010, Abergel2010}. 

With \Planck, the whole sky emission of thermal dust is mapped from the submillimetre to the millimetre range, with angular resolution comparable to {\it IRAS\/}.  We have complete and unbiased surveys of molecular clouds not only in terms of spatial coverage, but also in terms of spatial content of the data. Unlike ground or ballon observations, the maps contain all angular scales above the diffraction limit without spatial filtering.  This is extremely important for  Galactic science, because the interstellar matter contains a large range of intricate spatial scales.  Moreover, the emission is measured with unprecedented signal-to-noise ratio down to the faintest parts surrounding the bright and dense regions. Up to now, the rotational transitions of carbon monoxide (CO) were the main tracer of the interstellar matter in molecular clouds.  With \Planck\, thermal dust becomes a new tracer. The spectral coverage and the sensitivity allow for each line of sight precise measurements of the temperature, of the spectral emissivity index, and of the optical depth, independent of the excitation conditions. 

Molecular clouds present a wide range of physical conditions (illumination, density, star-forming activity), so are ideal targets for studying the emission properties of dust grains and their evolution. The goal of this paper is to discuss the results of the early analysis of HFI maps of the Taurus complex, which is one of the nearest giant molecular clouds \citep[$d= 140$\,pc, from ][]{Kenyon1994} with low-mass star-forming activity. 

First we present the HFI data and the ancillary observations used for this paper (Sect.\,\ref{Observations}). Then we describe the emission spectrum of thermal dust measured by HFI and {\it IRAS\/}, and discuss the validity of the measured spectra fitting with one single modified blackbody (Sect.\,\ref{Emission_spectrum}). Maps of the dust temperature and of the spectral emissivity index are analysed in Sect.\,\ref{map_analysis}. Then in Sect.\,\ref{Dust_opacity_per_NH} we discuss the evolution of the dust optical depth/column density conversion factor ($\tau/ N_{\rm H}$) from the atomic diffuse regions to the molecular regions. 

\section{Observations}
\label{Observations}
\subsection{HFI data}
\label{HFI data}

\begin{figure*}
\begin{center}
   \includegraphics[angle=90,width=7cm]{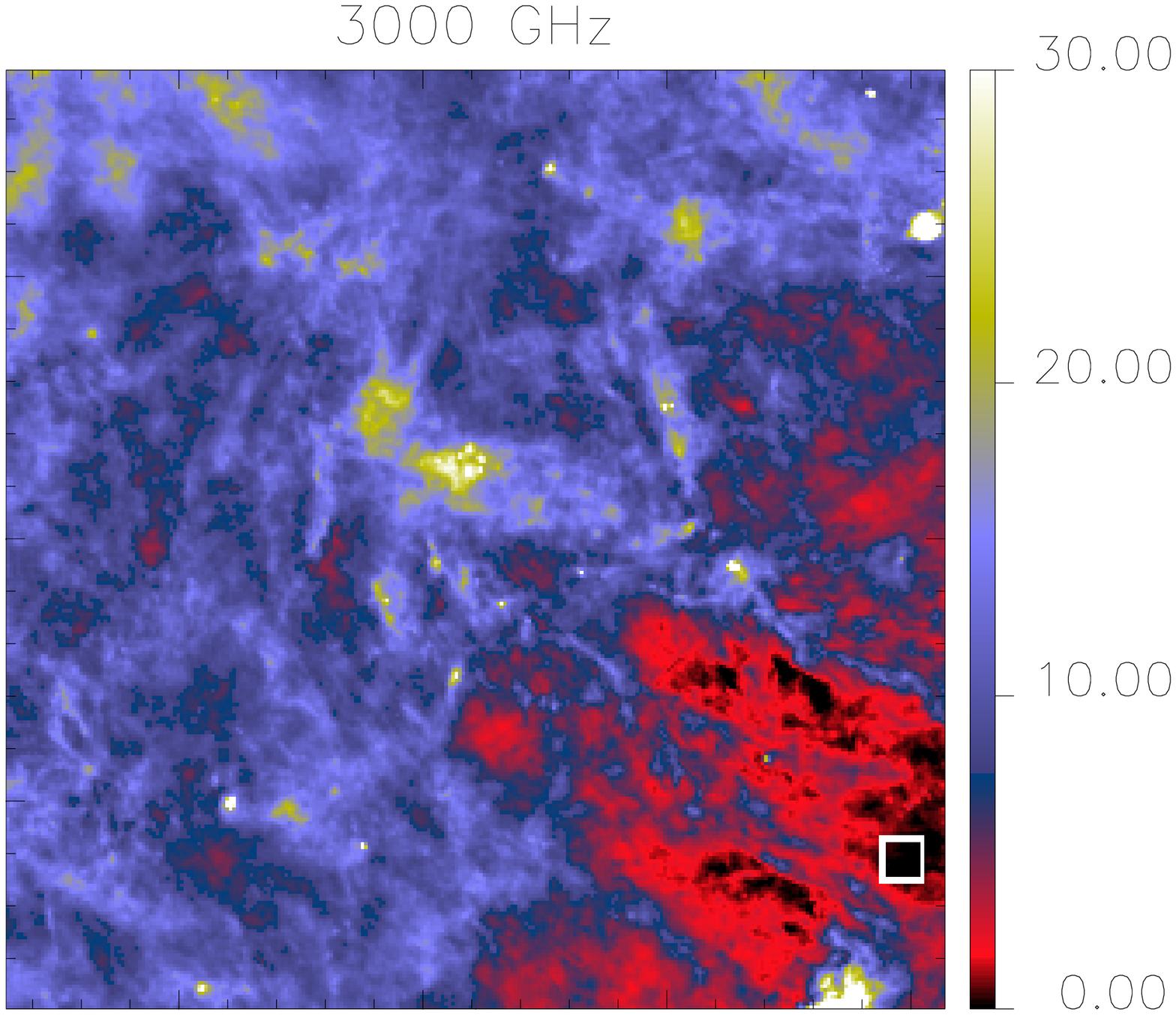}
   \includegraphics[angle=90,width=7cm]{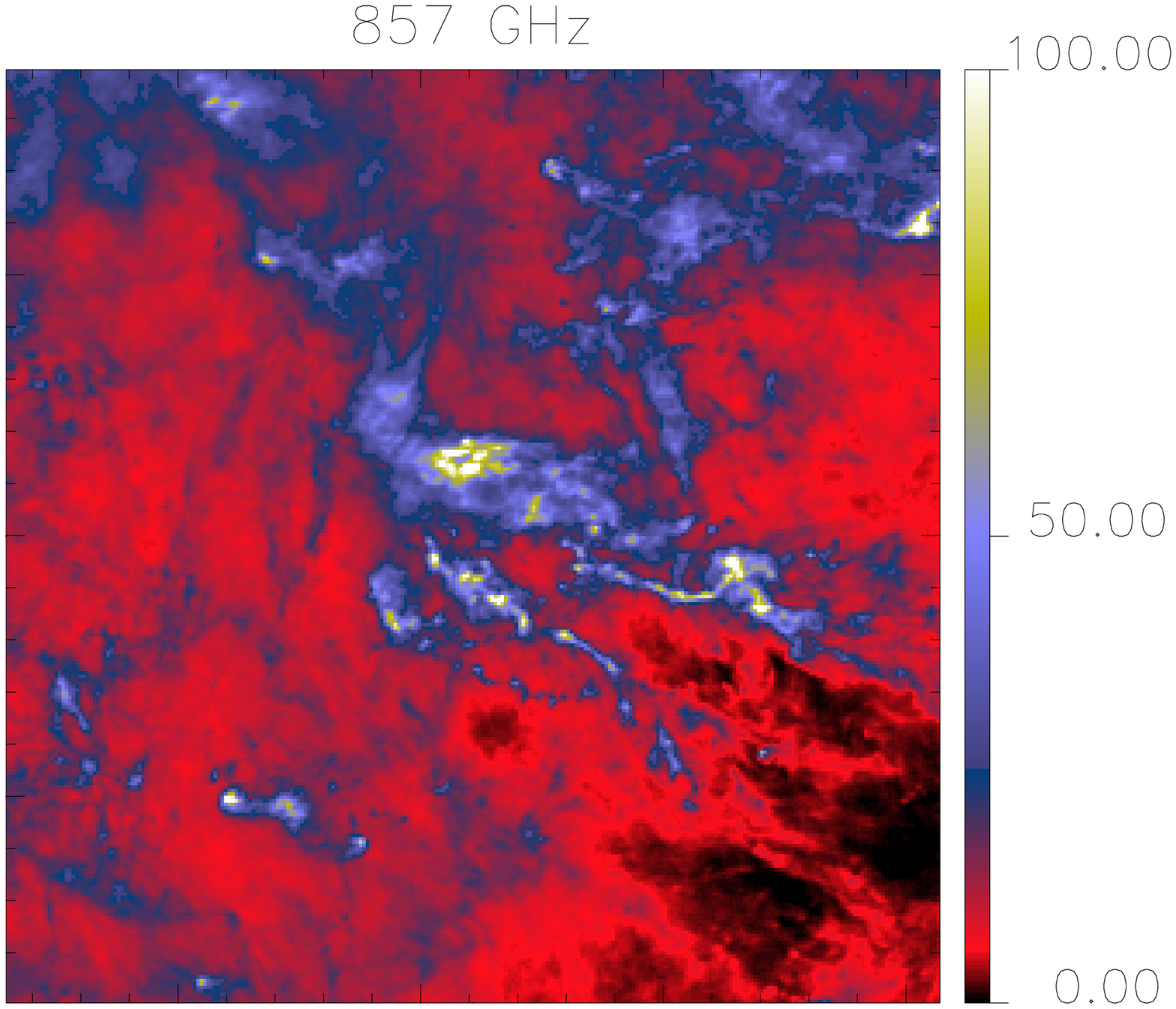}
  \vskip-0.7cm
   \includegraphics[angle=90,width=7cm]{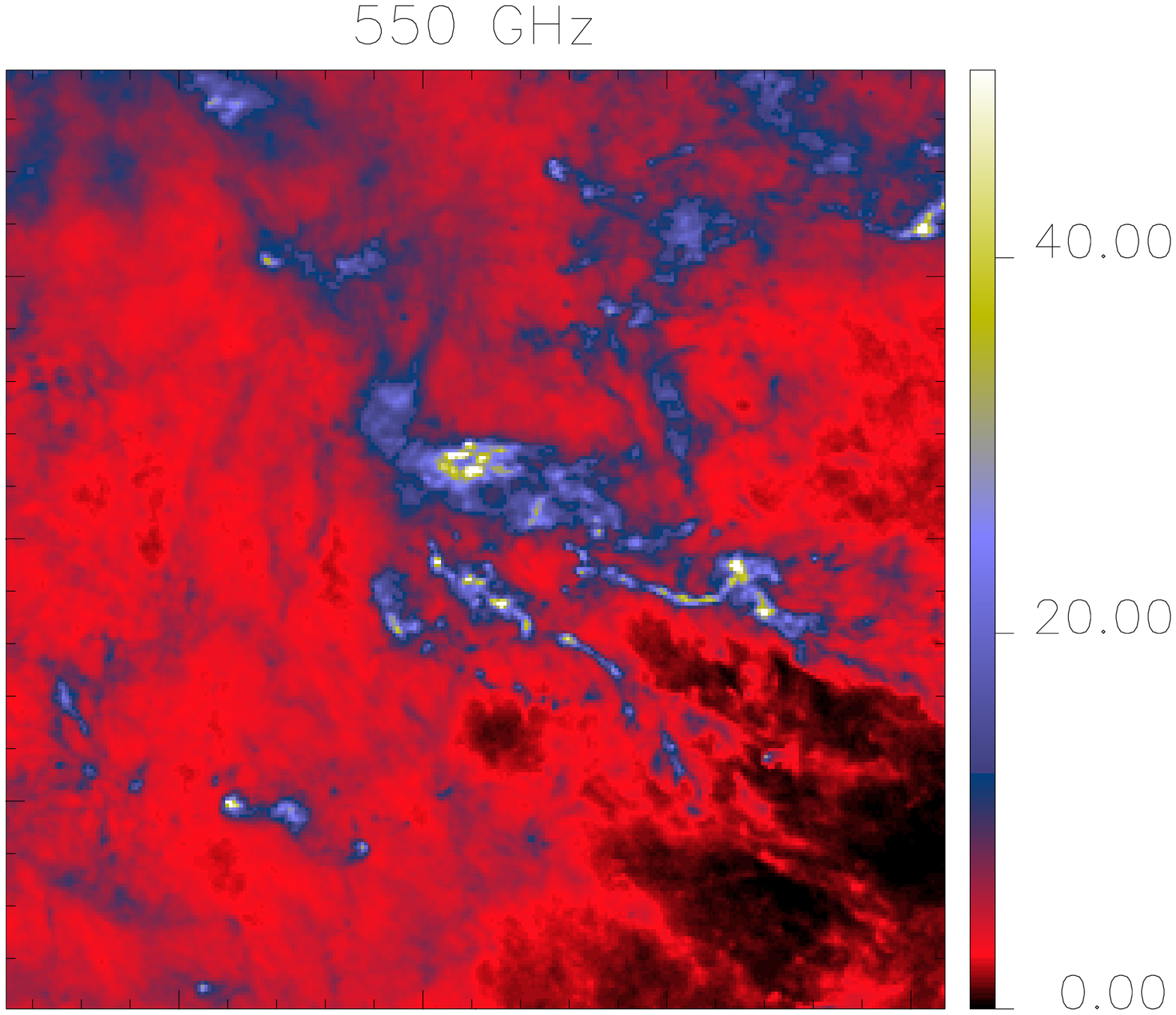}
  \includegraphics[angle=90,width=7cm]{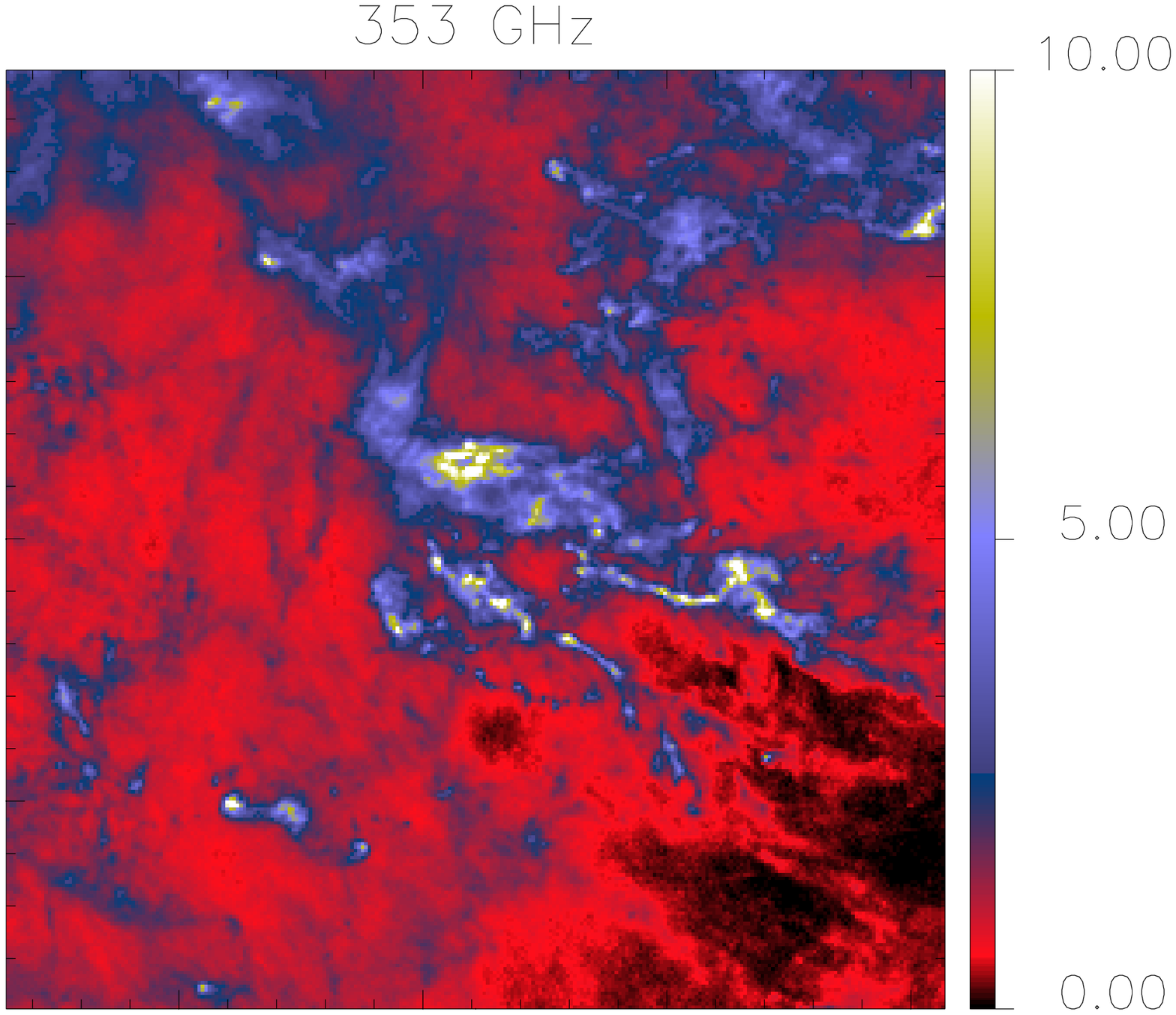}
 \vskip-0.7cm
   \includegraphics[angle=90,width=7cm]{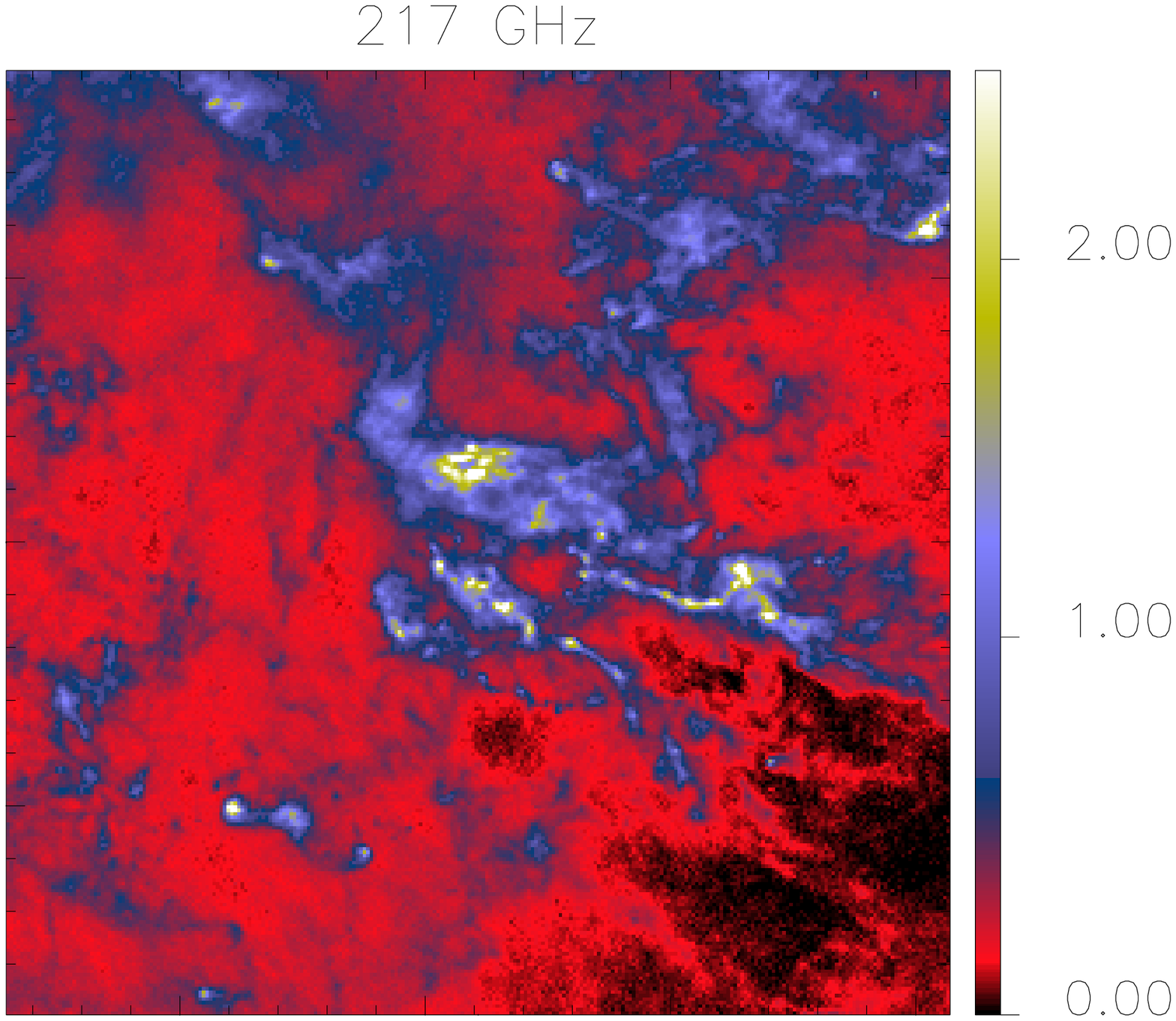}
   \includegraphics[angle=90,width=7cm]{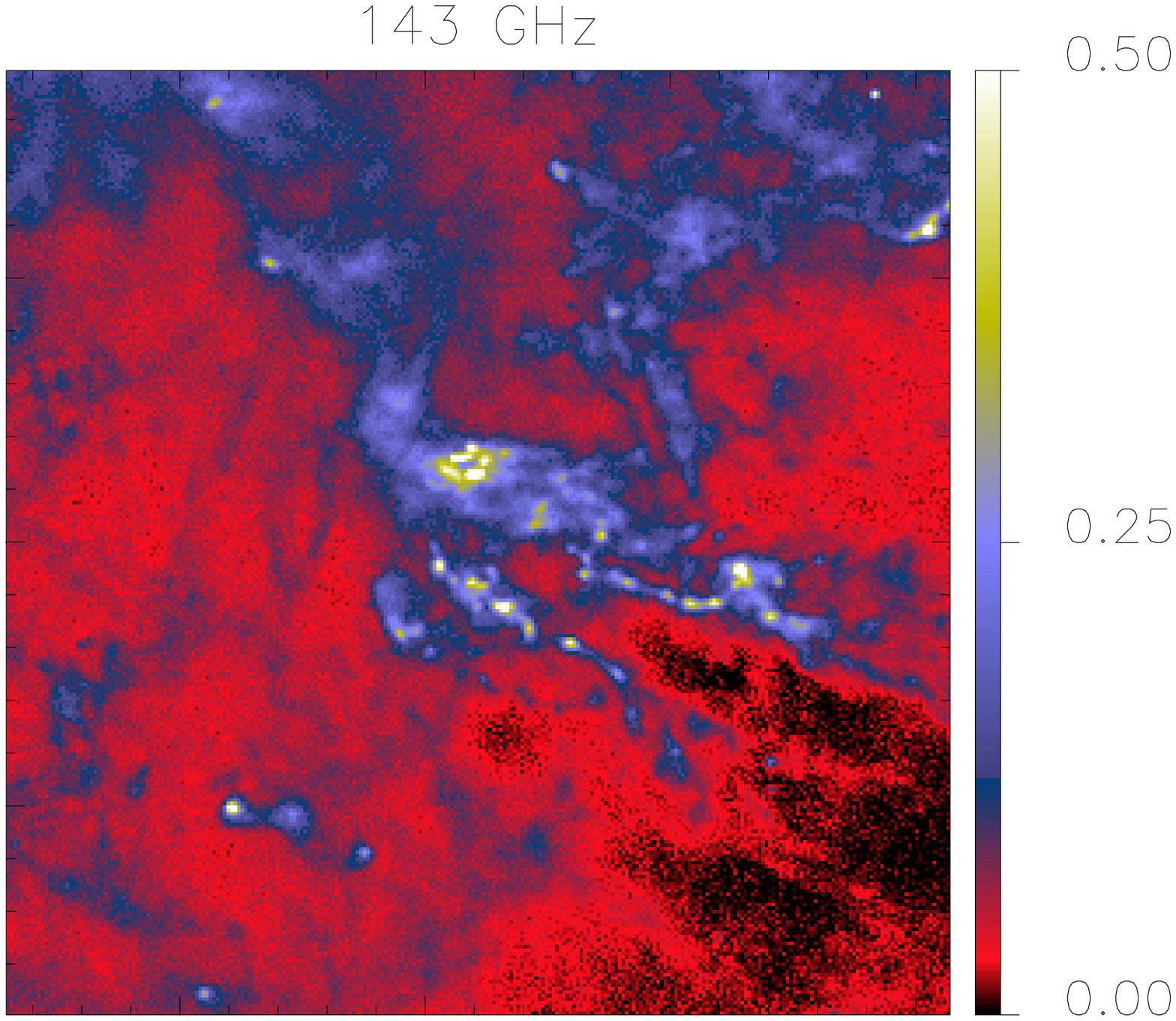}
  \vskip-0.7cm
  \includegraphics[angle=90,width=7cm]{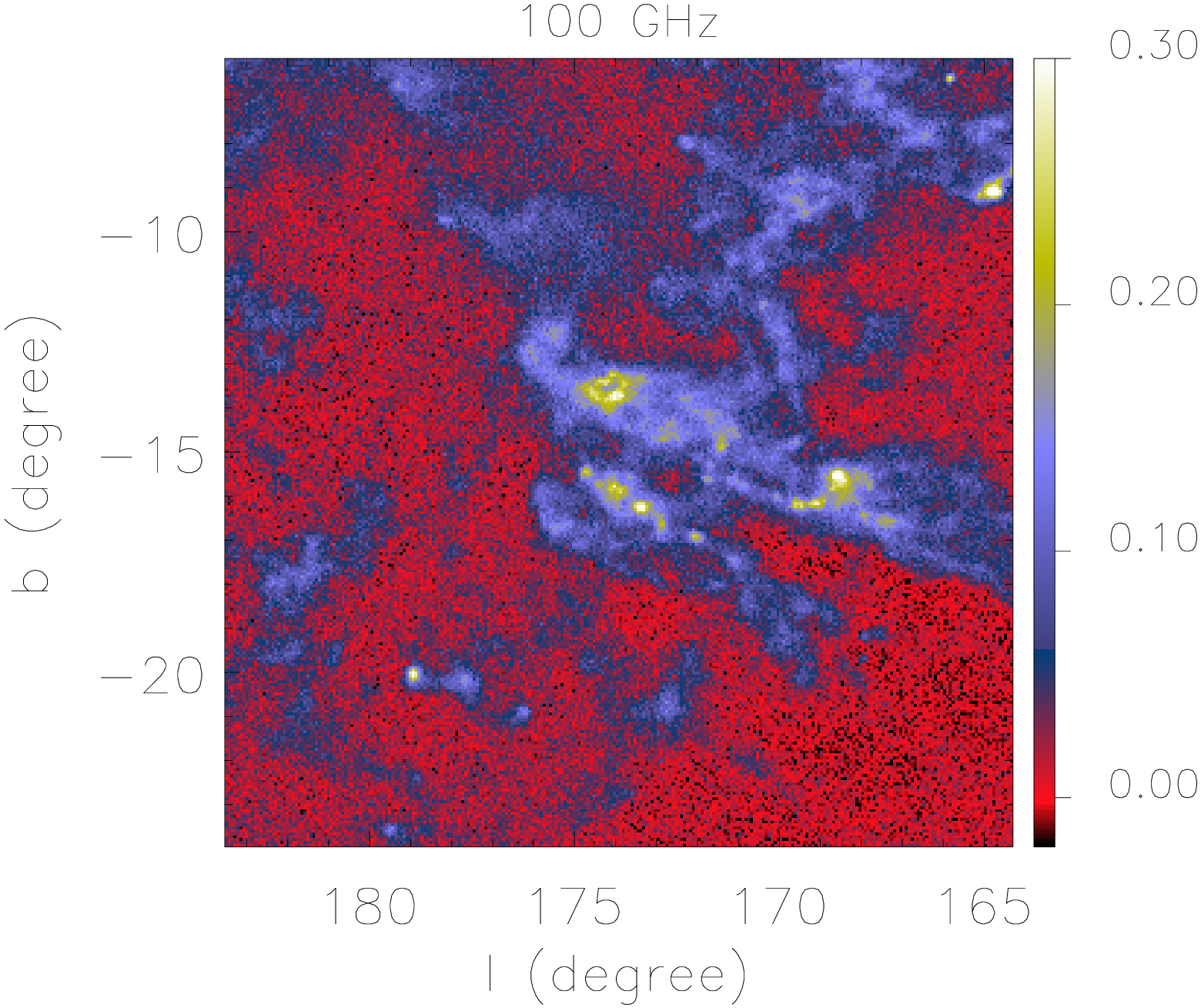}
 \vskip0.5cm
\caption{{\it IRAS\/} and HFI maps of the Taurus molecular cloud, in MJy\,sr$^{-1}$. The $48\arcmin \times48\arcmin$ reference window is seen on the {\it IRAS\/} map at 100\microns\ (3000\,GHz).  For all maps the average brightness computed within the reference window is subtracted.}
\label{fig_taurus_images}
\end{center}
\end{figure*}

We used the DR2 release of the HFI maps presented in Fig.\,\ref{fig_taurus_images} and begin with Healpix \citep{Gorski2005} with $N_{\rm side}$ = 2048 (pixel size 1\parcm7). The data processing and calibration are described in \cite{planck2011-1.7}. An important step is the removal of the CMB through a needlet internal linear combination method. 

The systematic calibration accuracy of HFI is summarized in Table\,1 (from \citealt{planck2011-1.5}). For the two bands at 857 and 545\,GHz, the gain calibration is performed using FIRAS data. The CMB dipole is used at lower frequencies. The systematic errors on the gain calibration are $7\%$ for the 857 and 545\,GHz bands, respectively (estimated using the dispersion in different regions of the sky), and about $2\%$ for the other bands. 

We estimated the statistical noise as follows \citep[see also Appendix\,B of][for details]{planck2011-7.12}. Two independent maps of the sky have been computed by the DPC from the first- and second-half ring of each pointing period. Because the coverage is identical for these two maps, the standard deviation of their half difference, $\sigma_{\rm HR}$, is equal to the standard deviation of the average map. For all bands, we compared the computed value of $\sigma_{\rm HR}$ with the standard deviation $\sigma_{\rm ref}$ of the DR2 map computed within a reference $48\arcmin\times48\arcmin$ window centered at $l=165\pdeg43$, $b=-21\pdeg06$, chosen in the lowest part of the map (white square in the first image of Fig.\,\ref{fig_taurus_images}). Obviously, the values of $\sigma_{\rm ref}$ give only upper limits on the statistical noise because of the contributions of CMB residuals, and of cosmic infrared background (CIB) and thermal dust fluctuations that increase with increasing frequencies. Table\,1 shows that for the 100, 143, and 217\,GHz bands, $\sigma_{\rm HR}$ is almost identical to $\sigma_{\rm ref}$. Therefore we conclude that
\begin{enumerate}
\item the standard deviation of CMB residuals is lower than the statistical noise in all bands; 
\item the standard deviation of the CIB anisotropies (CIBA) appears lower than the measured statistical noise in the 100, 143, and 217\,GHz bands (which is compatible with the CIBA measurements in \citealt{planck2011-6.6});  
\item the computed values of $\sigma_{\rm HR}$ give realistic estimates of the statistical noise. 
\end{enumerate}
For this early analysis, we used a constant statistical noise in each band, taken equal to the values of $\sigma_{\rm HR}$ in Table\,1.  
\begin{table*}  
\caption{Calibration accuracy, statistical noise, standard deviation within the reference window (white square in the \bf{3000\,GHz} \rm panel of Fig.\,\ref{fig_taurus_images}), median brightnesses, and average spectrum within the reference window. 
\label{tab_accuracy_noise} }
\begin{center}
\begin{tabular}{lllllllll}
\hline \\[-0.3cm]
\hline \\[-0.3cm]
Frequency (GHz)  & 100 & 143 & 217 & 353 & 545 & 857 & 3000\\
\hline \\[-0.3cm]
Calibration accuracy     &   2\%  & 2\% & 2\% & 2\% & 7\% & 7\% & 13.5\%\\
Statistical noise from $\sigma_{\rm HR}$ (MJy\,sr$^{-1}$)      & 0.014 & 0.0086 & 0.018 & 0.038 & 0.049 & 0.074 &  \\
% (MJy\,sr$^{-1}$)      & 0.013 & 0.0081 & 0.017 & 0.035 & 0.042 & 0.048 &  \\
%Mean value of the statistical noise in the reference window (MJy\,sr$^{-1}$)      & 0.013 & 0.0078 & 0.019 & 0.038 & 0.034 & 0.035 &  \\
$\sigma_{\rm ref}$   (MJy\,sr$^{-1}$)   &  0.013 & 0.0078 & 0.019 & 0.050 & 0.13& 0.35 &  0.37\\
Median brightness  (MJy\,sr$^{-1}$)   &  0.042 & 0.094 & 0.41 & 1.67 & 6.37& 17.46 &  13.9\\
Reference spectrum  (MJy\,sr$^{-1}$)   &  0.010 & 0.027 & 0.11 & 0.43 & 1.48 & 4.1 &  4.8\\
\hline \\[-0.3cm]
%\hline \\[-0.3cm]
\end{tabular}
 \end{center}
\end{table*}

\subsection{Ancillary data}
\label{ancillary_data}

We combined the \Planck\ maps with {\it IRAS\/} maps at 100\,$\mu$m (3000\,GHz), using the IRIS (Improved Reprocessing of the {\it IRAS\/} Survey) data computed by \cite{Miville2005}. The statistical noise of the 100\,$\mu$m maps is about 0.06\, MJy\,sr$^{-1}$ per pixel (pixel size of 1.8$\arcmin$), while the systematic error in the gain calibration from DIRBE is estimated to be 13.5\% (from \citealt{Miville2005}). For the Taurus molecular cloud, the 100\,$\mu$m brightness is in the range 1--20\,MJy\,sr$^{-1}$, so the statistical noise translates into relative errors in the range 0.03--6\%, well below the systematic error on the gain. 

The atomic gas was traced on large scales using the \ion{H}{i} data at 21\,cm taken with the Leiden/Dwingeloo 25-m telescope with an angular resolution of 36$\arcmin$ by \cite{Hartmann1997}. The velocity spacing was 1.03\,km\,s$^{-1}$, and the local standard of rest (LSR) velocity range was $-450 < V_{LSR} < 400$\,km\,s$^{-1}$. The data were corrected for contamination from stray light radiation to the 0.07\,K sensitivity level \citep{Hartmann1996}. 

The large-scale survey in the $^{12}$CO ($J = 1\rightarrow0$) emission line taken by \cite{Dame2001} with the CfA telescope was used to trace the molecular gas. The beam width was $8.8\arcmin\,\pm0.2\arcmin$. For the observations of the Taurus molecular cloud, the sampling distance was equal to 7.5$\arcmin$, the channel width is 0.65\,km\,s$^{-1}$ and the channel rms noise was 0.25\,\hbox{K}. The data cubes were transformed into the velocity-integrated intensity of the line ($W_{\rm CO}$) by integrating the velocity range where the CO emission is significantly detected using the moment method proposed by \cite{Dame2001}. The statistical noise level of the $W_{\rm CO}$ map is typically $1.2$\,K\,km\,s$^{-1}$. 

In order to analyse the central region of the Taurus molecular cloud with the $^{13}$CO $J = 1\rightarrow0$ emission lines, we also used the FCRAO survey of 98\,deg$^2$ conducted with an angular resolution 47$^{\prime\prime}$ \citep{Narayanan2008, Goldsmith2008}. We applied the error beam scaling factor recently proposed by \cite{Pineda2010}, so that the intensities of the CfA and FCRAO surveys of the $^{12}$CO emission line are compatible. The statistical noise per pixel (with pixel size of 0.33$\arcmin$) of the velocity-integrated intensity maps is about $0.4$\,K\,km\,s$^{-1}$ for the   $^{13}$CO line. 

\begin{figure*}
\begin{center}
\includegraphics[angle=90,width=7cm]{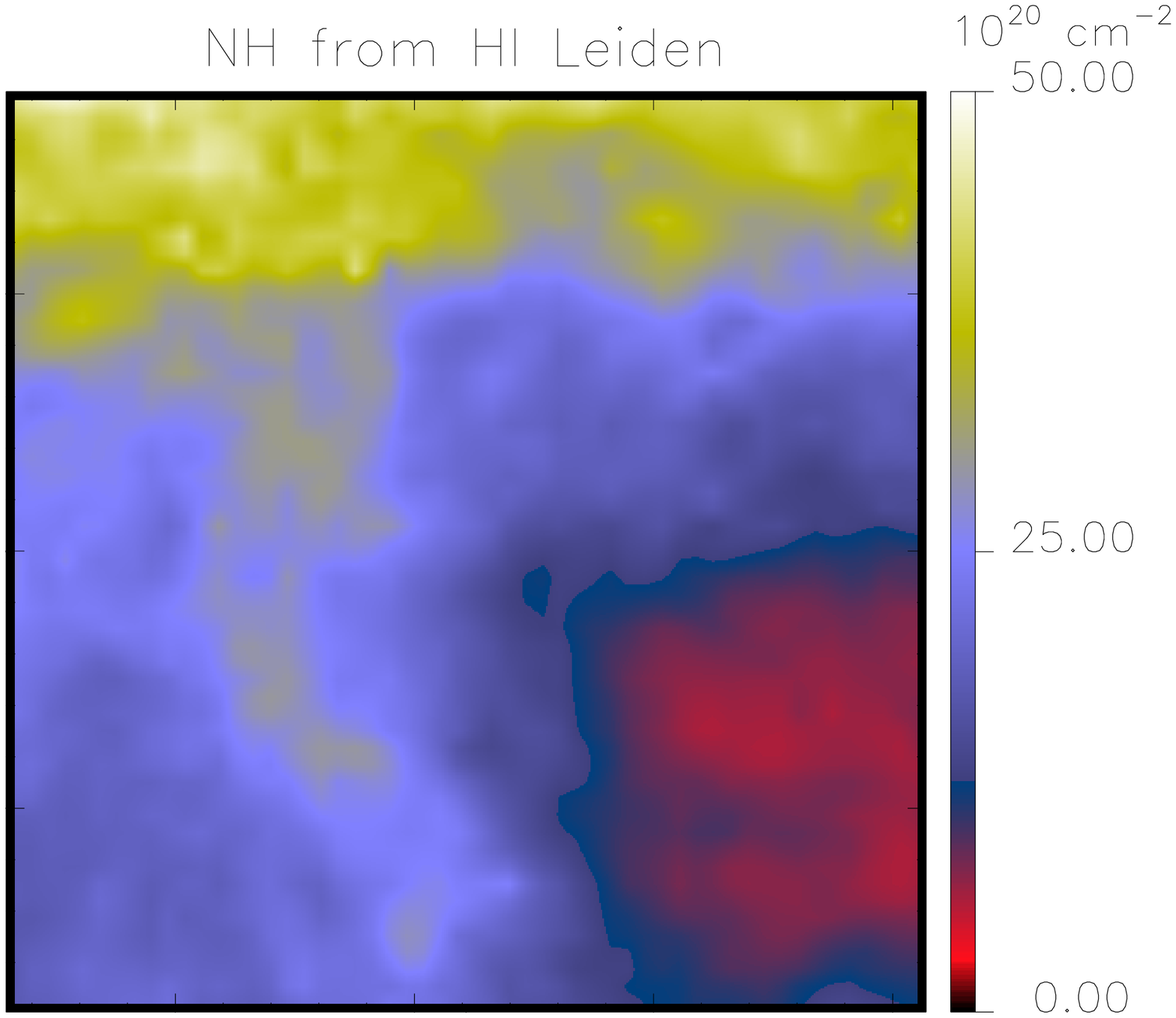}
     \includegraphics[angle=90,width=7cm]{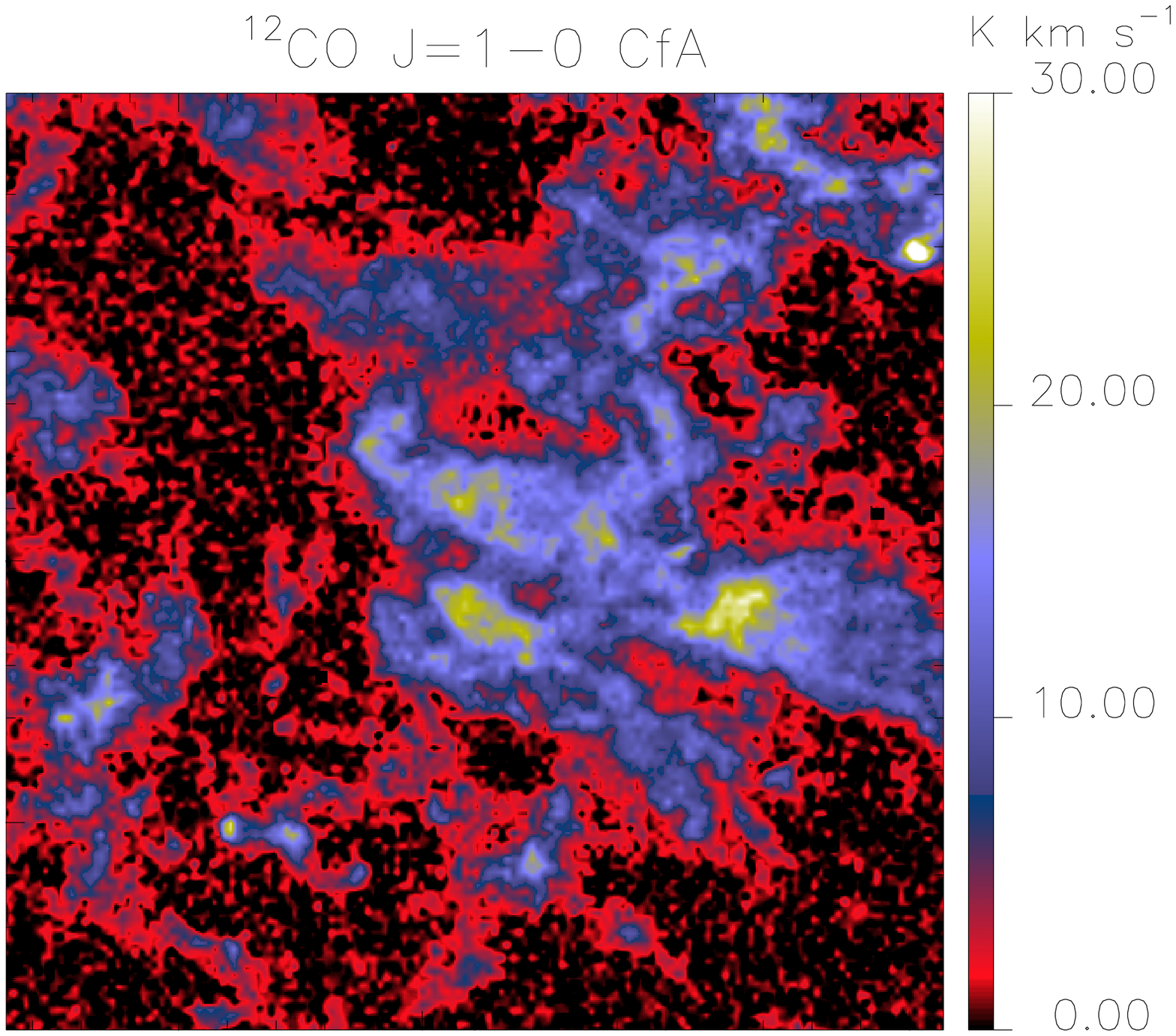}
 \vskip-0.6cm
\includegraphics[angle=90,width=7cm]{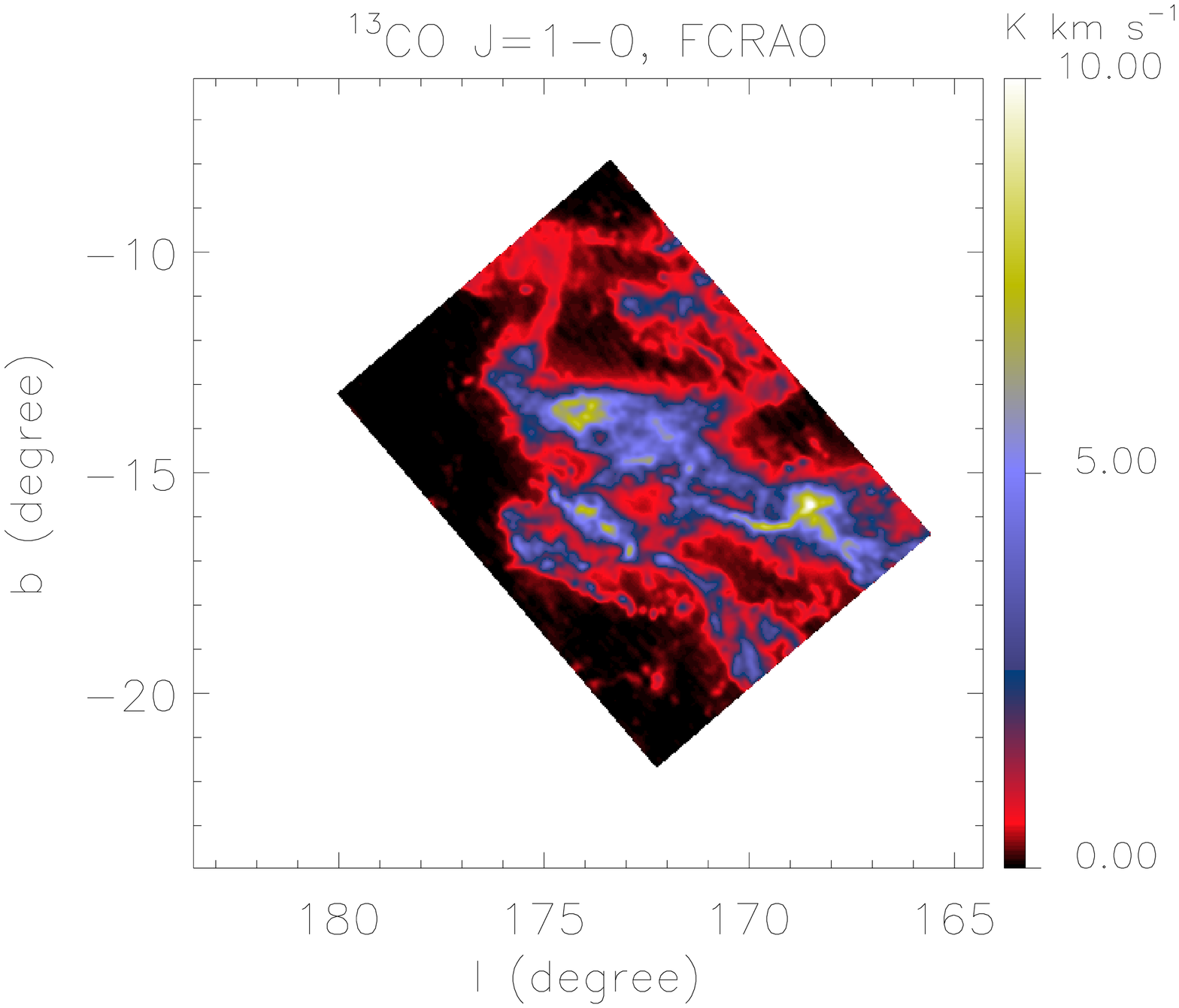}
     \includegraphics[angle=90,width=7cm]{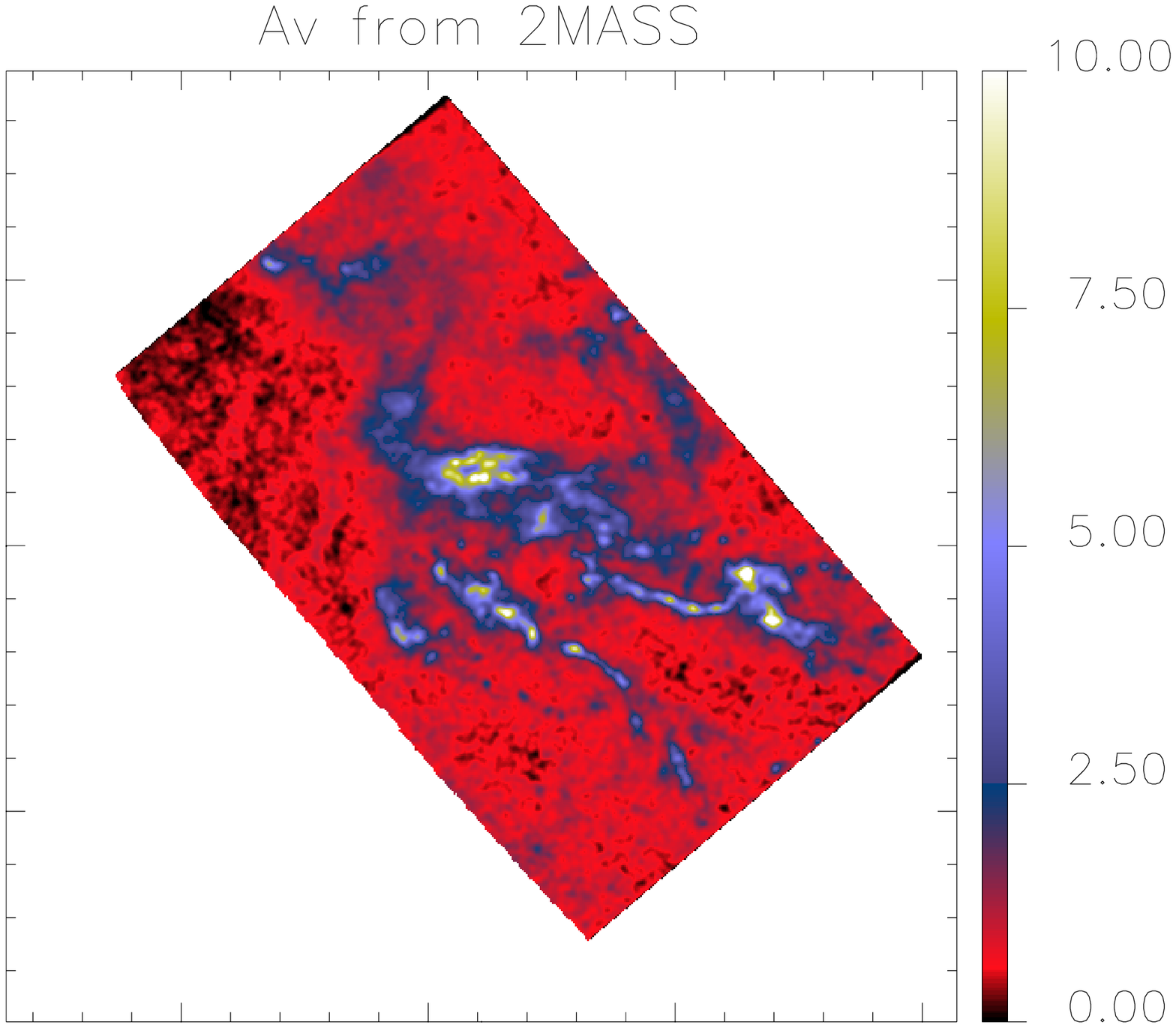}
  \vskip0.3cm
      \caption{
     Upper left panel: column density derived from the  \ion{H}{i} data at 21\,cm \citep{Hartmann1997}. Upper right panel:    $^{12}$CO ($J = 1\rightarrow0$) velocity integrated emission line \citep{Dame2001}. Lower left panel: $^{13}$CO $J = 1\rightarrow0$ velocity integrated emission line \citep{Narayanan2008, Goldsmith2008}. Lower right panel: NIR extinction using 2MASS \citep{Pineda2010}. The $^{13}$CO and NIR extinction maps are smoothed at the angular resolution of the 143\,GHz \hbox{HFI} band (FWHM: 7.08$\arcmin$). 
     \label{fig_taurus_ancillary_maps}}
\end{center}
\end{figure*}

The column density can also be traced from the near-infrared (NIR) extinction using the 2MASS point source catalogue (e.g., \citealt{Dobashi2005, Pineda2010}). For comparison with HFI data, we used the extinction map of the central part of the Taurus molecular cloud recently created by \cite{Pineda2010} 
and shown in Fig.\,\ref{fig_taurus_ancillary_maps}. It is Nyquist-sampled with an angular resolution of 200$^{\prime\prime}$, and corrected for the contribution of atomic gas to the total extinction. %The infrared colours are converted to visible extinction $A_V$ or column densities $N_{\rm H}$ using the extinction curve of \cite{Weingartner2001ApJ...548..296W}, adopting the standard ratio of selective to total extinction $R_V$=3.1 for the diffuse ISM \citep{Savage1979}. 
Using the extinction curve of \cite{Weingartner2001ApJ...548..296W} and adopting the standard ratio of selective to total extinction $R_V$=3.1 for the diffuse ISM \citep{Savage1979}, the infrared colours are converted to visible extinction $A_V$ or column densities $N_{\rm H}$ ($= A_V\times1.87\times 10^{21}$ cm$^2$). 
%TO BE DISCUSSED IN THE FINAL DISCUSSION  At large densities (with typical extinction $A_V>3$) $R_V$ is expected to increase up to $\sim4.5$ due to grain growth by accretion and coagulation \citep{Whittet2001}. 
Because the number of background stars used to compute the extinction decreases with increasing extinction, the error in $A_V$ increases from about 0.2\,mag at low extinctions ($A_V = $ 0--1\,mag) to 0.5\,mag at high extinctions ($A_V\sim10\,$mag), with an average error of 0.29\,mag. For higher extinctions ($A_V > 10$\,mag) the extinction map only gives lower limits.       

\section{Emission spectrum of thermal dust}
\label{Emission_spectrum}
\subsection{Reference spectrum}
\label{Reference_spectrum}
The combination of {\it IRAS\/} and \Planck\ data provides the spectral energy distribution (SED) from 3000\,GHz (100\microns) to 100\,GHz (3\,mm) for each pixel of the maps. The CMB has been removed from the maps we use. The maps could contain some CMB residual, but with an amplitude lower than the statistical noise even in the low frequency channels (see Sect.\,\ref{HFI data}). 

\begin{figure}
  % \centering
      \includegraphics[angle=90,width=\columnwidth]{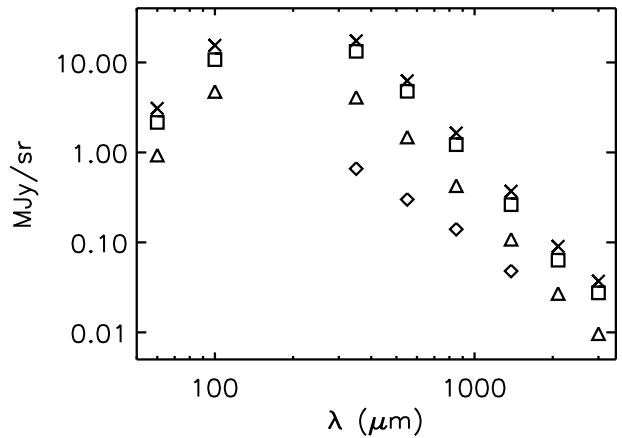}
\vskip -4mm
      \caption{Spectrum of one pixel in the non-molecular region: crosses, total brightness $I_{\rm tot}$; triangles, average brightness $I_{\rm ref}$ within the reference $48\arcmin\times48\arcmin$ window (seen on the 3000\,GHz panel of Fig.\,\ref{fig_taurus_images}); squares, $I_{\rm tot} - I_{\rm ref}$; and diamonds, CIB spectrum at the central frequency of the HFI filters from 857 to 217\,GHz, from \cite{Fixsen1998}.}
     \label{fig_plot_pixels_Taurus_ref}
\end{figure}

In all bands, the maps contain Galactic and non-Galactic emission that is not associated with the Taurus complex. Therefore, we subtracted for all maps the average brightness computed within the reference $48\arcmin\times48\arcmin$ window, chosen in the faintest region (white square in the first image of Fig.\,\ref{fig_taurus_images} and central coordinates given above in Sect.\,\ref{HFI data}) of the map.  This is illustrated for one pixel in Fig.\,\ref{fig_plot_pixels_Taurus_ref}. The average spectrum in the reference window is also shown in Fig.\,\ref{fig_plot_pixels_Taurus_ref}, together with the CIB spectrum (from FIRAS data by Fixsen et al.\,1998) at the central frequencies of the HFI filters. We see that the reference spectrum is mainly caused by Galactic atomic emission (there is no detected emission within the $^{12}$CO $J = 1\rightarrow0$ line), since the CIB contributes not more than 5--10\%. 

In order to derive the dust optical depth per column density in the atomic phase (Sect.\,\ref{Optical_depth_atomic}), the emission measured within the same reference window will be subtracted from the \ion{H}{i} data. 

\subsection{Choice of the spectral bands}
\label{spectral_bands}
We focus our analysis on the emission of dust particles in equilibrium between thermal emission and absorption of UV and 
visible photons from incident radiation. Therefore, we did not use the {\it IRAS\/} maps at 12, 25, and 60\,$\mu$m because of the contribution of small dust particles transiently heated each time they absorb a UV/visible photon. In all spectra presented in this paper, we left the 60\,$\mu$m data points in the figures to illustrate this contribution, but these data points were not used to analyse the \hbox{SED}s. 
In the 100\,$\mu$m band, the contribution of these small particles is expected to be lower than $10\%$ if the intensity of the interstellar radiation field is of the order of the value in the local diffuse ISM \citep[e.g., ][]{Compiegne2011}, as is the case in the Taurus molecular complex. This contribution is in any case lower than the gain calibration error of {\it IRAS\/} at 100\,$\mu$m (13.5\%, see Sect.\,\ref{ancillary_data}). 

For this early analysis, we did not use the data taken at 100 and 217\,GHz to analyse the thermal dust emission because they are expected to be contaminated in the molecular regions by rotational $J = 1\rightarrow0$ and $J=2\rightarrow1$ $^{12}$CO emission lines and by $^{13}$CO emission lines with a lower amplitude. Higher $J$-lines in the other bands have a lower relative amplitude and are neglected for this early analysis. We also neglected the emission of molecular lines tracing the densest regions.
Therefore, in a first step 
%this first analysis dedicated to the thermal dust emission 
only the bands at 3000, 857, 545, 353, and 143\,GHz (100, 350, 545, 850, and 2100\,$\mu$m) were used to analyse the thermal dust emission. %We neglect the possible contamination of the  J=3-2 lines to the 353\,GHz band. 

\begin{figure}
%\centering
% \hskip -7mm
    \includegraphics[angle=90,width=\columnwidth]{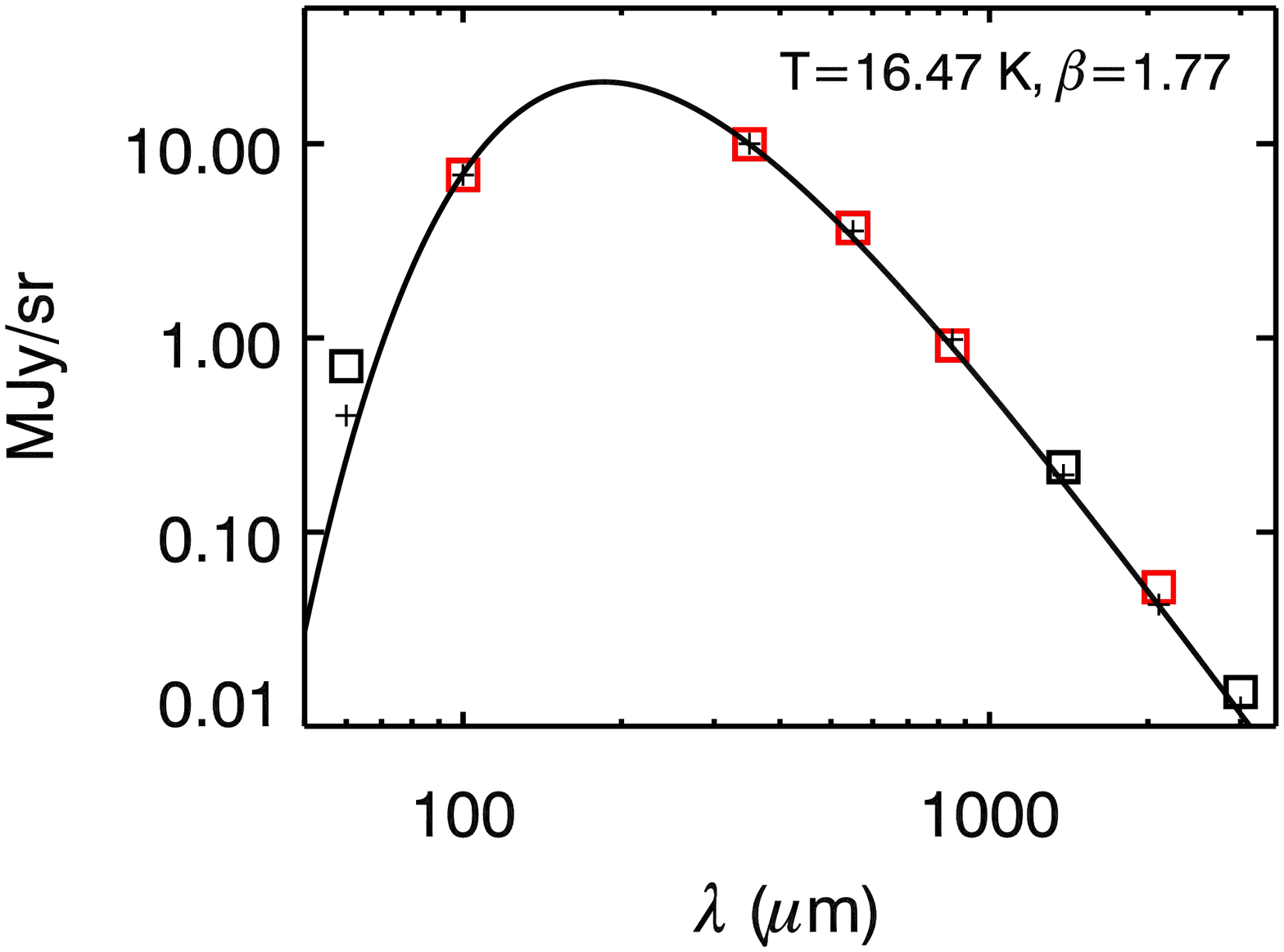}
\vskip -7mm
     \includegraphics[angle=90,width=\columnwidth]{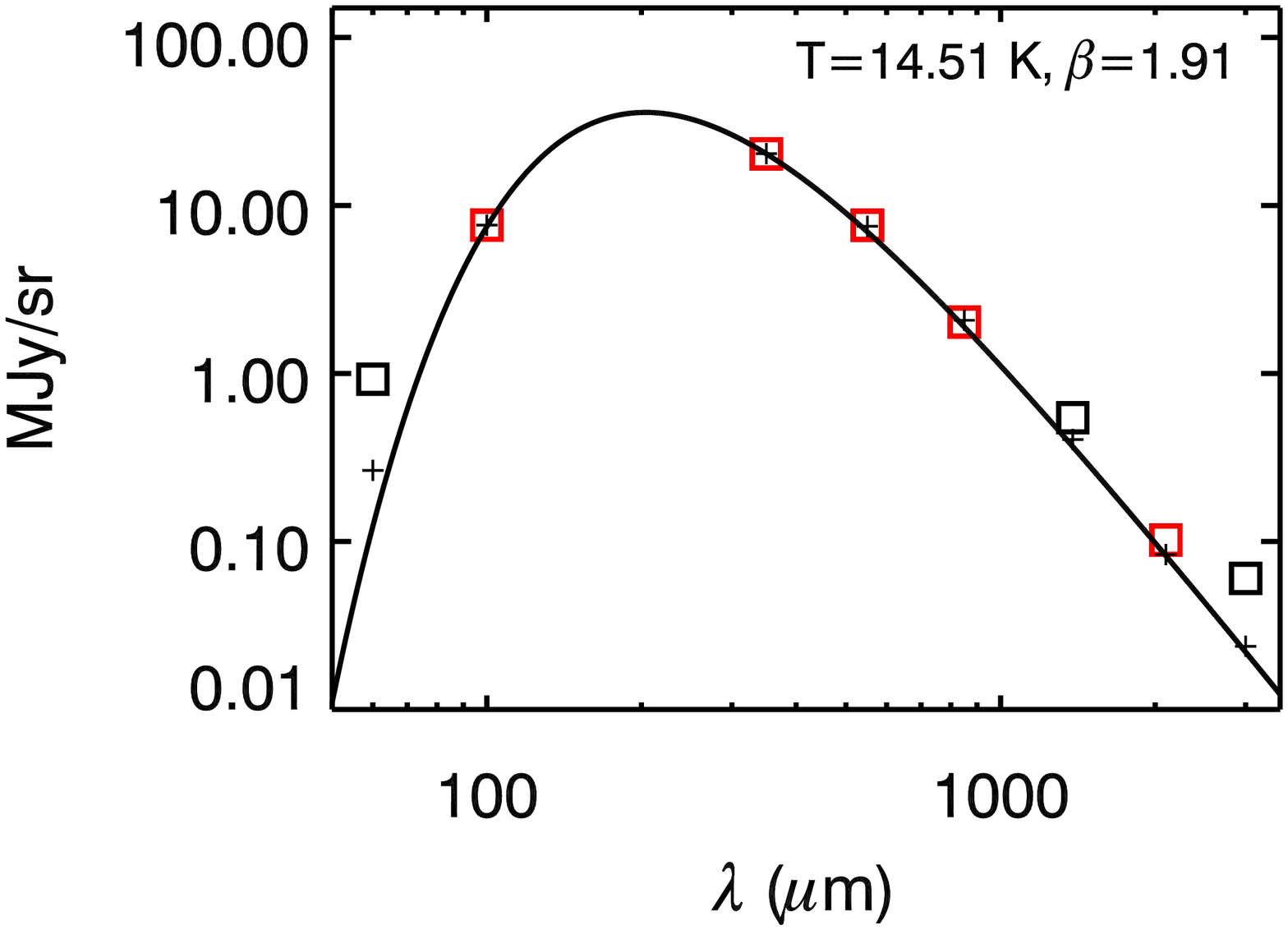}
\vskip -2mm
     \caption{SED of two pixels (top: non-molecular region, bottom: molecular region). The squares are data, the solid line is the fitted model, and the crosses are the fitted model integrated within the bands. The fits are performed using the 100\,$\mu$m, 857, 545, 353 and 143\,GHz bands (red squares), and using the statistical noise discussed in Sect.\,\ref{Observations}, which is too low to be visible on the figure. Significant excess in the 217 and 100\,GHz bands caused by $^{12}$CO and $^{13}$CO emissions are detected in the molecular spectrum. %On the other hand, a significant negative residual is observed at 353\,GHz. 
The 60\,$\mu$m data points are not used to analyse the \hbox{SED} because of the contribution of small dust particles transiently heated each time they absorb a UV/visible photon.
     \label{fig_plot_pixels_Taurus}}
\end{figure}

In practice, we smoothed all maps to the angular resolution of the 143\,GHz data (FWHM: 7.08$\arcmin$), assuming isotropic Gaussian beams and taking the FWHM from \cite{Miville2005} for {\it IRAS\/} and from \cite{planck2011-1.5} for \hbox{HFI}. 
We did not take into account the ellipticity of the PSF \citep{planck2011-1.5}. This may introduce some error for the detailed analysis of individual structures, but our goal is not to to derive any quantitative results for individual structures, but to extract from a pixel by pixel analysis some quantitative information on the emission that emerges from the different phases. The 100\,GHz map will be used to compute the residuals to the fits (see Sect.\,\ref{Analysis_residuals}), and is left at its original resolution (\getsymbol{HFI:FWHM:Mars:100GHz:units}). Figure \ref{fig_plot_pixels_Taurus} shows the spectra taken at positions in the non-molecular region and in the molecular region. % together with fitting of the data points at 100\,$\mu$m, 857\,GHz, 545\,GHz and 143\,GHz. 

\subsection{Principle of fitting}
\label{SEDfitting}
In this early analysis, the fitting function is a single and optically thin modified blackbody:
\begin{equation}
\label{eq_spec}
I_{\nu} ~=~
 \tau_{\nu_{0}} \times
\left(\frac{\nu}{\nu_{0}}\right)^{\beta}\times B_{\nu}(T),% \times N_{\rm H}
\end{equation}
where $\tau_{\nu_0}$ is the dust optical depth at frequency $\nu_0$, $\beta$ is the
spectral emissivity index, $B_{\nu}$ is the Planck
function, and $T$ the dust temperature. All quantitative values of the dust optical depth will be given at the frequency $\nu_0=1200$\,GHz (250\microns) to be comparable to previous analyses, and we define $\tau_{\nu_{\rm 0}}=\tau_{250}$. 

The three computed parameters for each pixel are $T$, $\beta$, and $\tau_{250}$. We used the IDL MPFIT routine, which performs weighted least-squares curve fitting of the data \citep{Markwardt2009} taking into account the noise (statistical noise or calibration uncertainty) for each spectral band. We applied colour-correction factors computed using  version~1 of the transmission curves \citep{planck2011-1.5}.

The goal of this adjustment is to reduce the properties of the SED measured for each line of sight to the set of three parameters $T, \beta$, and $\tau_{250}$. The temperature, spectral emissivity index, and optical depth of the emitting dust particles along the line of sight can obviously vary, so the fitted values of the three parameters, while representing some average properties, cannot give the complete picture of the dust particles located along that line of sight. Moreover, we assumed that the spectral emissivity index $\beta$ of the measured spectra is constant from far-infrared (FIR) to millimetre wavelengths. The spectral emissivity index of dust emission may vary due to temperature-dependent mechanisms at low temperatures, including free-charge carrier processes, two-phonon difference processes, and absorption mechanisms in two-level systems \citep{1996ApJ...462.1026A,2007A&A...468..171M}. Moreover, the measured spectra can be broadened around the peak of the modified blackbody in the submillimeter (submm) because of the contribution of dust at different temperatures. Thus, the spectral emissivity index $\beta$ of the measured spectra could increase from the submm to the millimeter, as illustrated by \cite{2009ApJ...696.2234S}. 

\subsection{Example of spectra. Systematic errors on the parameters derived from the fit}
\label{example_spectra}

Figure\,\ref{fig_plot_pixels_Taurus} is an illustration of the fitting for two pixels, one taken at a position with detected CO emission (within the $^{12}$CO $J = 1\rightarrow0$ line), the other at a position with no detected CO emission. We used for the fitting the statistical noise for HFI and {\it IRAS\/} data discussed in Sect.\,\ref{Observations}, taking into account the smoothing of the map at the angular resolution of the 143\,GHz band. 

Simulations have been performed to understand and to quantify the propagation of the calibration errors, the statistical noise, and the CIBA in the determination of $T, \beta$, and $\tau_{250}$ (see Appendix\,\ref{simulations} for details). The calibration errors propagate into 
systematic errors in $T, \beta$, and $\tau_{250}$ of about 0.7\,K, 0.07, and $18\%$, respectively.
We have also shown that the statistical noise and the CIBA propagates into statistical noise levels in $T$, $\beta$, and $\tau_{250}$ in the range 0.1--1\,K, 0.025--0.25 and 2--20\%, respectively, depending on the 100\micron\ brightness (10--1\,MJy\,sr$^{-1}$). Moreover, both for systematic errors and statistical noises, the three parameters are strongly correlated or anti-correlated (Fig.\,\ref{fig_simu_beta_T_opacity}). 

The two spectra of Fig.\,\ref{fig_plot_pixels_Taurus} show that to first order, a single modified blackbody gives a reasonable representation of the SED measured by {\it IRAS\/} and \hbox{HFI}. Obviously by increasing the number of free parameters in our fit (e.g., by using two modified blackbodies with different temperatures and spectral indices) it is possible to improve the fits significantly, but this is not our goal in this early paper. %Positive residuals in the molecular spectrum and in the 217 and 100\,GHz bands are due to $^{12}$CO and $^{13}$CO emissions. 
We used exactly the same method to fit the spectra for all pixels of the map to derive the temperature map and the spectral emissivity index map shown in Fig.\,\ref{fig_T_and_beta_map}.

\subsection{Analysis of the residuals}
\label{Analysis_residuals}
\begin{figure*}
\begin{center}
\hglue 1cm\includegraphics[angle=90,width=7cm]{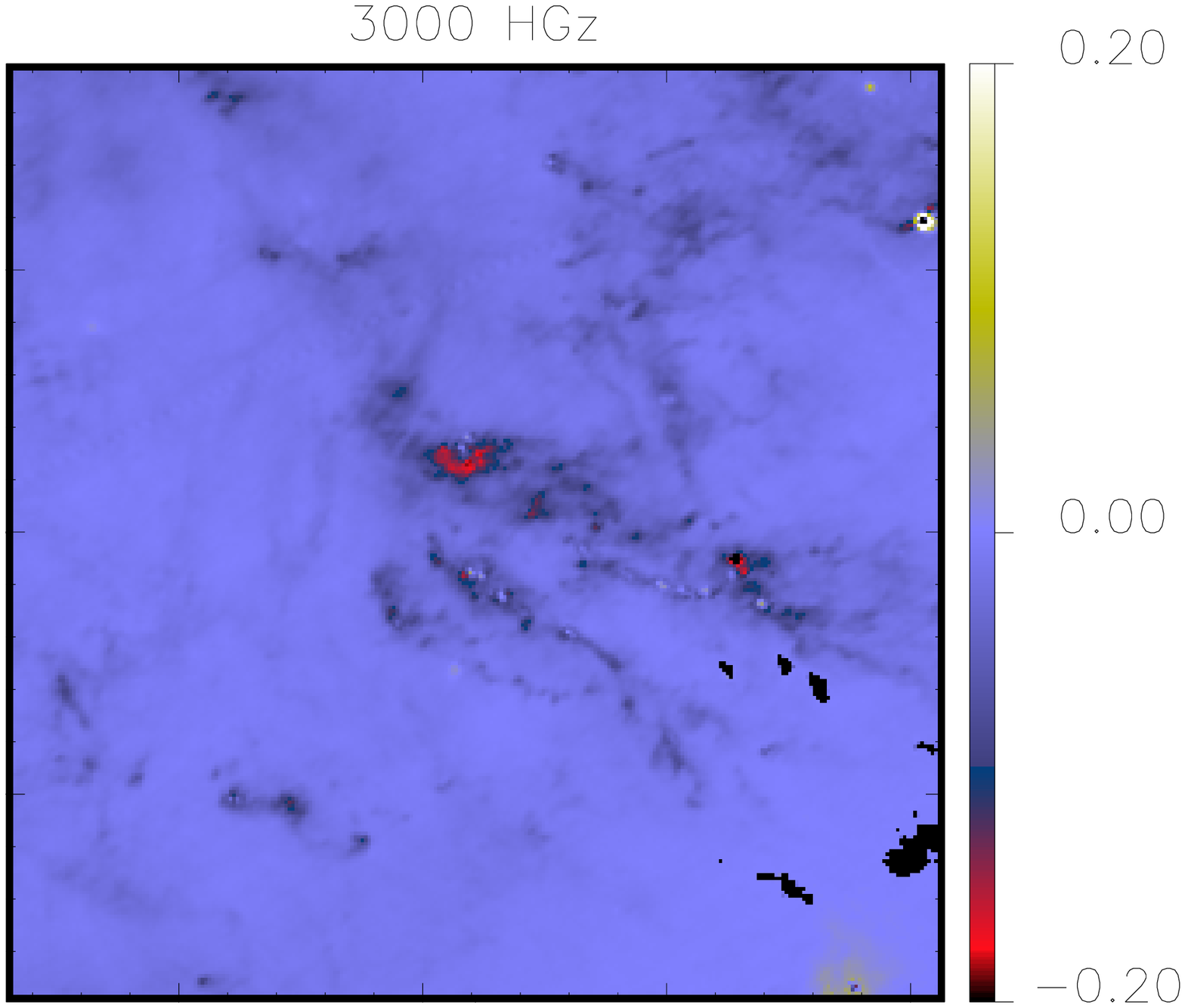}
     \includegraphics[angle=90,width=7cm]{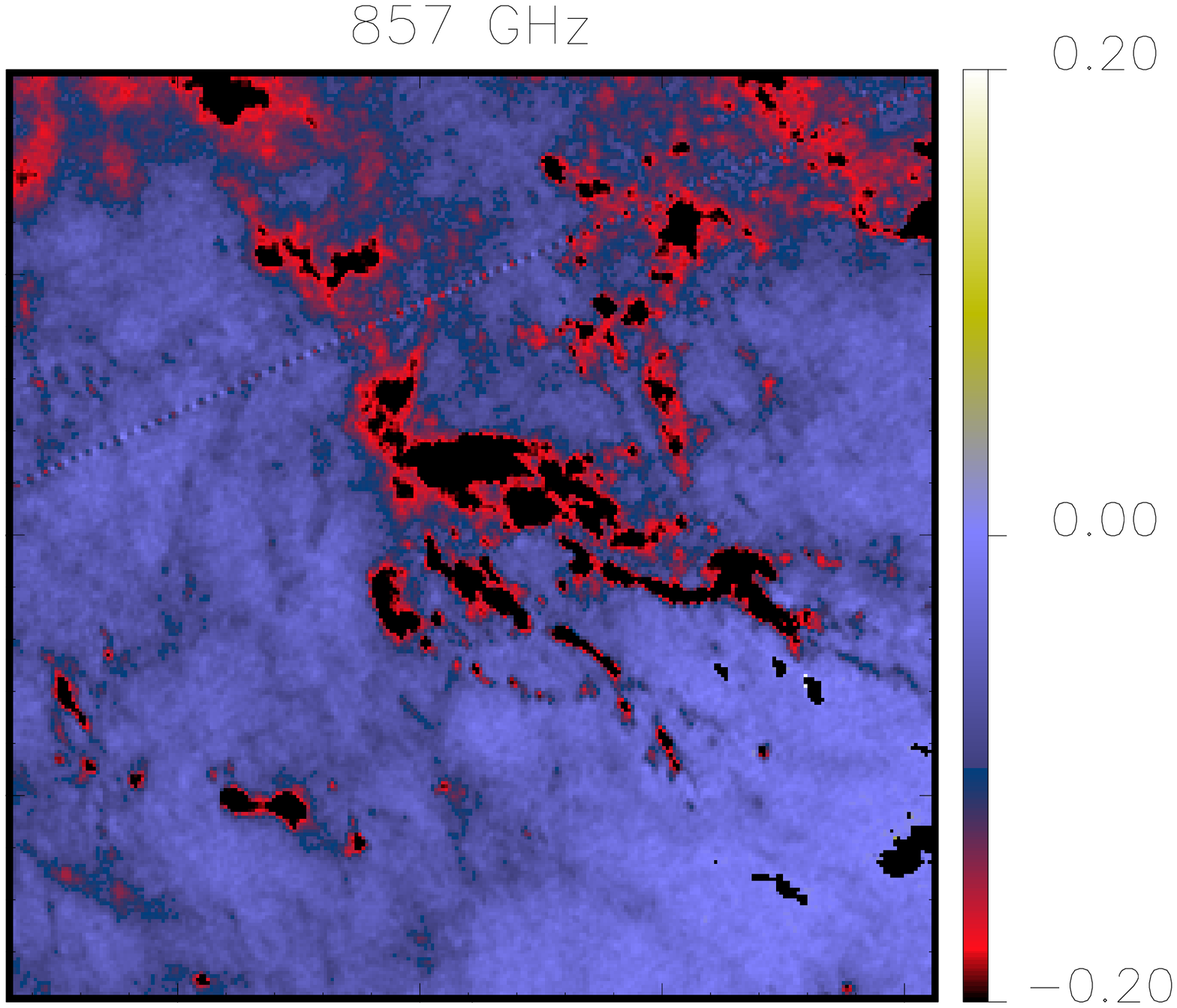}
 \vskip-0.8cm
\hglue 1cm\includegraphics[angle=90,width=7cm]{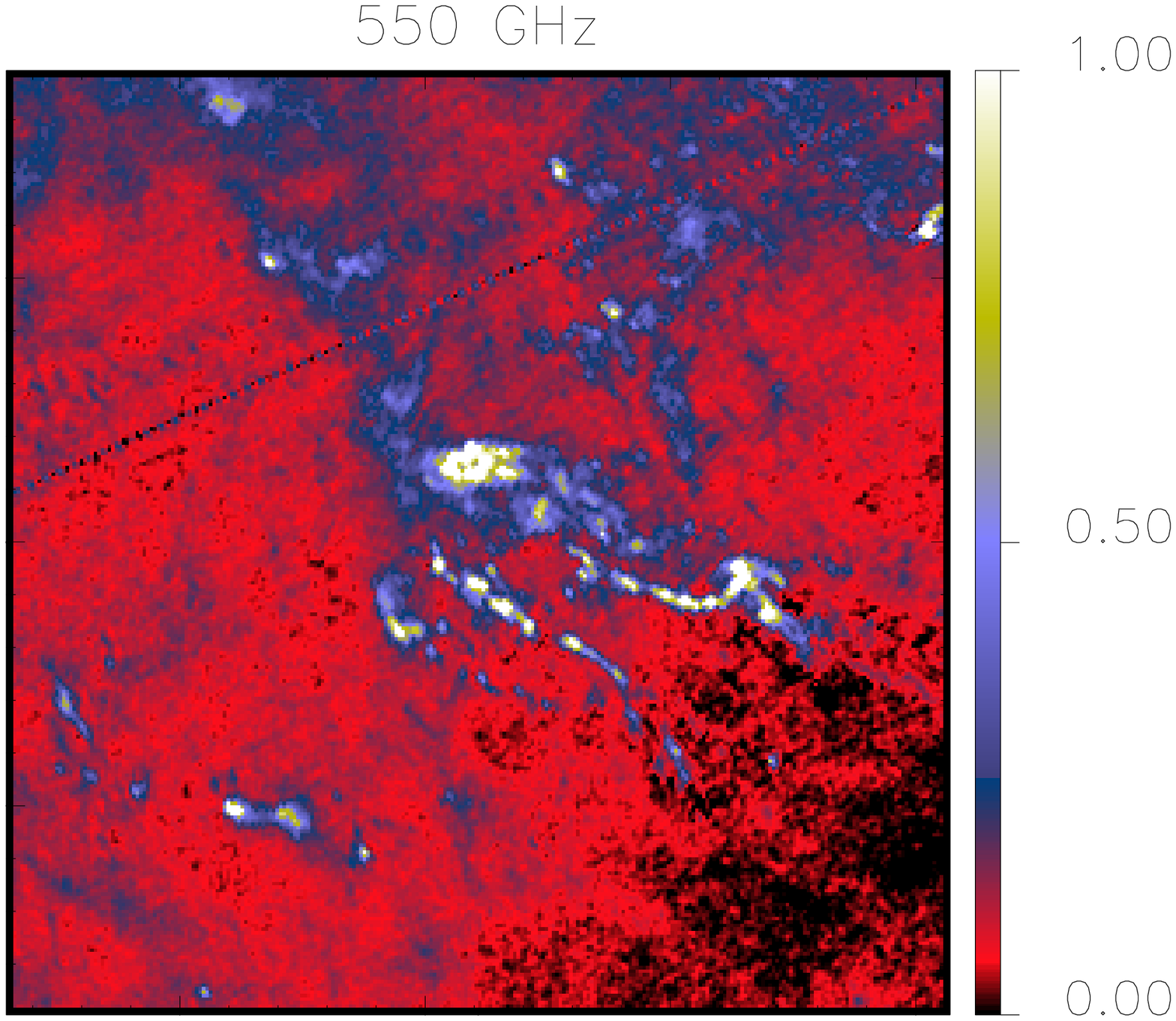}
     \includegraphics[angle=90,width=7cm]{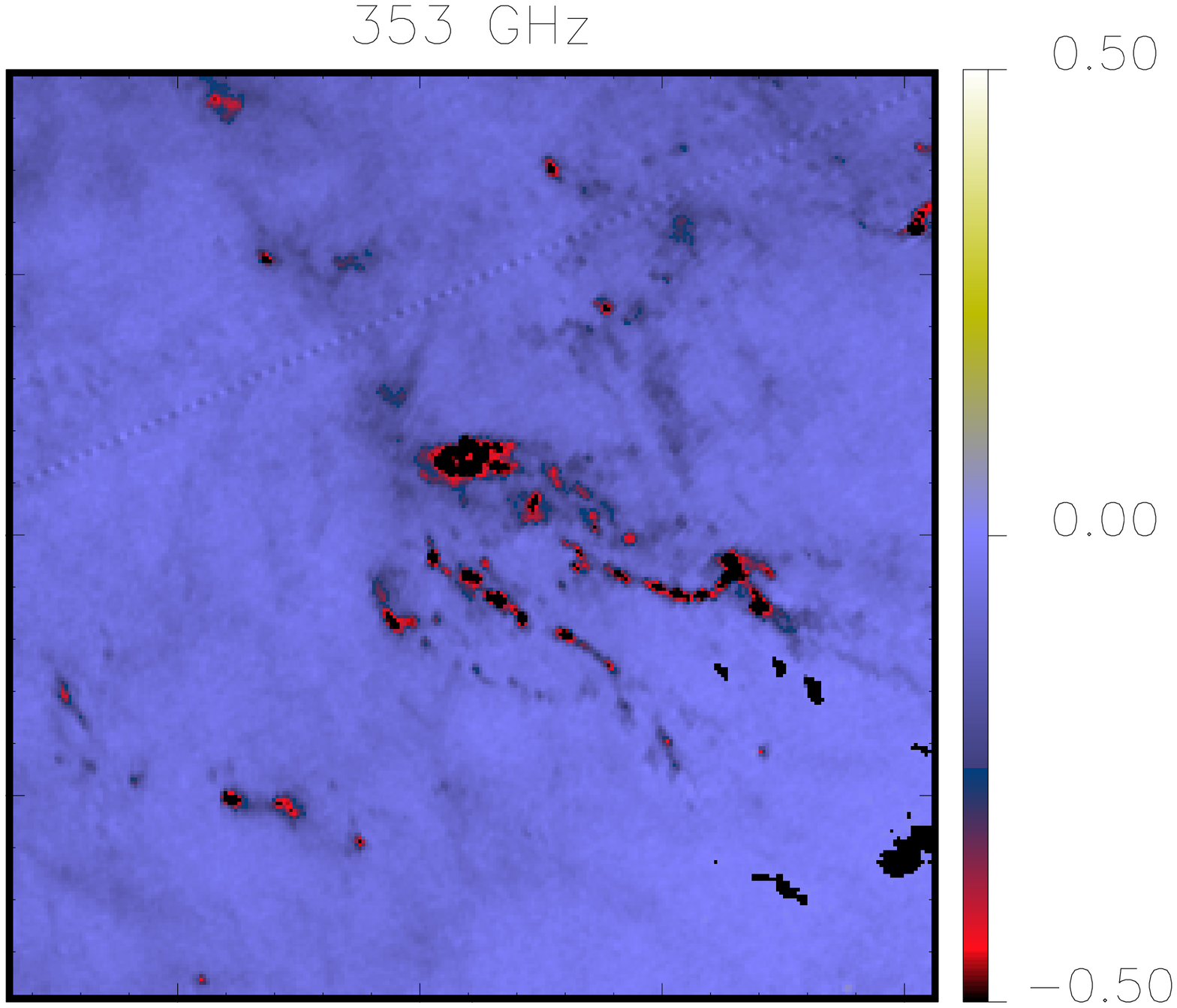}
  \vskip-0.8cm
\hglue 1cm\includegraphics[angle=90,width=7cm]{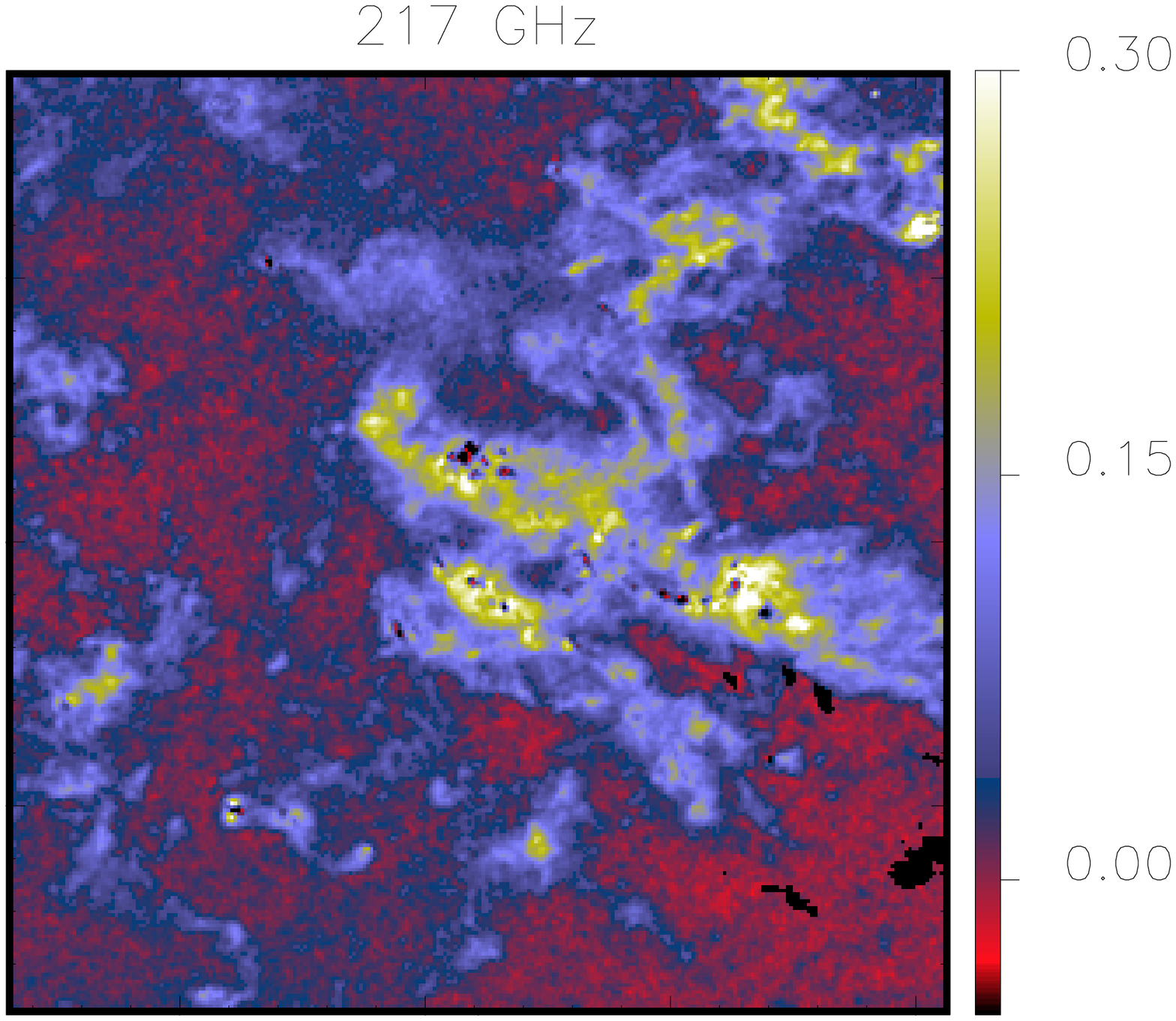}
     \includegraphics[angle=90,width=7cm]{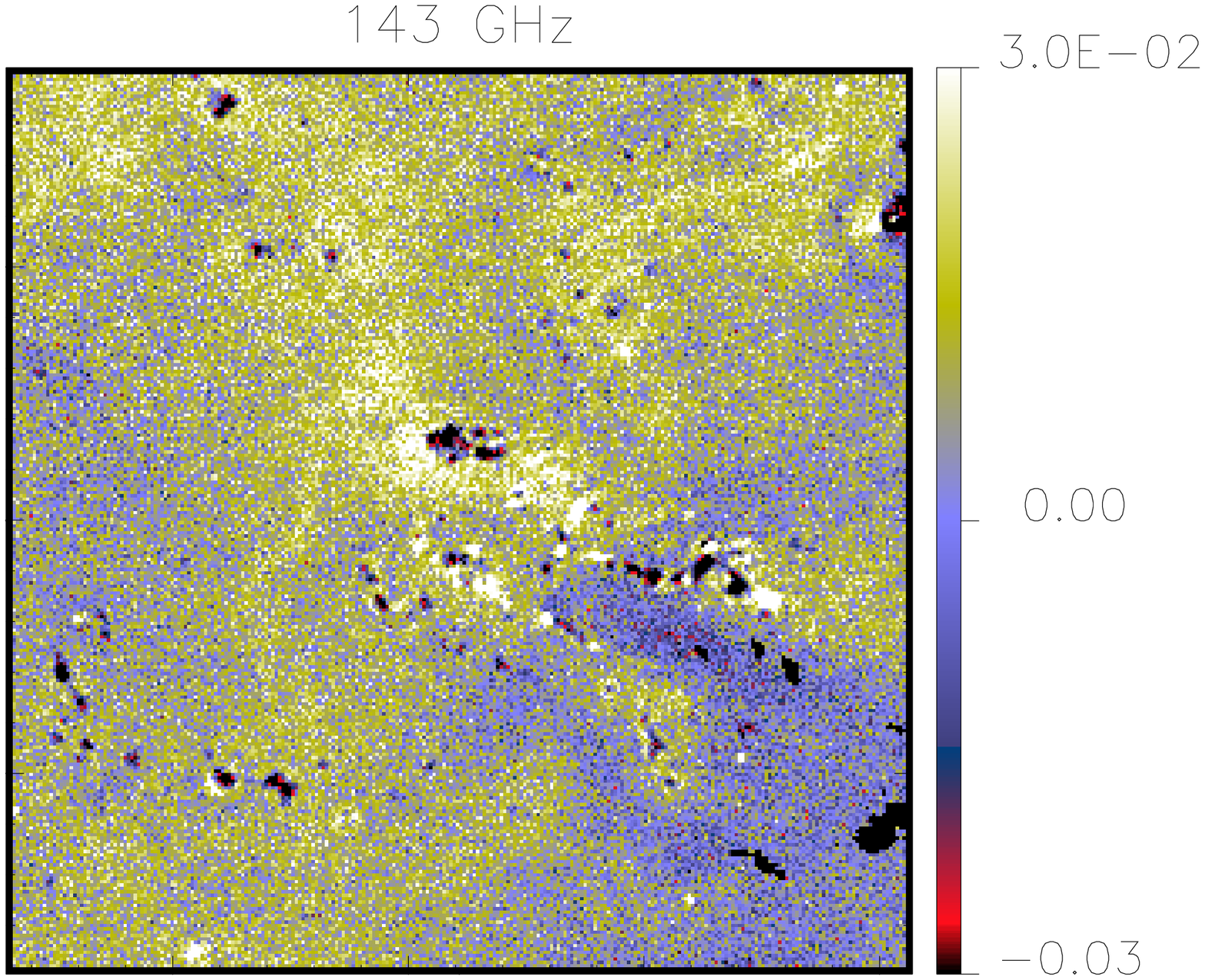}
  \vskip-0.8cm
\hglue 1cm\includegraphics[angle=90,width=7cm]{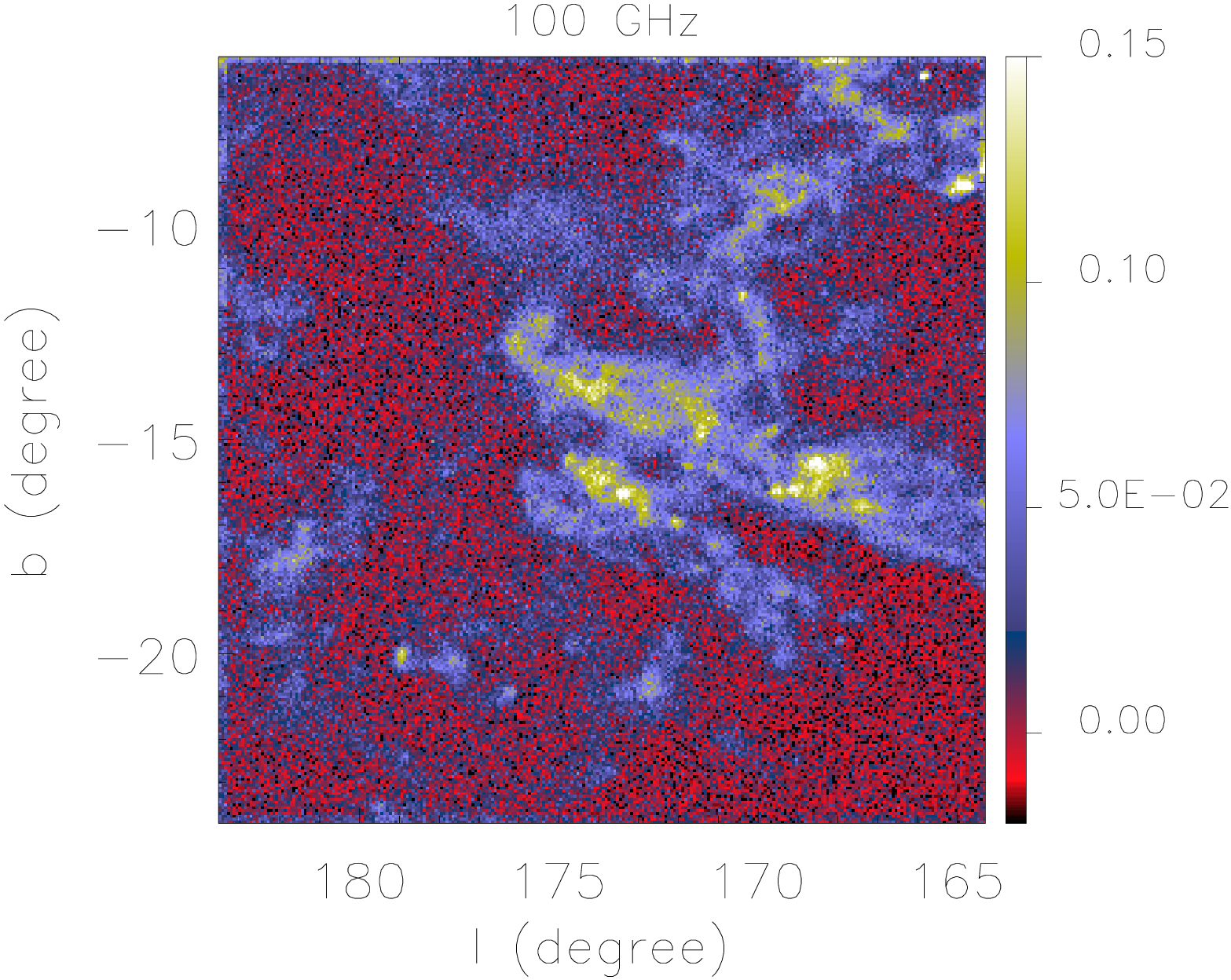}
    \includegraphics[angle=90,width=7cm]{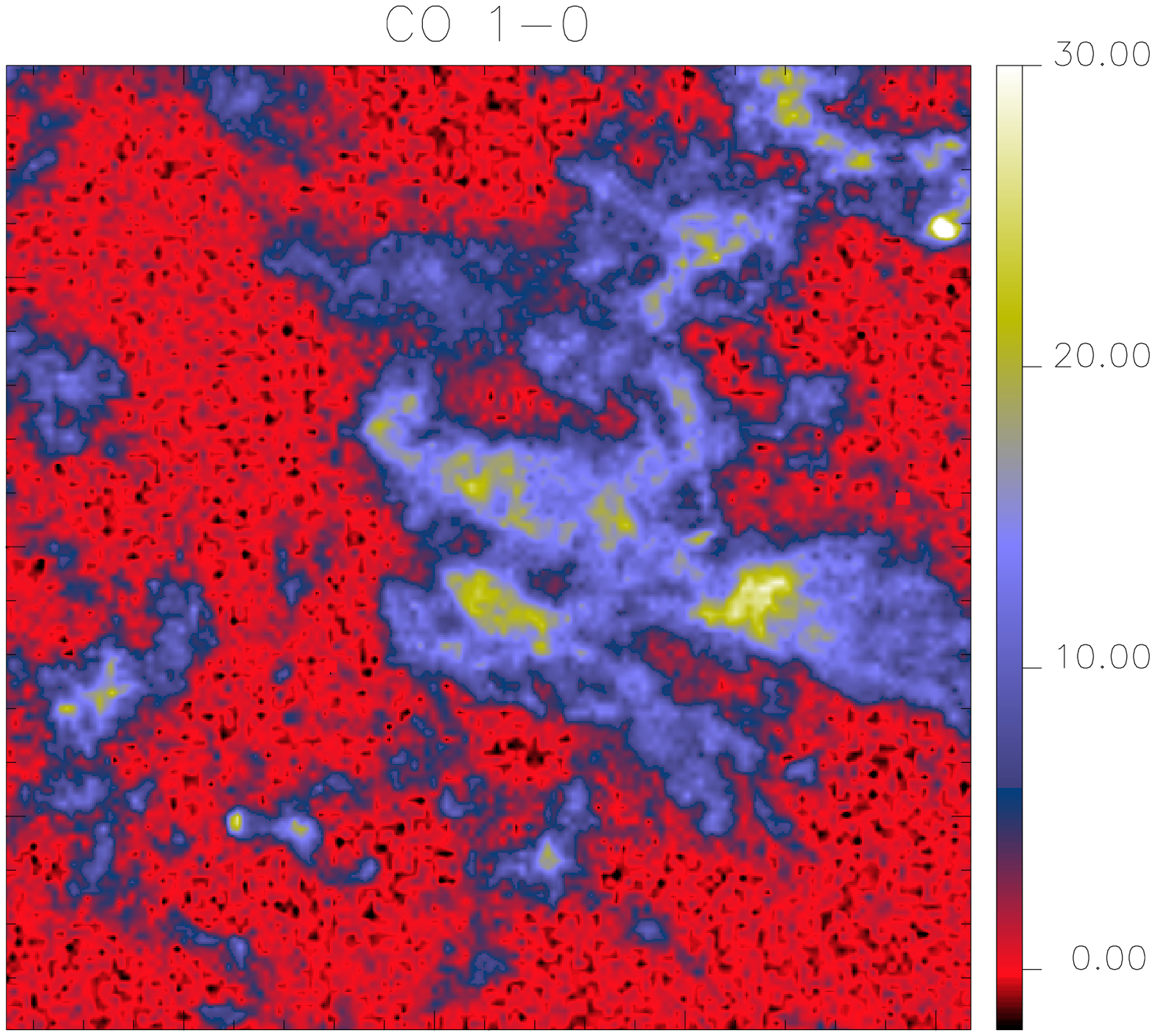}
  \vskip0.4cm
\caption{Same as Fig\,\ref{fig_taurus_images} for the fit residuals. The bottom right image is the CO $J = 1\rightarrow0$ integrated emission from \cite{Dame2001}. The units are K\,km\,s$^{-1}$.}
\label{fig_images_residuals}
\end{center}
\end{figure*}

\begin{figure*}
\begin{center}
\hglue 1cm\includegraphics[angle=90,width=7cm]{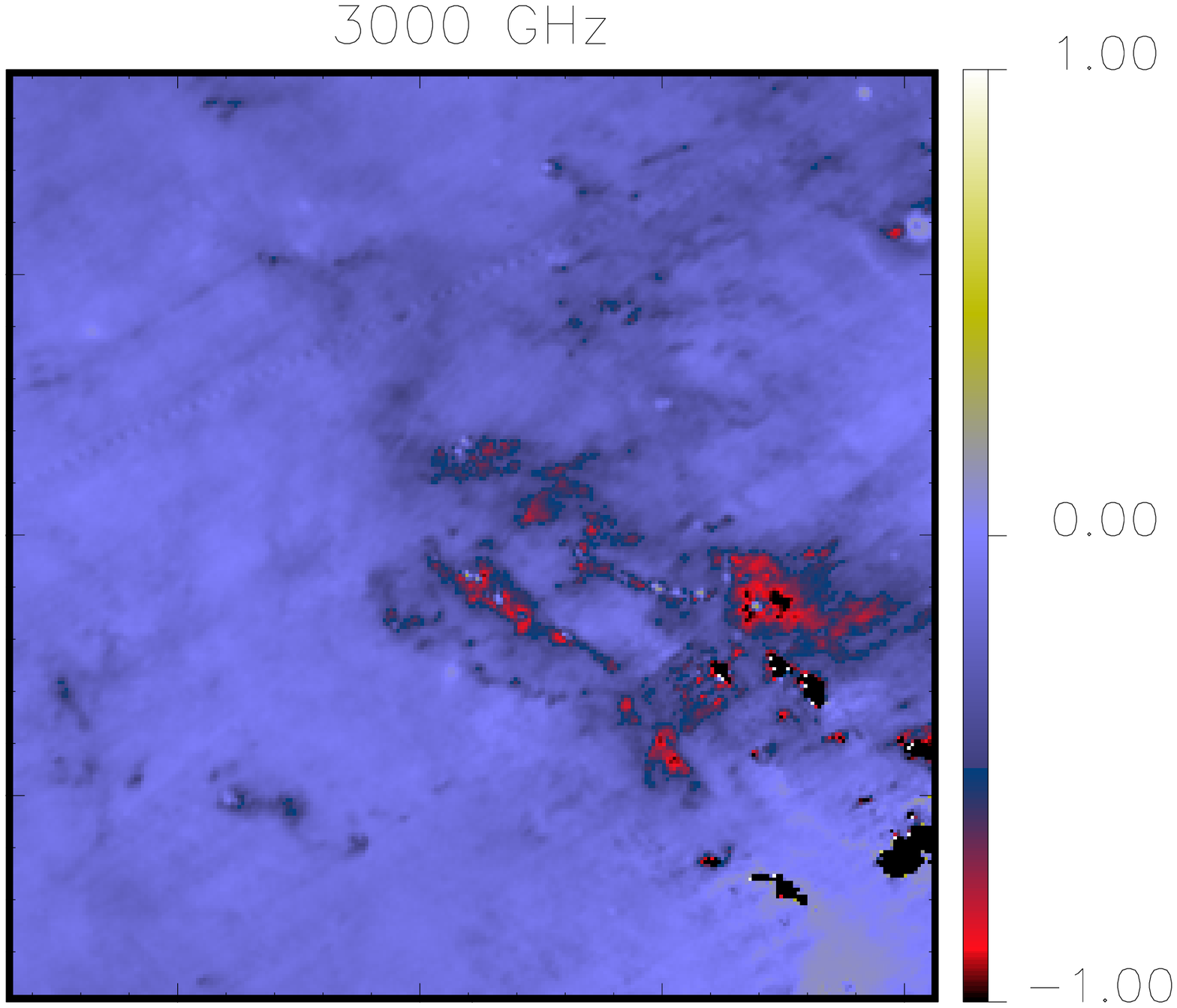}
     \includegraphics[angle=90,width=7cm]{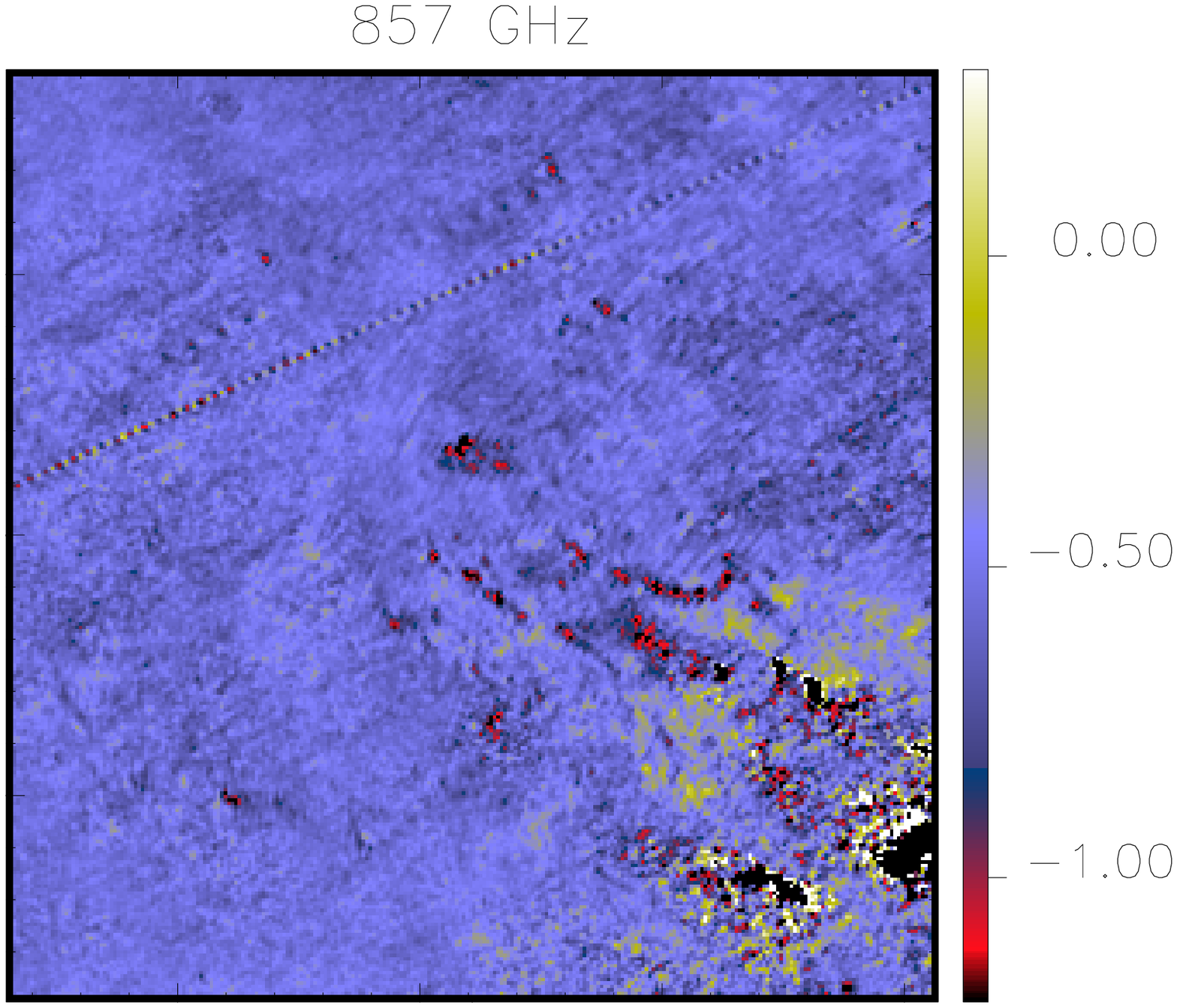}
 \vskip-0.8cm
\hglue 1cm\includegraphics[angle=90,width=7cm]{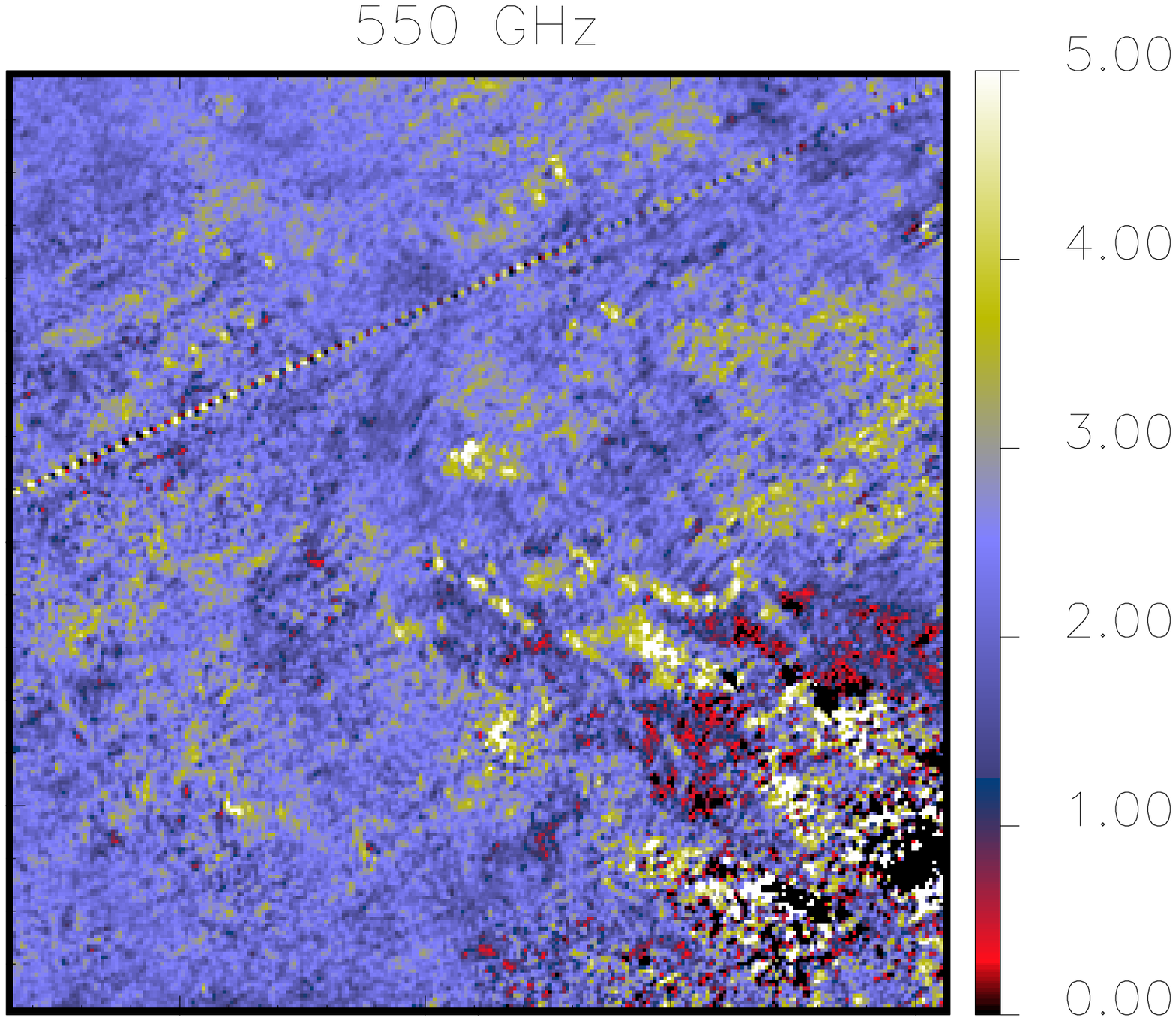}
     \includegraphics[angle=90,width=7cm]{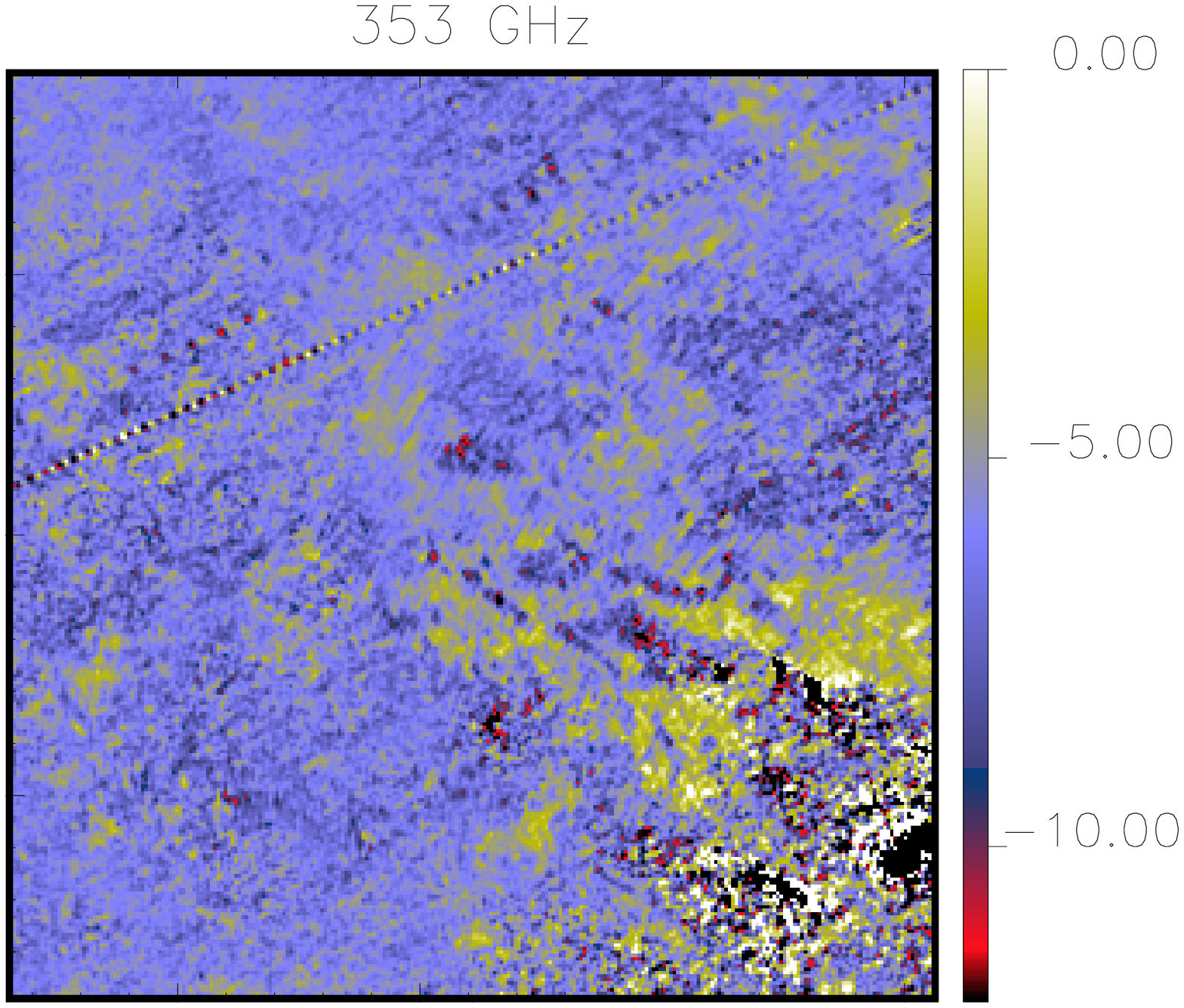}
  \vskip-0.8cm
\hglue 1cm\includegraphics[angle=90,width=7cm]{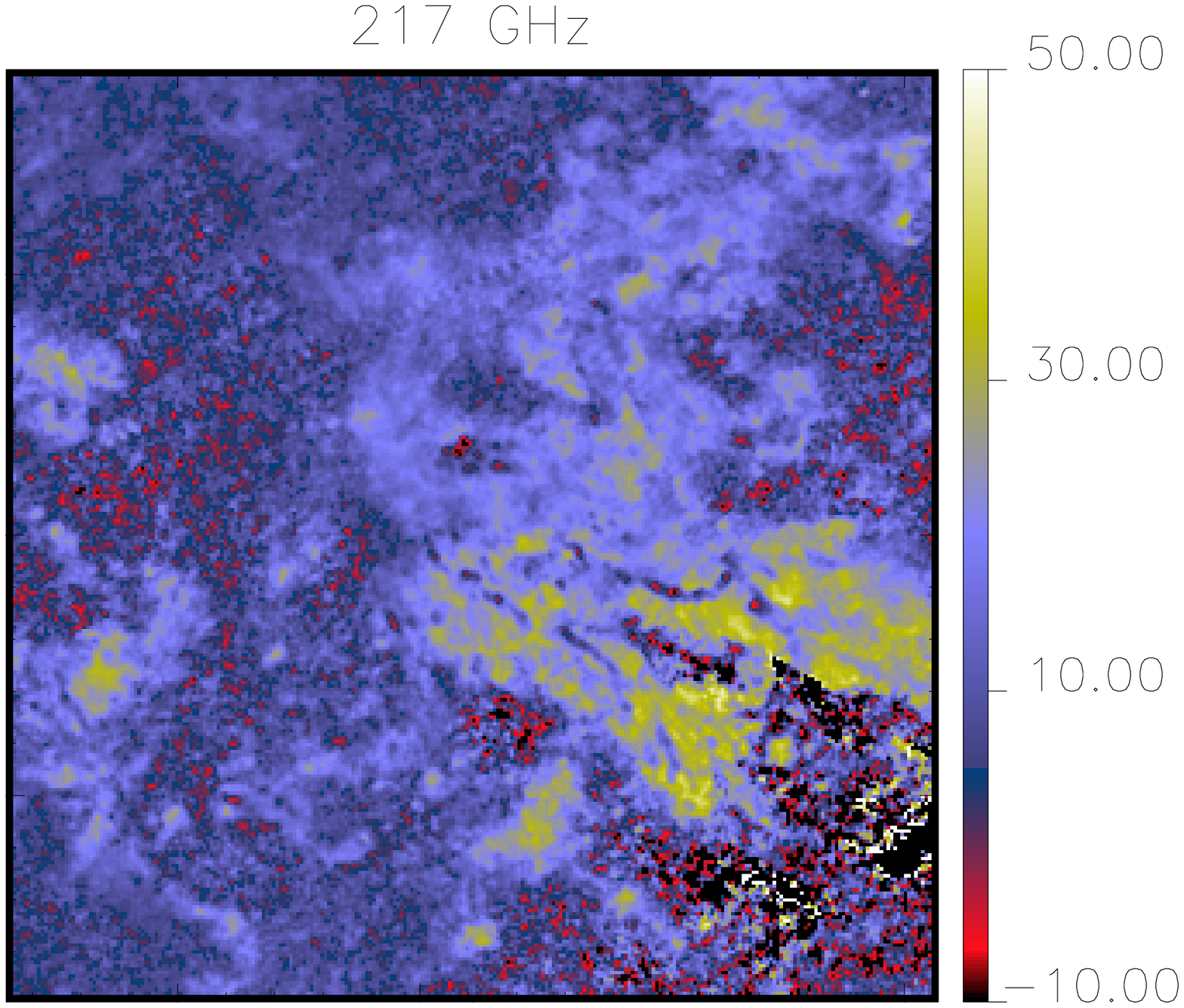}
     \includegraphics[angle=90,width=7cm]{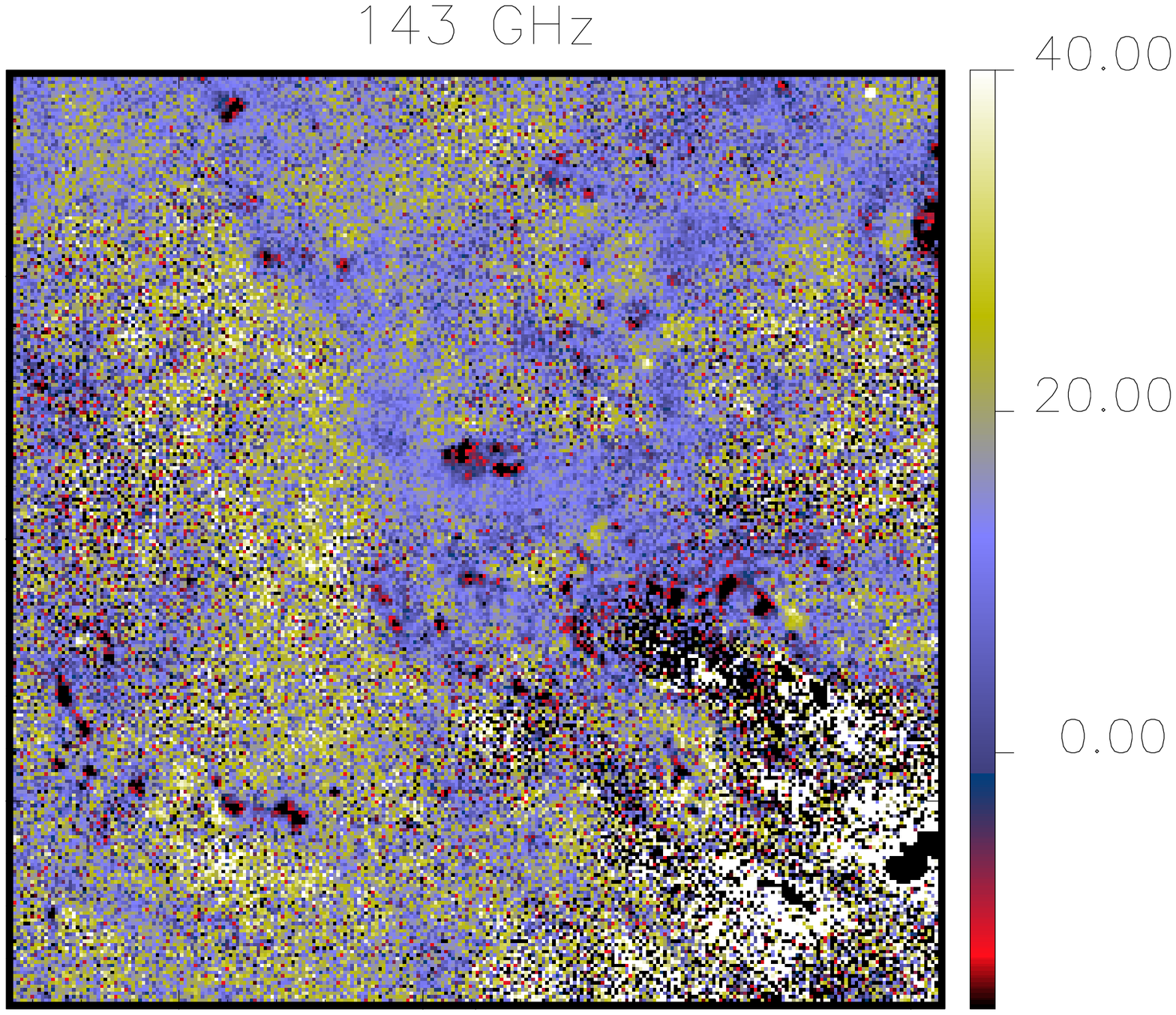}
  \vskip-0.8cm
\hglue 1cm\includegraphics[angle=90,width=7cm]{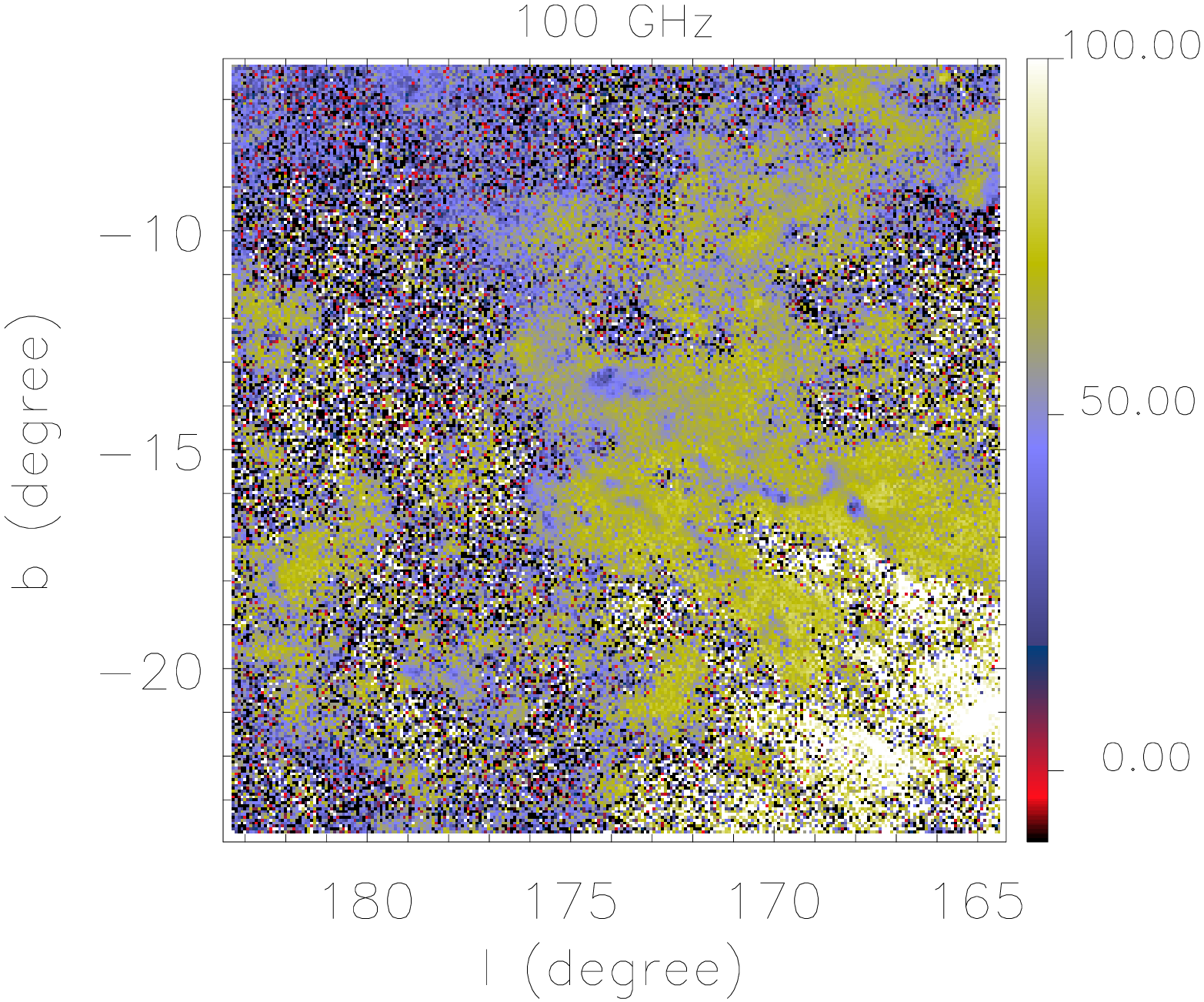}
 \vskip0.4cm
     \caption{Same as Fig\,\ref{fig_taurus_images} for the relative fit residuals. Units here are percentages.}
\label{fig_relative_residuals}
\end{center}
\end{figure*}

The fit residuals allow us to assess the limitations of using a single modified blackbody to fit the data. As seen in Sect.\,\ref{spectral_bands}, fits were made to the 3000, 857, 545, 353, and 143\,GHz data all smoothed to the angular resolution of 143\,GHz. The residual map at 100\,GHz was computed from the difference between the synthetic spectra smoothed at the 100\,GHz resolution and the data at 100\,GHz. The residual map of the fitting in all bands is shown in Fig.\,\ref{fig_images_residuals}. Figure\,\ref{fig_relative_residuals} shows the relative residual map (residual map divided by  measured map).  %Figure\,\ref{fig_Taurus_histo_relative_residuals} shows the histogram of the relative residuals in the molecular regions.  

%\begin{figure}
%\includegraphics[angle=90,width=\columnwidth]{Fig_histo_all_relative_residuals.eps}
%\includegraphics[angle=90,width=\columnwidth]{Fig_histo_all_relative_residuals_only_353_143GHz.eps}
%\vskip -4mm
%\caption{Top: Histogram of the relative residuals in the molecular regions at 100\,GHz (green), 143\,GHz (red), 217\,GHz (orange), and 353\,GHz (blue), for data fitting performed using the bands at 3000, 857, 545, 353, and 143\,GHz. The relative residuals at 3000, 857, and 545\,GHz are below 1--3\%.\quad Bottom: The same at 100\,GHz (green) and  217\,GHz (orange) for data fitting performed using only the bands at 353 and 143\,GHz (in that case the residuals at 353\,GHz and 143\,GHz are obviously equal to 0).}
%\label{fig_Taurus_histo_relative_residuals}
%\end{figure}

\subsubsection{Residuals for the five fitted bands}
We see in Fig.\,\ref{fig_relative_residuals} that the relative residuals are distributed around zero and below 1--3\% for the 3000, 857, and 545\,Hz bands. 
%, so they are compatible with the calibration errors (Sect.\,\ref{HFI data}) . 
On the other hand, the residuals at 353\,GHz are systematically negative, with a median relative amplitude about $-7\%$, while the residuals at 143\,GHz are systematically positive, with a median relative amplitude about $+13\%$. %(see also the histograms on Fig.\,\ref{fig_Taurus_histo_relative_residuals}). 
As can be seen in Figs.\,\ref{fig_images_residuals} and \ref{fig_relative_residuals}, these residuals are spatially correlated with the measured brightness, so they are related to the emission spectrum of the dust particles located in the complex.  

We have seen in Sect.\,\ref{HFI data} (Table\,1, from \citealt{planck2011-1.5}) that the calibration errors are estimated to be 13.5\% at 3000\,GHz (100\microns), $7\%$ at 857 and 545\,GHz, and 2\% at 353 and 143\,GHz. These numbers translate into systematic errors of about $5\%$ and $4\%$ on the 353 and 143\,GHz brightnesses (computed from the fit of the five bands), respectively. We conclude in this early analysis that the negative residuals at 353\,GHz (with a median amplitude about $-7\%$) and the positive residuals at 143\,GHz (with a median amplitude about $+13\%$) are not caused by calibration errors. The negative residuals at 353\,GHz could be caused by a broadening of the measured spectra around the peak of the modified blackbody (3000--545\,GHz spectral range) because of the contribution of dust at different temperatures along the line of sight. On the other hand, the positive residuals at 143\,GHz suggest a flattening of the emission spectra at low frequencies.
\begin{figure*}
\begin{center}
%\vskip-.5cm
\includegraphics[angle=90,width=7cm]{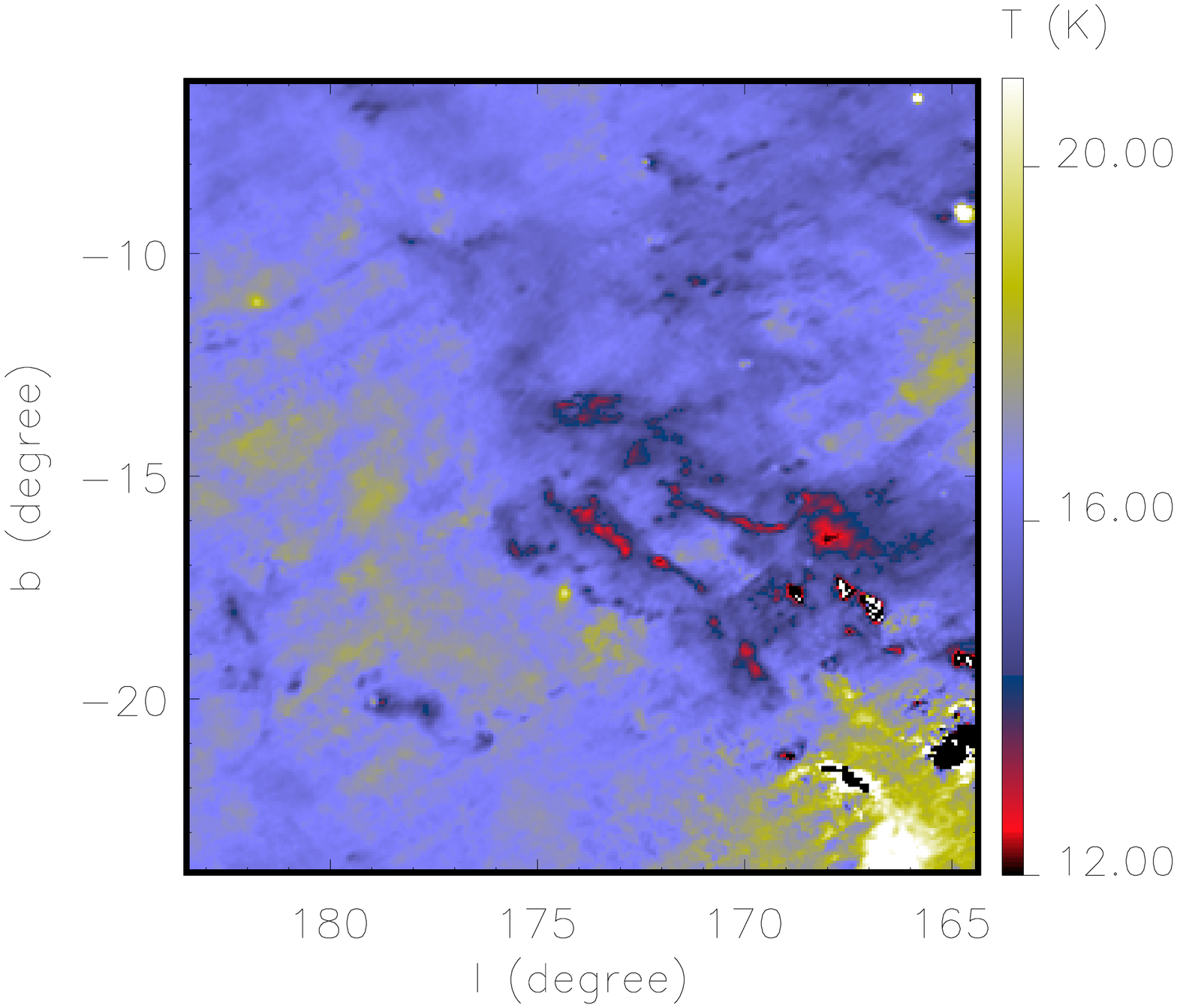}
\includegraphics[angle=90,width=7cm]{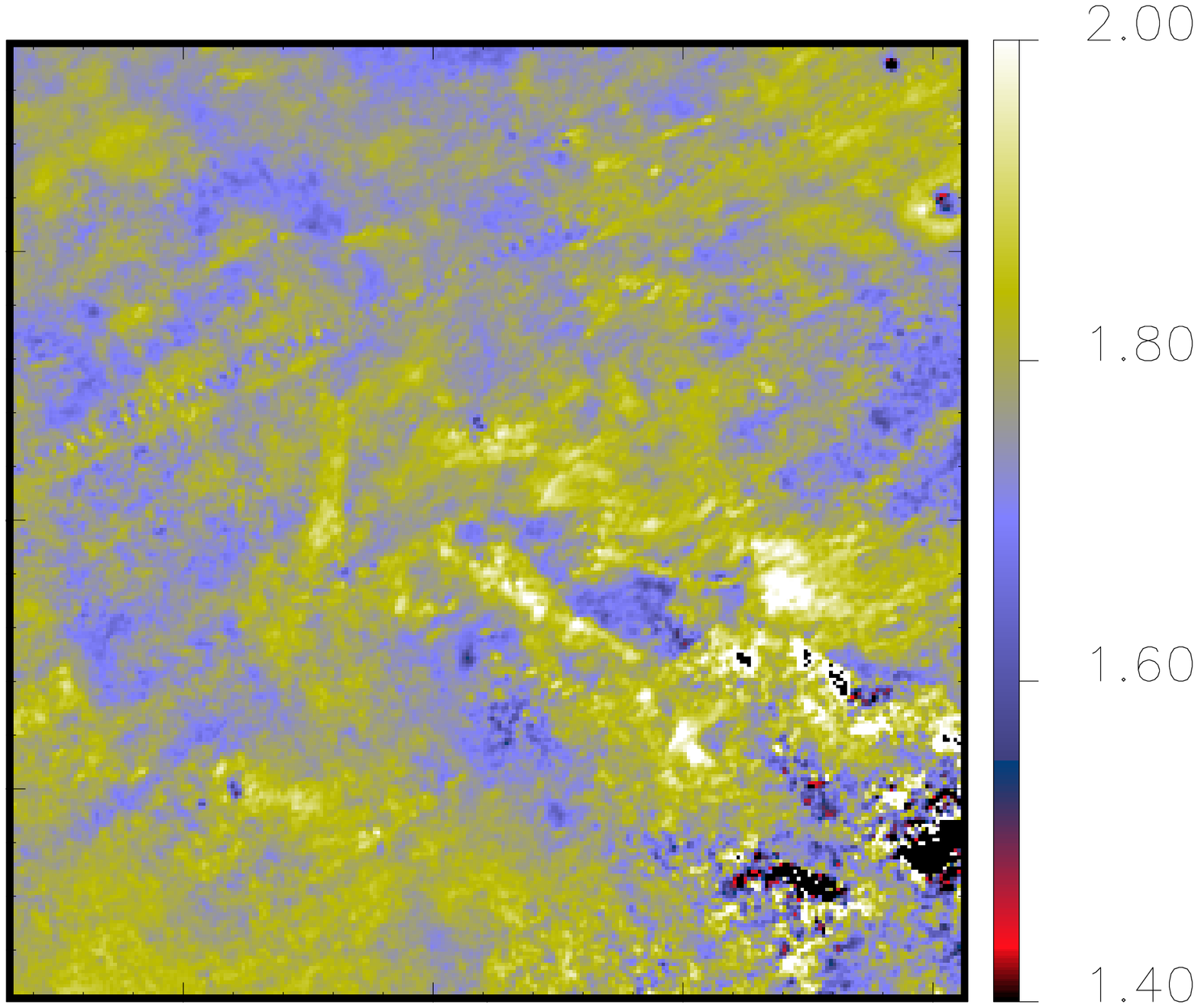}
    \caption{Left panel: dust temperature map. Right panel: spectral emissivity index map.}
\label{fig_T_and_beta_map}
\end{center}
   \vskip-0.7cm
\end{figure*}
\begin{figure*}
\centering
   \hskip -10mm
     \includegraphics[angle=90,width=9cm]{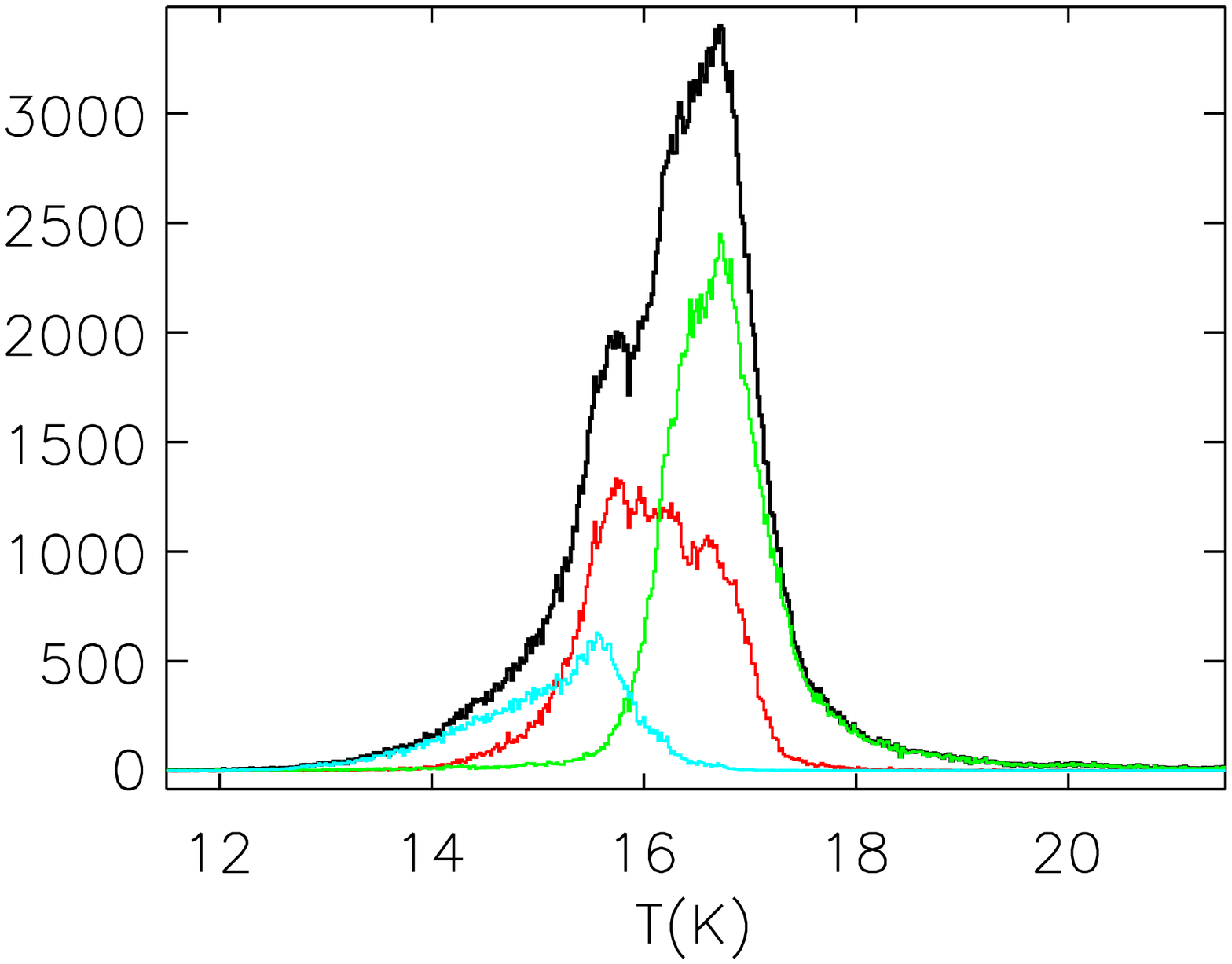}
\hskip -7mm
      \includegraphics[angle=90,width=9cm]{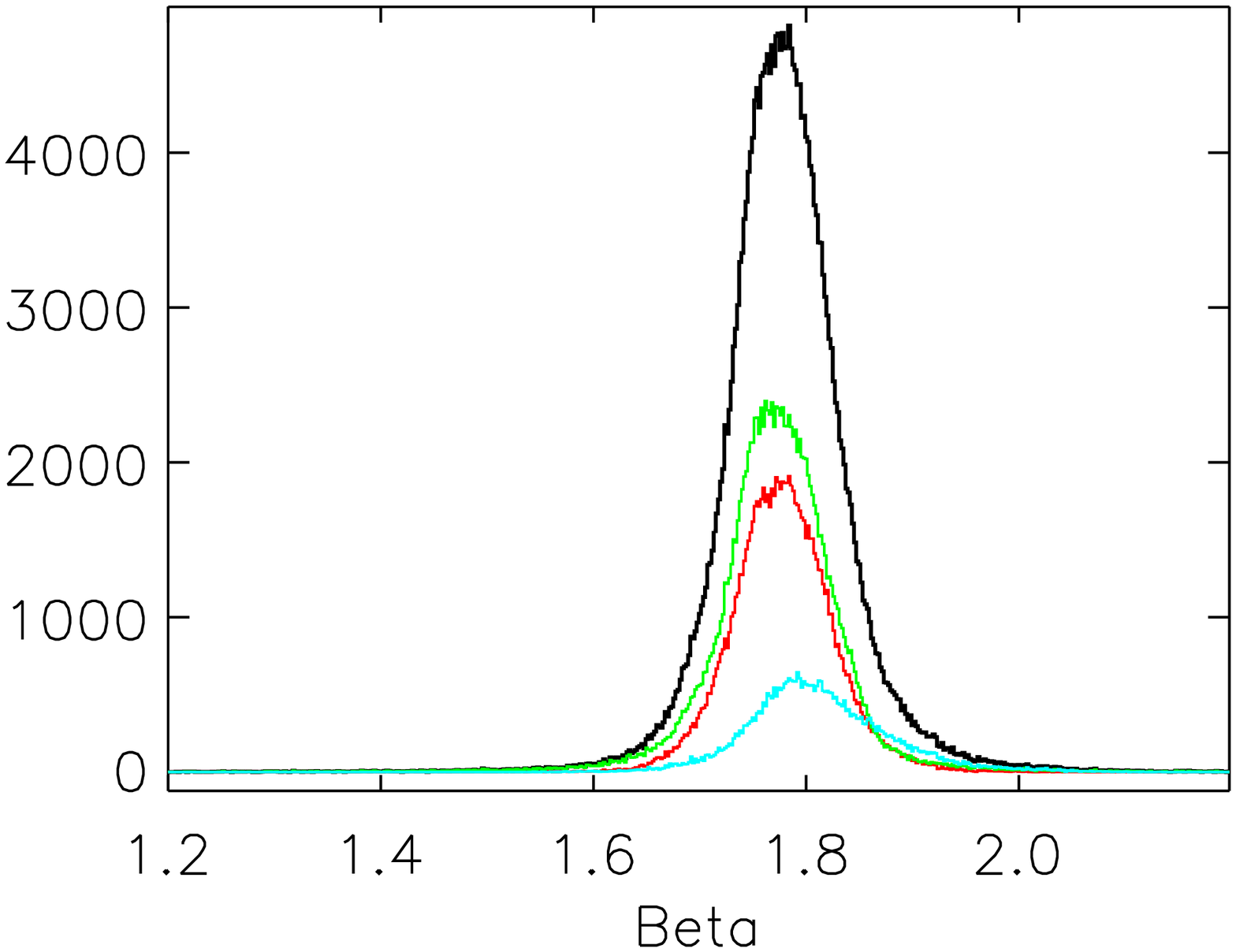}
\vskip -4mm
    \caption{Dust temperature and spectral emissivity index histograms: black, all pixels; green, pixels without detected  $^{12}$CO emission; red, pixels with detected  $^{12}$CO emission but no detected $^{13}$CO emission in the central molecular region covered by \cite{Pineda2010}; and blue, pixels in the central molecular region with detected  $^{13}$CO emission.}
\label{fig_T_and_beta_histo_map}
\end{figure*}
\subsubsection{Residuals at 217 and 100\,GHz}

As shown in Fig.\,\ref{fig_images_residuals}, there is a striking spatial correlation between the residual maps at 100\,GHz and 217\,GHz and the map of the integrated emission of the $^{12}$CO $J = 1\rightarrow0$ emission line. This confirms that 100\,GHz and 217\,GHz residuals are dominated by a contamination by CO molecular lines (mainly $^{12}$CO, but also \hbox{$^{13}$CO} and other isotopes and molecules in the densest regions). However, we have 
seen above that the measured dust spectra are more complex than a single modified blackbody from 3000\,GHz to 100\,GHz, so the residuals computed at 100\,GHz and 217\,GHz in this early analysis and shown in Fig.\,\ref{fig_images_residuals} are only indicative of the CO emission. 

\section{Analysis of the dust temperature and spectral emissivity index maps.}
\label{map_analysis}

%\subsection{Temperature maps.}
% no We derive upper limits of the errors on the temperature and spectral emissivity index maps by computing the standard deviations in different windows where these two parameters appear flat. This not possible for the opacity maps. 
\subsection{Dust temperature map}
The dust temperature map is shown in Fig.\,\ref{fig_T_and_beta_map}. The black regions are artifacts: they correspond to low values of the temperature (below 12\,K) caused by statistical noise and CIBA in the faint regions (with $I\,(100\,\mu$m$)<1$\,MJy\,sr$^{-1}$, as also illustrated in our simulations presented in Fig.\,\ref{fig_simu_beta_T_vs_100}). Excluding these regions and some residual stripping in the faintest regions with amplitudes around $0.15\,$\,K,  most of the variations seen on the temperature map are real because their amplitude is higher than the noise computed from our simulations (about 0.1--1\,K for  $I\,(100\,\mu$m$)=10-1\,$MJy\,sr$^{-1}$, from Fig\,\ref{fig_simu_beta_T_vs_100}).

At least three regions with different temperature distributions can be identified in the temperature map (Fig.\,\ref{fig_T_and_beta_map}) and its histogram (Fig.\,\ref{fig_T_and_beta_histo_map}): 
\begin{itemize}
\item the outer parts of the molecular cloud (with no detected $^{12}$CO emission), with temperatures $\sim16$--$17.5$\,K; 
\item the molecular cloud with detected $^{12}$CO emission but no detected $^{13}$CO emission, with temperatures $\sim15$--$17$\,K;
\item the densest parts of the molecular cloud with detected $^{13}$CO emission which coincide with the well-known dense filaments,  with temperatures $\sim13$--$16$\,K. 
\end{itemize}
Most of the differences between the {\it IRAS\/} map at 100\,$\mu$m and the HFI maps (Fig.\,\ref{fig_taurus_images}) are caused by temperature variations. Except in diffuse regions with uniform illumination and constant dust temperature, this demonstrates that {\it IRAS\/} maps at 100\microns\ used alone cannot properly probe the gas column density. Comparable temperature variations have also been found by \cite{Flagey2009} in the central region of our field (average temperature about 14.5\,K with a dispersion of 1\,K) by combining \textit{Spitzer} 160\microns\ and {\it IRAS\/} 100\microns\, maps and assuming a constant value of $\beta$ equal to 2. 

\begin{figure}
\centering
\hglue 5mm\includegraphics[angle=90,width=7.cm]{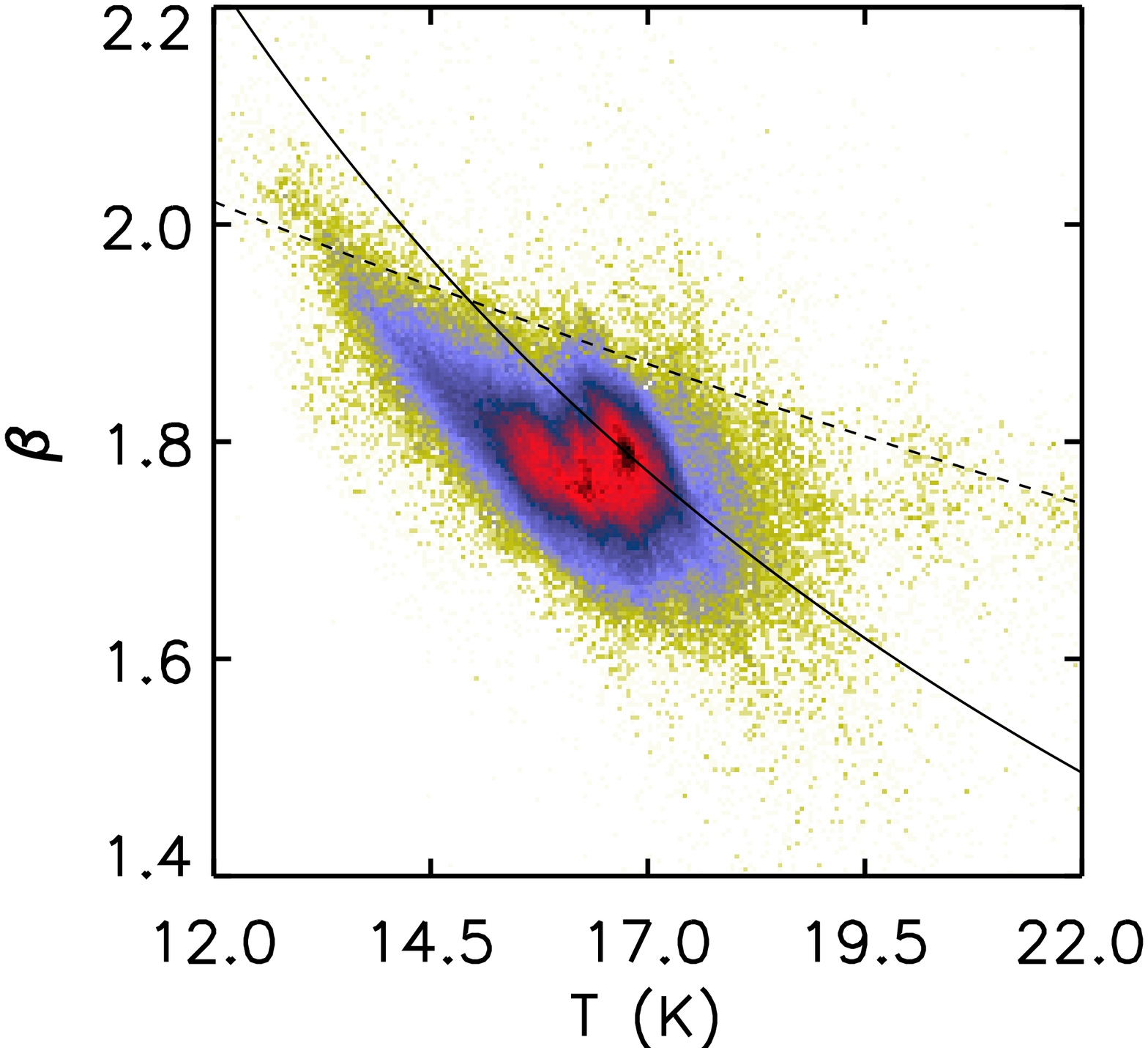}
\hglue 5mm\includegraphics[angle=90,width=7.cm]{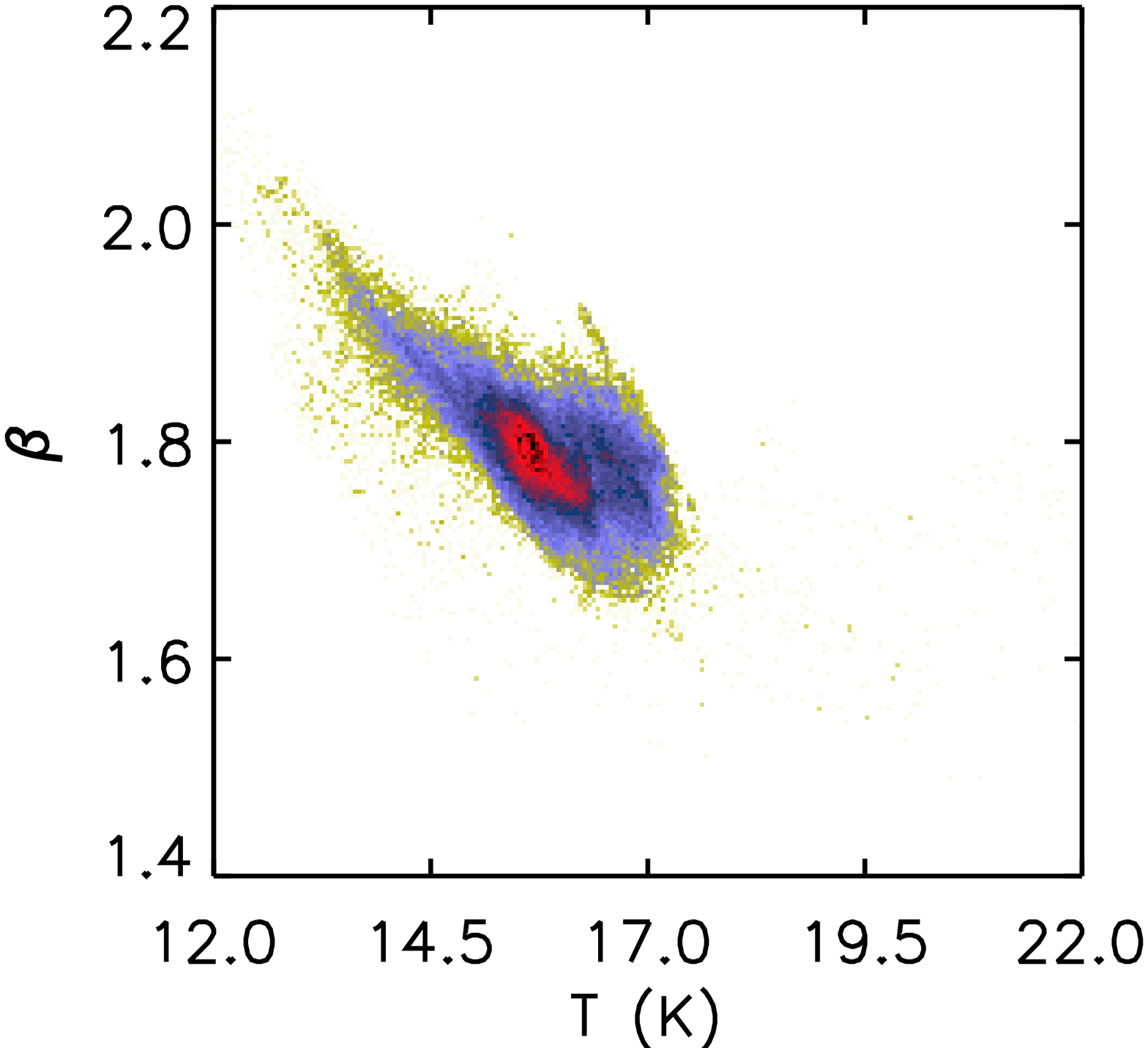}
\vskip5mm
\caption{Correlation between the spectral emissivity index and the dust temperature. Upper panel: for all pixels. Lower panel: for pixels in the molecular region with detected CO emission and $I\,(100\,\mu$m$)>10$\,MJy\,sr$^{-1}$.  %The thick lines is the fit to the data with the power-law $\beta=A\,T^\gamma$ suggested by \cite{2008A&A...481..411D}. 
The solid and dashed lines show the relations deduced from {\em Archeops}  \citep{2008A&A...481..411D} and PRONAOS \citep{2003A&A...404L..11D}, respectively.}
\label{fig_T_vs_beta}
\end{figure}

\subsection{Spectral emissivity index map}
The spectral emissivity index presents a symmetric distribution (Fig.\,\ref{fig_T_and_beta_histo_map}) with average and median values both equal to $1.78$ and a standard deviation $\sigma_{\beta}=0.08$. 
The standard deviation caused by statistical noise and CIBA estimated from our simulations is in the range 0.025--0.25 for $I\,(100\,\mu$m$)=10-1$\,MJy\,sr$^{-1}$ (Fig\,\ref{fig_simu_beta_T_vs_100}). 
Moreover, we can see in Fig.\,\ref{fig_T_and_beta_map} and more explicitly in the first panel of Fig.\,\ref{fig_T_vs_beta} that there is a some anti-correlation between $T$ and $\beta$. The higher values of $\beta$ ($>1.8$) are found in the coldest (about 14\,K) structures. 

We show in Appendix\,\ref{simulations} and it was also discussed in detail by \cite{2009ApJ...696..676S,2009ApJ...696.2234S} that the instrumental errors always produce intrinsic anti-correlation between $T$ and $\beta$. A significant fraction of the anti-correlation seen in Fig.\,\ref{fig_T_vs_beta} could be caused by noise. However, high values of $\beta$ generally correspond to bright regions. This is illustrated in the second panel of Fig.\,\ref{fig_T_vs_beta}, which shows the $T-\beta$ correlation diagram for pixels in the molecular region with detected CO emission and $I\,(100\,\mu$m$)>10$\,MJy\,sr$^{-1}$: the $T-\beta$ anti-correlation is still visible, with an amplitude significantly higher to what is expected from the contribution of the statistical noise and the CIBA computed from simulations (green points in the first panel of Fig.\,\ref{fig_simu_beta_T_opacity}). 

Previous observations of thermal dust emission at FIR to millimetre wavelengths in a variety of Galactic regions indicate an anti-correlation between the fitted values of the spectral emissivity index $\beta$ and the dust temperature, from PRONAOS data 
%\citep{1998ApJ...496..267R,1999A&A...347..640B,Stepnik2003,2001ApJ...553..604D,2002A&A...392..691D,2003A&A...404L..11D},  
for $T=12-20$\,K \citep{2003A&A...404L..11D},  
{\em Archeops} data for $T=7-27$\,K \citep{2008A&A...481..411D}, and {\em Herschel} data for $T=10-30$\,K \citep[e.g.,][]{2010A&A...518L..99A,2010A&A...520L...8P}. We see in Fig.\,\ref{fig_T_vs_beta} that the $T-\beta$ anti-correlation we find is steeper than the relation deduced from PRONAOS, but comparable to the relation deduced from {\em Archeops}. This $T-\beta$ anti-correlation is for line-of-sight-averaged data and reveals an intrinsic property of the emission spectrum of the dust particles \citep[e.g.,][]{2007A&A...468..171M, 2005ApJ...633..272B}. The amplitude of the anti-correlation may be higher for the emission spectrum of the dust particles than for line-of-sight-averaged data because in the general case there is a combination of dust temperatures in each line of sight. 

\subsection{Comparison with other Planck results and dust models}

As shown in \cite{planck2011-7.12}, both FIRAS and HFI+{\it IRAS\/} spectra of the local diffuse ISM can be well-fit by a modified blackbody with $T=17.9$\,K and $\beta= 1.8$. Morever, median values of $T=17.7\,$K and $\beta= 1.8$ are also found at $b >10\deg$ in the all-sky analysis using the 3000, 857, and 545\,GHz bands \citep{planck2011-7.0}. These two results are fully compatible with our distribution of spectral emissivity indices (Fig.\,\ref{fig_T_and_beta_histo_map}) centred on 1.78 with a standard deviation of 0.08. We also saw in Sect.\,\ref{example_spectra} that the systematic error in $\beta$ is estimated to be $0.07$.  

\begin{figure*}
      \includegraphics[angle=0,width=8cm]{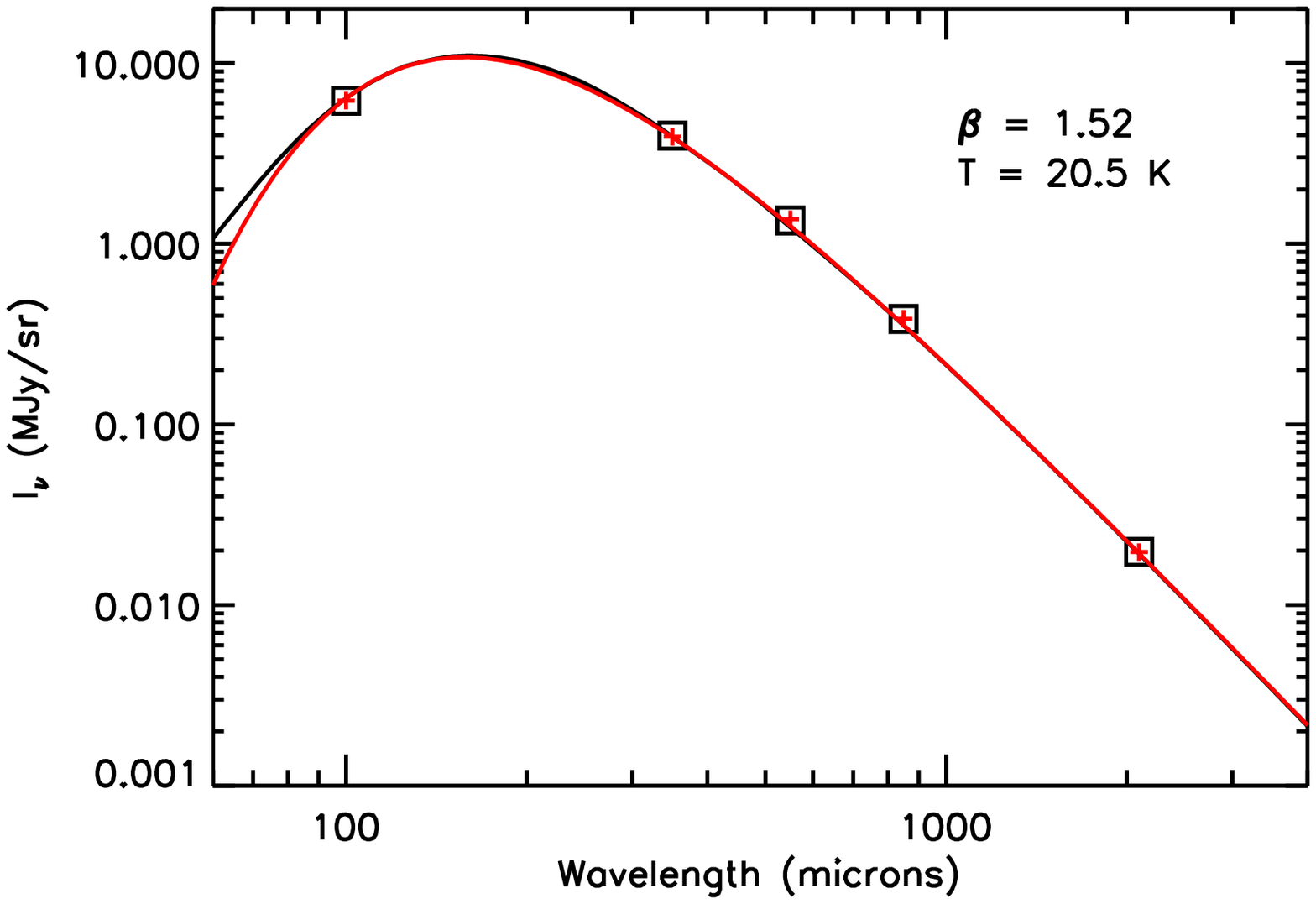}
     \includegraphics[angle=0,width=8cm]{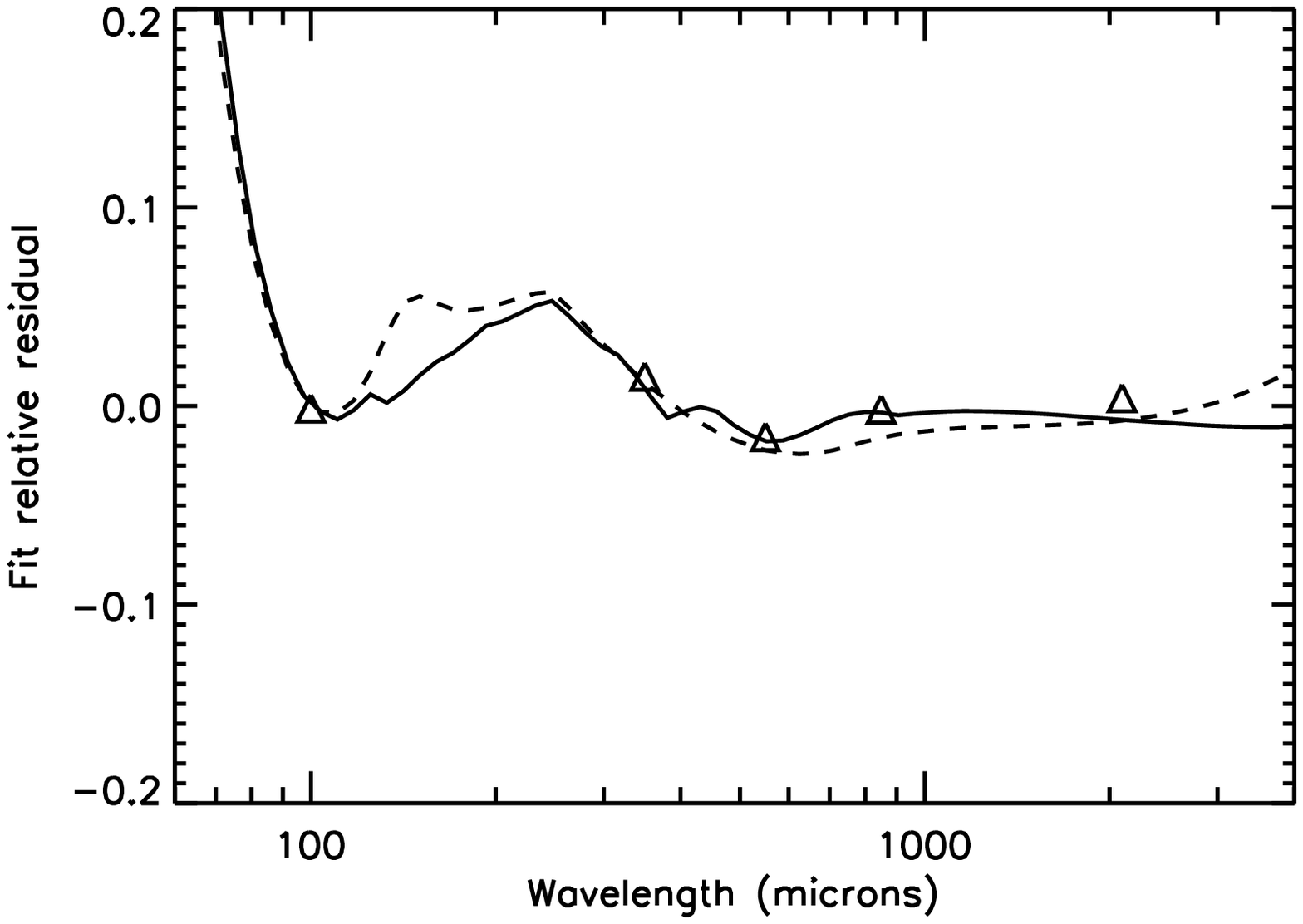}
\caption{Left panel: fit of the DustEM model of \cite{Compiegne2011} for the diffuse ISM heated by the standard interstellar radiation field (ISRF) of \cite{Mathis1983}. The black solid line is the model, the squares are the model in the photometric bands at 100\,$\mu$m, 857, 545 and 143\,GHz, and the red solid line is the fitted spectrum. Right panel: relative residuals of the fit. The solid line is the continuous model and the triangles are the model in the photometric bands. The dashed line shows the relative residuals of the fit for the \cite{Draine2007} model. The increase of the residuals at wavelengths below 100\,$\mu$m is caused by the contribution of transiently heated small particles.}
\label{fig_dust_model}
\end{figure*}

\begin{figure*}
\begin{center}
\includegraphics[angle=90,width=7cm]{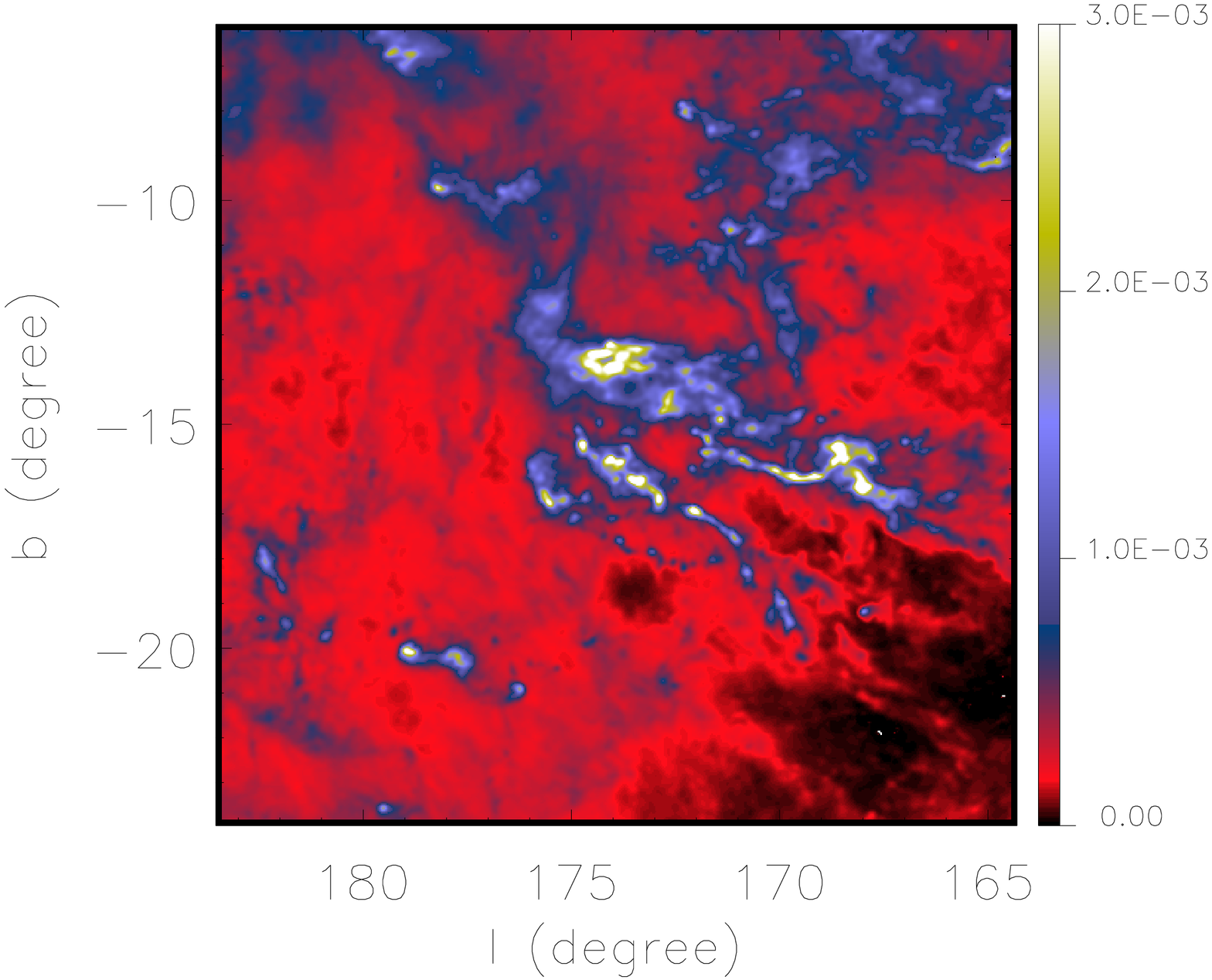}
\includegraphics[angle=90,width=7cm]{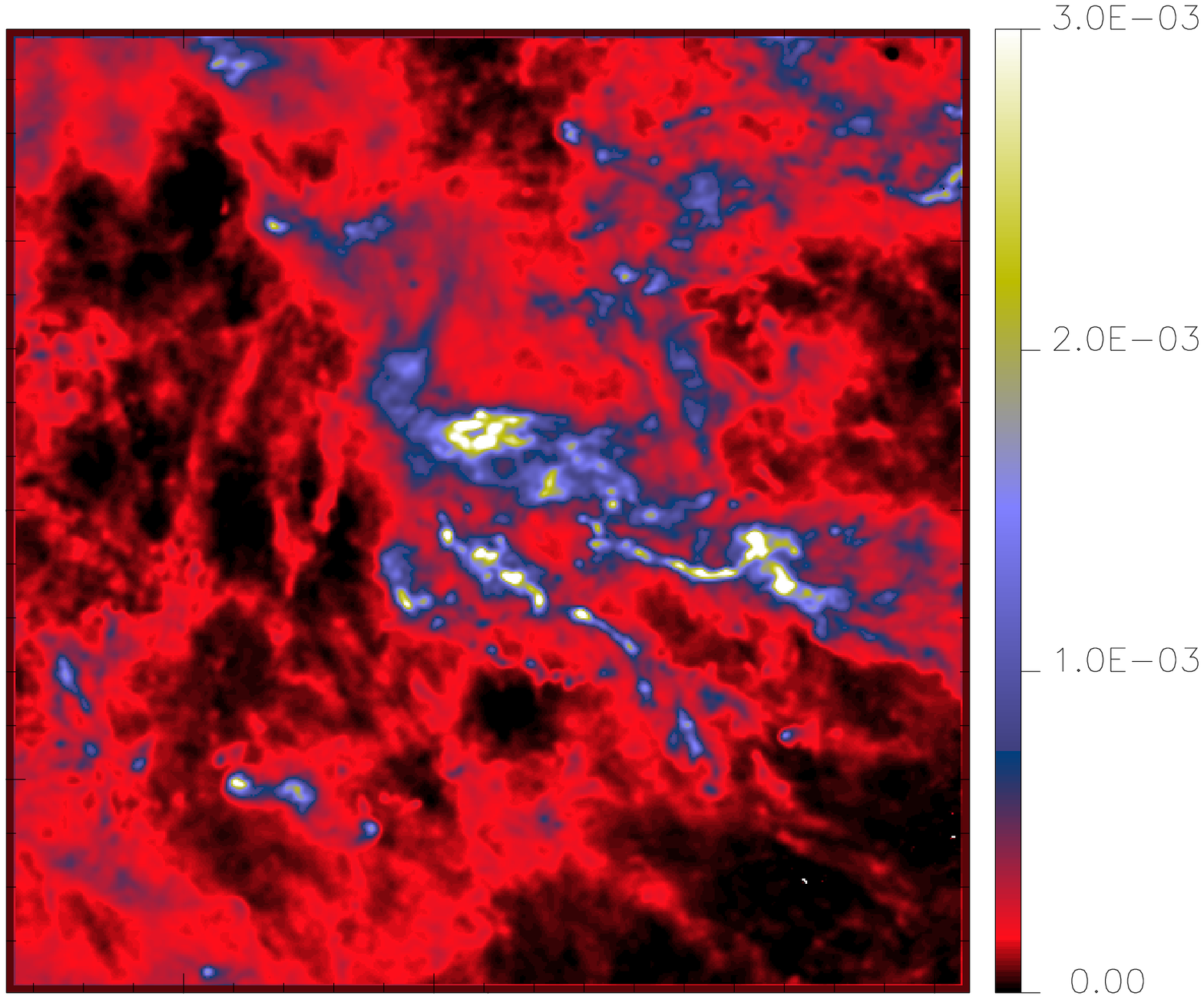}
\vskip 0.3cm
 \caption{Maps of the dust optical depth at 250\microns\,(1200\,GHz). Left panel: total optical depth derived from the pixel-by-pixel fit of the HFI and IRAS data. Right panel : optical depth map of the molecular phase alone (the optical depth associated with the atomic phase has been removed, see Sect.\,\ref{Optical_depth_molecular}). 
 \label{fig_tau_map}}
\end{center}
\end{figure*}

%The sensitivity of HFI is such that the temperature and the spectral emissivity index of the measured spectra are determined with an unprecedented precision, which bring new constrains for dust models. 
We compare here our results to the dust models of \cite{Draine2007} and \cite{Compiegne2011}, which were computed with the \hbox{DustEM} tool (http://www.ias.u-psud.fr/DUSTEM). 
As far as the properties of thermal dust are concerned, silicates are the same in both models, whereas the carbon grains are taken to be graphite \citep{Draine2007}, or hydrogenated amorphous carbon \citep{Compiegne2011}. We first computed dust emission models for all grain types in the diffuse ISM, i.e., heated by the standard interstellar radiation field (ISRF) of \cite{Mathis1983}. Next, we generated simulated data points by taking the flux densities of our model spectra in the bands at 100\,$\mu$m, 857, 545, 353 and 143\,GHz. We then applied our fitting procedure to these simulated band flux densities. Fitting the model of 
\cite{Draine2007} yields $T\simeq19.6$ K and $\beta \simeq1.67$. The latter value is intermediate between that obtained from fits for silicates (1.57) and graphite (1.93) alone. Fitting the model of \cite{Compiegne2011} yields $T\simeq 20.5$ K and $\beta \simeq$ 1.52 (see Fig.\,\ref{fig_dust_model}). Individual fits for silicates and amorphous carbon give similar values of $\beta$. We note that the highest residuals from a single modified blackbody are found between 150 and 300\,$\mu$m, in the spectral gap between {\it IRAS\/} and HFI (Fig.\,\ref{fig_dust_model}). 

For both the \cite{Draine2007} and \cite{Compiegne2011} models, the fitted values of $\beta$ are slightly below the central value found in the Taurus molecular cloud (1.78, with a systematical error of 0.07). However, it is worth noting that to account for the FIRAS spectrum of the diffuse ISM, \cite{LD01_II} decreased $\beta$ from 2.1 to 1.6 in the absorption efficiency of silicates at wavelengths above 250\,$\mu$m. If we do not apply this correction and simply extrapolate the absorption efficiency of silicates as a single power law with $\beta=2.1$ for $\lambda\geq 30\;\mu$m, our fit provides $T\simeq 19.4$\,K and $\beta\simeq 1.73$ for the model of \cite{Compiegne2011}, and $T\simeq 18.2$\,K and $\beta\simeq 1.97$ for the model of \cite{Draine2007}. The value of $\beta$ obtained with the \cite{Compiegne2011} model is then compatible 
with the observations. Finally, the dust temperature derived from the \cite{Compiegne2011} model is significantly higher than the value of 17.9\,K found by \cite{planck2011-7.12} in diffuse clouds: this is because the amorphous carbon of \cite{Compiegne2011} absorbs the ISRF in the near-IR (between 1 and 10\,$\mu$m) more efficiently than graphite or silicates.

This first comparison of the \Planck\ data and recent dust models shows that the emission of thermal dust can be represented as a first approximation by a single modified blackbody with a constant spectral emissivity index $\beta$.

\section{Dust optical depth per unit column density}
\label{Dust_opacity_per_NH}

We have seen in the previous section that the spectral coverage of \Planck/HFI allows for the measurement of the dust temperature and of the spectral emissivity index $\beta$ for each line of sight. The third adjustable parameter is the dust optical depth at 250\microns, $\tau_{250}$. To allow a direct comparison of dust optical depth results with the other \Planck\ Early Results papers, we fitted the data using only the bands at 3000, 857, and 545\,GHz, holding $\beta$ fixed at 1.8. Compared to fits with five bands (including 353 and 143\,GHz), the fitted optical depths do not change by more than about $10\%$.  

For the first time we have an unbiased map of the optical depth of thermal dust in a molecular complex from the most diffuse regions to the densest parts. The optical depth map in Fig.\,\ref{fig_tau_map} has a dynamic range greater than 100. The statistical error and the systematic noise in $\tau_{250}$ are estimated to be about 1--10\% and 12\%, respectively, using the method described in Appendix\,\ref{simulations} (but with a fixed value of $\beta=1.8$, and considering fitting of only the three bands at 3000, 857, 545\,GHz).  

\subsection{Independent tracers of the column density}
\label{Independent_NH}

\subsubsection{Atomic phase}
\label{Independent_NH_atomic}

The column density of the atomic gas is traced using the  \ion{H}{i} data at 21\,cm taken with the Leiden/Dwingeloo 25-m telescope with an angular resolution of 36$\arcmin$\ \citep{Hartmann1997}. 
In the optically thin hypothesis, the velocity-integrated emission can be converted to column density using the classical factor 1\,K$^{-1}$\,km$^{-1}\,{\rm s} =1.81\times\,10^{18}\,$cm$^{-2}$. However, the \ion{H}{i} emission is subject to self-absorption in the cold neutral medium (CNM), and $N_{\rm H}$ can be underestimated. An exact calculation requires knowledge of the density profiles of the CNM components along each line of sight. This estimate has been performed by \cite{Heiles2003} by measuring the emission/absorption of the 21-cm line against a number of continuum sources. Correction factors around 1.25 are found in the Taurus/Perseus region. In our case, the precise correction pixel by pixel for each line of sight is not possible. We tested the simple correction method discussed in \cite{planck2011-7.12}, which assumes a constant spin temperature $T_{\rm S}$ (this is not really justified because $T_{\rm S}$ varies from the warm neutral media (WNM) to the CNM). Correction factors of about 1.1--1.5 are obtained for $T_{\rm S}=100\,$K. In this early analysis we decided to apply a constant multiplicative correction factor of 1.25, with a conservative uncertainty of 20\%, to derive from the \ion{H}{i} data the column density map for the atomic phase (Fig.\,\ref{fig_taurus_ancillary_maps}). 
In any case, we checked that a different choice for the correction method does not affect the quantitative analysis of the dust optical depth per unit column density in the molecular phase (see below). 

The Taurus molecular cloud is 15\deg\ from the Galactic Plane and contains large-scale emission associated with background atomic gas with LSR velocities from $-50$\,km\,s$^{-1}$ to $0$\,km\,s$^{-1}$.  This explains the north-south Galactic gradient in the HFI and {\it IRAS\/} maps (Fig.\,\ref{fig_taurus_images}) and also in the dust optical depth map (Fig.\,\ref{fig_tau_map}). An east-west gradient caused by a filamentary structure that extends away from the Galactic plane and crosses the eastern part of Taurus is also detected, with velocity in the same range as the velocity in the $^{12}$CO $J = 1\rightarrow0$ emission line, i.e., from $\sim0$\,km\,s$^{-1}$ to $\sim15$\,km\,s$^{-1}$ \citep{Narayanan2008}. %Both gradients are detected in the {\it IRAS\/} and HFI maps (Fig.\,\ref{fig_taurus_images}), and as a consequence in the the opacity map (left pannel of Fig.\,\ref{fig_tau_map}).  
\subsubsection{Molecular phase}

The column density associated with the molecular regions can be traced with the NIR extinction map of \citealt{Pineda2010} (Fig.\,\ref{fig_taurus_ancillary_maps}), which gives the extinction caused by dust associated with the non-atomic component.  Special care is taken by \cite{Pineda2010} to remove the overall extinction associated with \ion{H}{i} located between the  background stars used and the Earth (0.3\,mag), and also the extinction associated with the widespread  \ion{H}{i} emission (0.12\,mag).   
  
The $J = 1\rightarrow0$ line of $^{12}$CO is a widely-used tracer of the molecular phase. % used for the decomposition of the large scale emission of our Galaxy detected by Planck (2 references). 
However, this line is sensitive to variations in abundance (depletion in densest regions, formation and destruction), excitation conditions, and radiative transfer effects (the line is generally optically thick), which 
explains the difference between maps of integrated $^{12}$CO\,($J = 1\rightarrow0$) emission (\citealt{Dame2001}; Fig.\,\ref{fig_taurus_ancillary_maps}) and our dust optical depth map (Fig.\,\ref{fig_tau_map}). A detailed comparison between the NIR extinction and the integrated emission of the $^{12}$CO\,($J = 1\rightarrow0$) line has been presented by \cite{Pineda2010}. 
In this paper, our strategy is to use the NIR extinction map of \cite{Pineda2010} as a quantitative tracer of the column density of the molecular phase. We also used the $^{12}$CO\,($J = 1\rightarrow0$) integrated emission map of \cite{Dame2001}, which covers the whole complex, to define regions containing (or not containing) molecular material. 

\begin{figure}
\begin{center}
%\vskip 5.8cm
\hglue -1cm\includegraphics[angle=0,width=9cm]{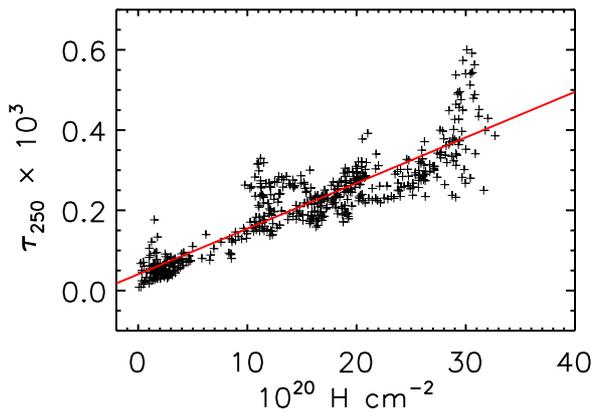}
%\vskip -6.8cm
     \caption{Dust optical depth at 250\microns\, as a function of the atomic column density, for pixels with no detected CO emission. The red line shows the result of the linear regression: $\tau_{250}= 1.14 \times 10^{-25} \times N_{\rm H}+  4.1\times 10^{-5}$.   
     \label{Fig_tau_vs_HI}}
\end{center}
\end{figure}

\subsection{Dust optical depth per unit column density in the atomic medium}
\label{Optical_depth_atomic}

To measure the dust optical depth at 250\microns\, per unit column density $\tau_{250}/ N_{\rm H}$ in the atomic phase, we used the column density map derived from \ion{H}{i} data presented in Fig.\,\ref{fig_taurus_ancillary_maps}. We subtracted from this map the averaged column density computed within the reference window (see Sect.\,\ref{Reference_spectrum}). Then we smoothed our dust optical depth map (Fig.\,\ref{fig_tau_map}) to the angular resolution of the \ion{H}{i} data (FWHM 36$\arcmin$) assuming Gaussian beams. We also selected all pixels with no detected CO emission, using the $^{12}$CO $J = 1\rightarrow0$ map  smoothed to the \ion{H}{i} resolution and with the criterion $W_{\rm CO}<0.5\,$K\,km\,s$^{-1}$. 

The dust optical depth at 250\microns\, $\tau_{250}$ as a function of the atomic column density for the selected pixels is shown in Fig.\,\ref{Fig_tau_vs_HI}. The relationship presents some dispersion, which may be owing to the uncertainties in the \ion{H}{i} opacity correction (estimated to be 20\,\%, see Sect.\,\ref{Independent_NH_atomic}), to the statistical noise in the computed values of $\tau_{250}$ (about 1--10\%, see above), and also to the contribution of some molecular material not detected on the $^{12}$CO survey. However, the relationship appears linear over the full range of column density from $1 \times 10^{20}$ cm$^{-2}$ to $3 \times 10^{21}$ cm$^{-2}$, with a slope of $\tau_{250}/ N_{\rm H}= 1.14\pm0.2 \times 10^{-25}$ cm$^2$. The uncertainty of $\tau_{250}/ N_{\rm H}$ takes into account the uncertainty in the opacity correction of \ion{H}{i} data, the statistical noise and the systematic error in the dust optical depth map (12\%). 

We conclude that the value of $\tau_{250}/ N_{\rm H}$ computed in the atomic medium in the Taurus region appears to be consistent with the standard value for the diffuse ISM,  $1\times10^{-25}$ cm$^2$ \citep{Boulanger1996}. %Actually, we have no  \ion{H}{i} data at the same angular resolution as HFI in order to analyse the spatial variation of $\tau/ N_{\rm H}$ in the atomic medium. 
\begin{figure}
\begin{center}
\vskip -2.5cm
\hglue -1cm\includegraphics[angle=90,width=9cm]{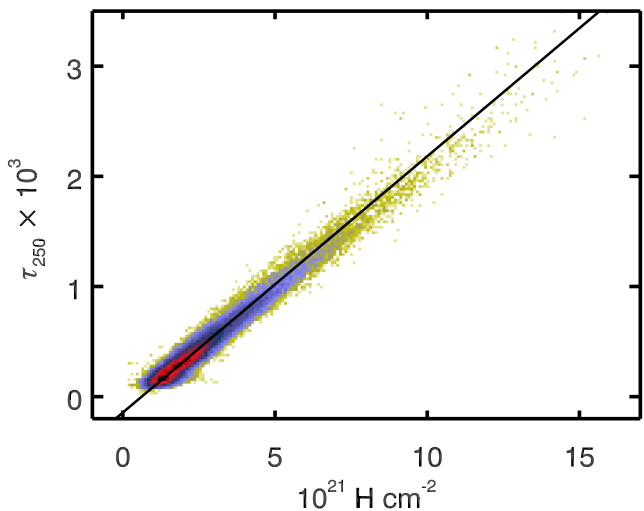}
\vskip 0.7cm
     \caption{Dust optical depth at 250\microns\, as a function of the column density $N_{\rm H}$ computed from the NIR extinction map of \citealt{Pineda2010} (shown on the lower right panel of our Fig.\,\ref{fig_taurus_ancillary_maps}), for pixels with detected CO emission ($W\,($CO$)>3$\,K\,km\,s$^{-1}$). The black line shows the result of the linear regression: $\tau_{250}= 2.32 \times 10^{-25} \times N_{\rm H} - 1.44\times 10^{-4}$.   
     \label{Fig_tau_vs_NH_from_2MASS}}
\end{center}
\end{figure}

\subsection{Dust optical depth per unit column density in the molecular phase}
\label{Optical_depth_molecular}

In order to measure $\tau_{250}/ N_{\rm H}$ in the molecular phase, we need to subtract the dust optical depth associated with the atomic phase from the dust optical depth map. In this paper the atomic gas is traced using \ion{H}{i} data, which have an angular resolution of 36$\arcmin$, which is significantly lower than \Planck/\hbox{HFI}. However, the local densities are higher in the molecular phase than in the atomic one, so we can consider that in lines of sight containing detected CO emission the small scale spatial fluctuations in the dust optical depth maps are dominated by density fluctuations in the molecular phase. Consequently, first we computed the optical depth map of the dust associated to the atomic medium alone $\tau_{250, \rm HI}$ by multiplying the  \ion{H}{i} column density by the value of $\tau_{250}/ N_{\rm H}$ found in the previous section. Then we subtracted the $\tau_{250, \rm HI}$ map from the dust optical depth map to obtain the map of the optical depth of the dust associated to the molecular phase (Fig.\,\ref{fig_tau_map}). We see that the north-south and east-west gradients detectable in the total dust optical depth map have disappeared, as expected. 
%The small scale fluctuations in the "molecular phase" opacity map... 

The dust optical depth at 250\microns\ $\tau_{250}$ in the molecular phase 
%, for pixels with detected CO emission ($W\,($CO$)>3$\,K\,km\,s$^{-1}$), 
as a function of the column density $N_{\rm H}$ computed from the NIR extinction map of \citealt{Pineda2010} (shown in the lower right panel of our Fig.\,\ref{fig_taurus_ancillary_maps}) is presented in Fig.\,\ref{Fig_tau_vs_NH_from_2MASS}. The statistical and systematic errors on $\tau_{250}$ are estimated to be about 1--10\% and 12\%, respectively (see Sect.\,\ref{Dust_opacity_per_NH}). 
%The statistical error on $N_{\rm H}$ is on average of $0.5 \times 10^{21}$ cm$^2$ (0.29\,mag, see Sect.\,\ref{ancillary_data}). 
The statistical error on $N_{\rm H}$ is in the range $0.2-0.5 \times 10^{21}$\,cm$^2$ for $N_{\rm H}=0-20\times 10^{21}$\,cm$^2$ (see Sect.\,\ref{ancillary_data}, and taking into account the smoothing at the angular resolution of the 143\,GHz \hbox{HFI} band). 
The systematic error on $N_{\rm H}$ is additive and is mainly caused by the uncertainty on the zero level from the correction of the extinction associated to the atomic phase located between the background stars and the Earth, and is estimated to be $0.5-1\times10^{21}$ cm$^2$ by \citealt{Pineda2010}. 

We see in Fig.\,\ref{Fig_tau_vs_NH_from_2MASS} that the $\tau_{250}- N_{\rm H}$ relationship appears linear. The dispersion is mainly caused by the statistical errors on $N_{\rm H}$ ($0.2-0.5 \times 10^{21}$ cm$^2$) and $\tau_{250}$ (1--10\%).  
The slope gives a measurement of the dust optical depth per unit column density in the molecular phase $\tau_{250}/N_{\rm H}= 2.32 \pm0.3 \times 10^{-25}$. The uncertainty takes into account the statistical error on $\tau_{250}$ and $N_{\rm H}$, and the systematic error on $\tau_{250}$ (12\%). 

The $\tau_{250}$-$N_{\rm H}$ linear relationship presents a faint non-zero positive residual $N_{\rm H}\,(\tau_{250}=0)\simeq 6\times 10^{20}$\,cm$^2$ compatible with the uncertainty on the zero level of the $N_{\rm H}$ map.

%Finally, we obtain the map of $\tau/ N_{\rm H}$ in the molecular phase shown on Fig.\,\ref{Fig_em_molecular_phase}, by dividing the molecular phase optical depth map  (Fig. \ref{fig_tau_map}) by the column density map computed from the NIR extinction map of \citealt{Pineda2010} (Fig.\,\ref{fig_taurus_ancillary_maps}). 

\subsection{Discussion}
\subsubsection{Previous observations}

The possibility of a higher value of $\tau_{250}/ N_{\rm H}$ in dense clouds than in the diffuse ISM has been a long-standing question \citep{O94,HMS95}. This was not unexpected, as dust grains must go through coagulation processes in dense environments, which tends to increase $\tau/ N_{\rm H}$ \citep[e.g.,][]{O94,S95}. Low grain temperatures ($\sim13$\,K) have been observed also with the PRONAOS balloon in relatively diffuse clouds, the translucent Polaris Flare \citep{B99} and one molecular filament (for $A_V\gsim2$) in the Taurus molecular cloud \citep{Stepnik2003}. As demonstrated by the authors, this low temperature cannot be explained by the effect of extinction alone: one needs to increase the value of $\tau_{250}/ N_{\rm H}$ (with $N_{\rm H}$ derived from the NIR extinction) by a factor of about 3 compared to the standard value for the diffuse ISM of $1\times10^{-25}$ cm$^2$. 

An increase of $\tau_{250}/ N_{\rm H}$ in the FIR by a factor $\ge1.5$--$4$ (always with $N_{\rm H}$ derived from the NIR extinction) has been detected from ISOPHOT observations of several high-latitude translucent clouds \citep{CBLS01,DB03,R06,K06,L07} and TMC-2 \citep{DB05}, by \textit{Spitzer} observations of the Perseus molecular cloud \citep{S08} and of the Taurus molecular cloud \citep{Flagey2009}, and more rencentely by \textit{Herschel} observations of Galactic dense cores \citep{Juvela2011}. Because the decrease in temperature is generally observed to be associated with a decrease of the 60\,$\mu$m over 100\,$\mu$m intensity ratio, which traces the abundance of small grains relative to that of big grains, coagulation has often been invoked to explain this emissivity increase \citep{B99,Stepnik2003,CJB05}. 

However, the increase of $\tau_{250}/ N_{\rm H}$ with increasing column density is not systematically observed. No excess is detected in the Corona Australis molecular cloud from \textit{Spitzer}/MIPS or APEX/Laboca data using also the NIR extinction as a tracer of the column density \citep{Juvela2009}. Finally, contradictory results have  also been obtained using the $^{12}$CO\,($J = 1\rightarrow0$) emission line to trace the column density. For instance, \cite{RD10} found that the deviations between CO surface density and FIR emission (measured by \textit{Herschel}/SPIRE and \textit{Spitzer}/MIPS) are more probably caused by H$_2$ envelopes not traced by CO and therefore not accounted in the column density, than by gas-to-dust ratio or $\tau_{250}/ N_{\rm H}$ variations. On large scales, \cite{P09} conclude from DIRBE, Archeops, and WMAP data that the dust that gives an excess of $\tau_{250}/ N_{\rm H}$ in the submm may recover its diffuse ISM value in the millimetre. This contradicts current scenarios of dust coagulation, for which a constant emissivity increase over the whole FIR-submm wavelength range is predicted for aggregates of astronomical silicate grains. Dust optical constants in the FIR-submillimetre range, however, are not well constrained by laboratory studies.

In any case, instruments before \Planck \,never had the angular resolution, appropriate spectral coverage, sensitivity, and mapping capability to perform full and unbiased surveys of the thermal dust emission within individual complexes, from the most diffuse regions to the densest parts. The key observational questions are: where and on which angular scale do the dust properties evolve in the \hbox{ISM}. Resolving these questions is the first step in understanding the physical processes that regulate the optical properties of thermal dust.  

\subsubsection{Planck results}
We have seen in Sections \ref{Optical_depth_atomic} and \ref{Optical_depth_molecular} that the averaged value of the dust optical depth at 250\microns\ per unit column density $\tau_{250}/ N_{\rm H}$ increases from $1.14\pm0.2 \times 10^{-25}$ cm$^2$ in the atomic phase to $2.32 \pm0.3 \times 10^{-25}$ cm$^2$ in the molecular phase. It is interesting to note that the same value of $\tau_{250}/ N_{\rm H}$ was found by \cite{Flagey2009} in the molecular central region of the Taurus complex from {\it IRAS\/} 100\microns\ and \textit{Spitzer} 160\microns\ maps. 

In the molecular phase, the $\tau_{250}$-$N_{\rm H}$ relationship appears linear for $N_{\rm H}=0-15\times 10^{21}$ cm$^2$ (Fig.\,\ref{Fig_tau_vs_NH_from_2MASS}), so the value of  
$\tau_{250}/ N_{\rm H}$ does not appear to depend on the column density $N_{\rm H}$. However, the value of  
$\tau_{250}/ N_{\rm H}$ for dust particles located in dense regions could be higher, due to different effects: 
\begin{enumerate}
\item Because there is no embedded heating star, the distribution of dust temperatures along a ``cold'' line of sight, with a low value of the measured temperature, should be generally broader than along lines of sight with higher measured temperatures. Moreover, the measured temperatures for cold lines of sight 
are always warmer than the average temperature along that line of sight, so the values of $\tau_{250}$ for cold lines of sight may underestimate the average values for dust particles \footnote{For the same reason, we can also conclude that 
%unlike the observed inverse correlation of the dust spectral emissivity index with temperature, 
the inverse correlation of $\tau_{250}/ N_{\rm H}$ with $T$ is robust against the unavoidable combination of dust temperature on the line of sight, and reveals a real anti-correlation of $\tau_{250}/ N_{\rm H}$ and $T$ for the dust particles.} 
\citep[see also][]{CBLS01, L07, S08}. 
%As a consequence, the value of $\tau$-$N_{\rm H}$ for the dust particles located in dense regions could be higher than $2.3 \times 10^{-25}$. 
\item As seen in Sect.\,\ref{ancillary_data}, the NIR extinction map is converted into column density using the standard ratio of selective to total extinction $R_V=3.1$ corresponding to the diffuse ISM. However, $R_V$ is expected to increase at high densities (with typical extinction $A_V>3$), up to about 4.5 owing to grain growth by accretion and coagulation \citep{Whittet2001}. This increase lowers the column densities derived from the extinction. Therefore the computed values of $\tau_{250}/ N_{\rm H}$ could be systematically underestimated in dense regions. 
\end{enumerate}
The increase of $\tau_{250}/ N_{\rm H}$ by a factor around 2 between the atomic phase and the molecular phase obviously tends to decrease the equilibrium temperature of the dust particles. For constant intensity of the incident radiation field and at a first order, the temperature of the dust particles should follow the relationship $T \;\alpha \;(\tau_{250}/ N_{\rm H})^{-1/(4+\beta)}$, which gives for $\beta=1.8$ a decrease of 2\,K for dust at 17\,K. However, the radiation field is attenuated in dense regions, and radiative transfer effects must be considered for a complete analysis of the spatial variations of the dust temperature, which is beyond the scope of this paper. 

%We conclude that the $\tau/ N_{\rm H}$ systematically increases in the outer parts of the molecular phase, for A$_v>1$, by a factor at least equal to 2. %We confirm on the full complex the... 
\section{Conclusions}
 
Combined with {\it IRAS\/} maps at 100\,$\mu$m (3000\,GHz), HFI maps at 857, 545, 353, and 143\,GHz allow the precise measurement of the emission spectrum of thermal dust with  unprecedented sensitivity from the faintest atomic regions to the densest parts of the Taurus molecular complex. 

While the dust particles located along the lines of sight have no reason to be at the same temperature and may have different optical properties, we find that for each pixel of the map the measured spectra are reasonably fitted with a single modified blackbody, which gives one dust temperature, one spectral emissivity index, and one dust optical depth per pixel. %Systematic residuals are obtained at 353 and 143\,GHz with amplitudes $\sim-7\%$ and $\sim+13\%$, respectively, which indicates that the real shape of the measured spectra may be slightly more complex than a single modified blackbody. 
However, the modified blackbody can be slightly broadened around the peak of the spectrum because of the range of dust temperature along the line of sight, which explains the negative residuals found at  353\,GHz (around $-7\%$). On the other hand, the positive residuals at 143\,GHz (around $+13\%$) could be attributed to a slight decrease of the spectral emissivity index at low frequency. 

The dust temperature map we derive from the pixel-by-pixel fits provides a spectacular description of the cooling of the thermal dust across the whole complex from about 17.5\,K to about 13\,\hbox{K}. These variations can be caused by variations of both the excitation conditions and the optical properties of the dust particles. 

The spectral emissivity index map we derive presents significant spatial variations, from 1.6 to 2. The distribution is centred on $1.78$, with systematic error estimated to be 0.07. We have checked that the synthetic spectra computed with the post-\textit{Spitzer} dust models of \cite{Draine2007} and \cite{Compiegne2011} have almost identical values of their spectral emissivity indexes. Slightly higher values ($>1.8$) are found in the coldest (about 14\,K) structures, and we detect a $T-\beta$ anti-correlation that cannot be explained by the statistical noise and the CIBA.

We also derive a dust optical depth map with a very high dynamic range, which reveals the spatial distribution of the column density of the molecular complex from the densest molecular regions to the faint diffuse regions. Using the NIR extinction as an independant tracer of the column density, we report an increase of the measured dust optical depth at 250\microns\ per unit column density in the molecular phase by a factor of about 2 compared to the value found in the diffuse atomic \hbox{ISM}. The increase of optical depth per unit column density for the dust particles could be even higher in dense regions, owing to radiative transfer effects and the increase of $R_V$ in dense regions.

\begin{acknowledgements}
A description of the Planck Collaboration and a list of its members, indicating which technical or scientific activities they have been involved in, can be found at http://www.rssd.esa.int/Planck. We thank Gopal Narayanan and Jorge Pineda for providing FCRAO and NIR extinction data, and the referee, Paul Goldsmith, for very helpful comments.
\end{acknowledgements}

\bibliographystyle{aa}

%\bibliography{Planck_bib.bib, biblio_AA.bib}
\bibliography{biblio_AA.bib}
\raggedright

%%%%%%%%%%% APPENDICE %%%%%%%%%%%%%%%%%%%%%%%%%%%%%%%%%%%%%%%%%%%%%%%%%%%%%%%%%%%%%%

\appendix{}
\section{Error propagation - simulation of the fitting}
\label{simulations}
\subsection{Principle of the simulations}
\begin{figure}
\centering
\includegraphics[angle=90,width=8.7cm]{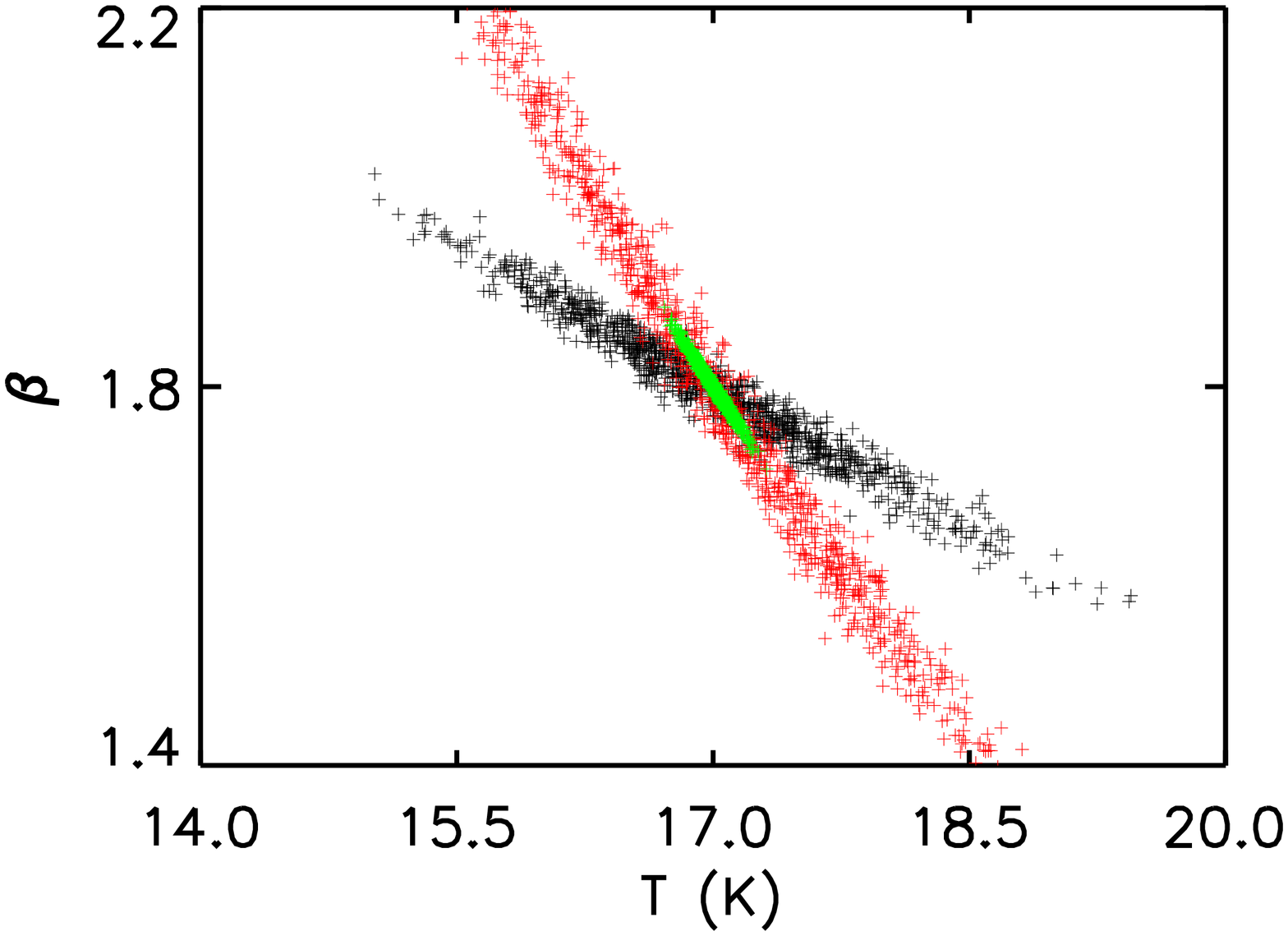}
\vskip -7mm
\includegraphics[angle=90,width=8.7cm]{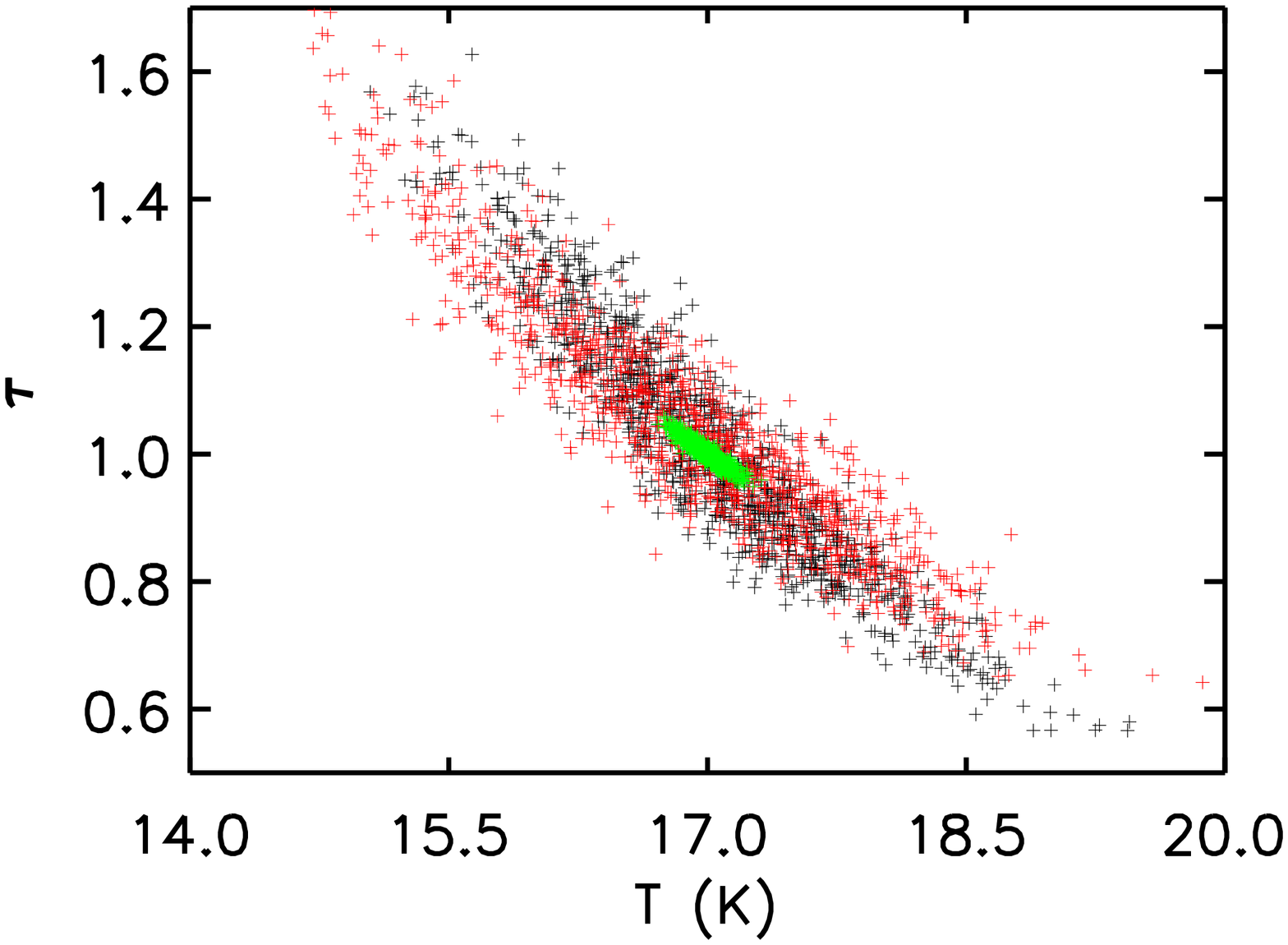}
\vskip -7mm
\includegraphics[angle=90,width=8.7cm]{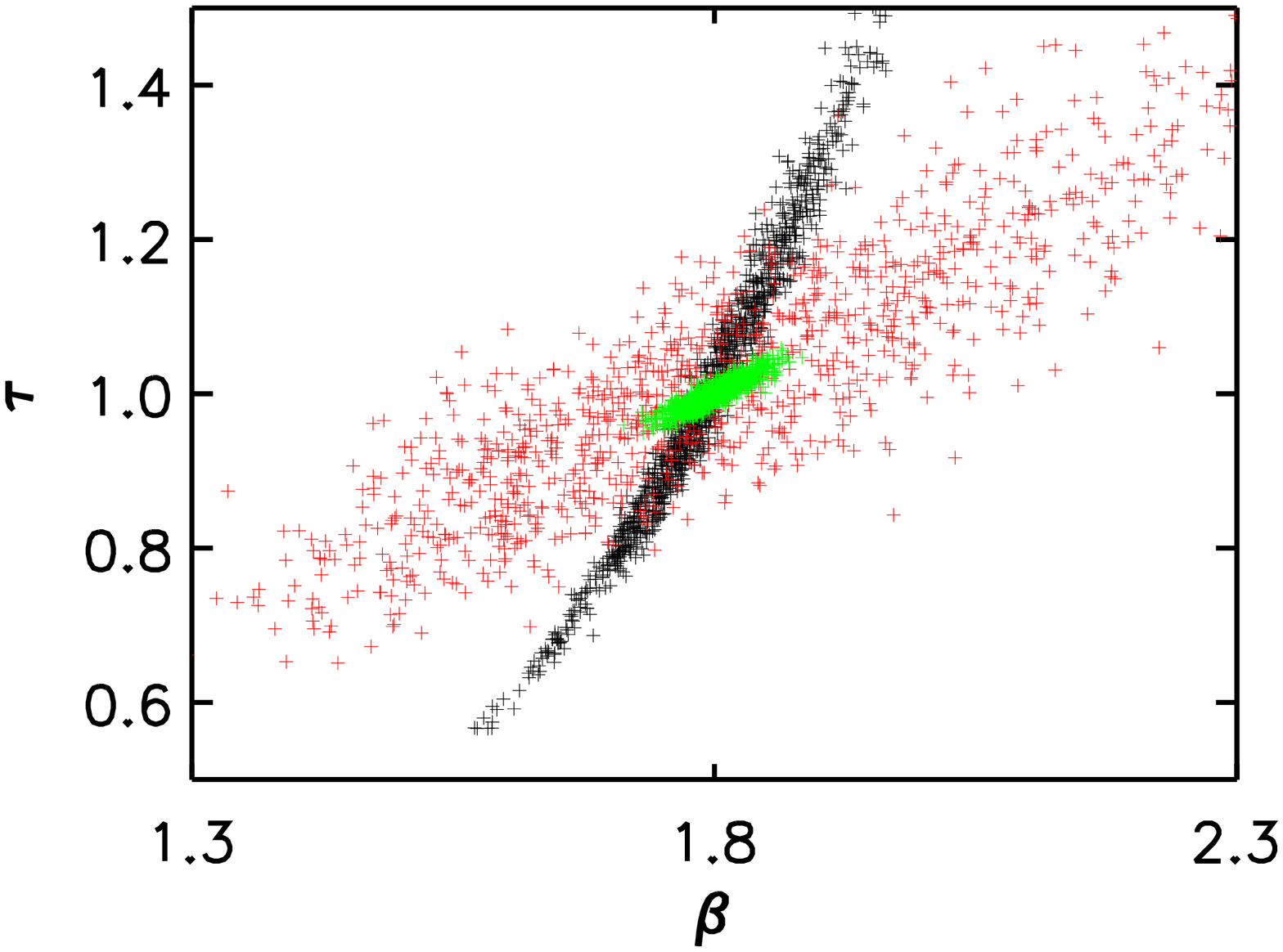}
\vskip -4mm
\caption{Correlation between the fitted values of $T$, $\beta$, and $\tau_{250}$ for 1000 simulated spectra at 100\microns\, 857, 545, and 143\,GHz (350, 545, and 2100\microns) with $T=17\,$K, $\beta=1.8$, and $\tau_{250}=1$: black, with systematic errors on the gains; red and green, with statistical noise and CIBA with $I\,(100\,\mu$m$)= 1$ and 10\,MJy sr$^{-1}$, respectively. Relative values are given for $\tau_{250}$.}
\label{fig_simu_beta_T_opacity}
\end{figure}

We performed Monte-Carlo simulations to understand the propagation of the calibration errors, the statistical noise, and the CIBA in the determination of the dust temperature $T$, the spectral emissivity index $\beta$, and the dust optical depth $\tau_{250}$ at 250\microns\ from the fit to a single modified blackbody of the five bands at 3000\,GHz (100\,$\mu$m), 857, 545, 353, and 143\,GHz (Sect.\,\ref{SEDfitting}). In practice, we computed 1000 synthetic spectra with the same values of $T$ and $\beta$ ($T=17$\,K and $\beta=$\,1.8) at the central frequency of the five bands. We take a fixed value of the dust optical depth ($\tau_{250}=1$) to study the calibration errors, and fixed values of the brightness at 100\,$\mu$m to study statistical noise and CIBA. Then we randomnly add some error for each band computed using a Gaussian distribution with a standard deviation $\sigma$ (and mean of 0) equal to the assumed noise. 
%For each band, we take the standard deviations discussed in section ???, for systematic and statistical errors.  
Finally, we apply our fitting procedure, and obtain for each simulation a set of 1000 values of $T, \beta$, and $\tau_{250}$. For the different simulations we conducted, the standard deviations obtained for each parameters are given in Table\,A.1,
%\cite{tab_simulation_RMS_fitted_parameters}
while Fig.\,\ref{fig_simu_beta_T_opacity} shows the correlations between the three parameters. 

\begin{table*} 
\caption{Standard deviations of the three parameters $T, \beta$ and $\tau_{250}$ adjusted for different simulations, with $T=17\,$K and $\beta=1.8$.  
\label{tab_simulation_RMS_fitted_parameters} 
}
\begin{center}
\begin{tabular}{llllllll}
\hline \\[-0.3cm]
\hline \\[-0.3cm]
%\multicolumn{1}{c}{Mode}  & \multicolumn{1}{c} Observing Time (hours)} \\
Simulation type  & $I\,(100\,\mu$m$)$ (MJy\,sr$^{-1}$) & $\sigma_{\rm T}$ (K) & $\sigma_{\beta}$ & $\sigma_{\tau}$  \\
\hline \\[-0.3cm]
Calibration errors    &   	& $\sim0.7$ & $\sim0.07$ & $\sim18\%$ \\
%without CIBA 
%Statistical errors      & 1 	& $\sim0.6$ & $\sim0.15$ & $\sim0.16\%$ \\
%Statistical errors      & 10   & $\sim0.06$ & $\sim0.015$ & $\sim1.5\%$ \\
%with CIBA 
Statistical noise and CIBA     & 1 	& $\sim1$ & $\sim0.25$ & $\sim20\%$ \\
Statistical noise and CIBA     & 10   & $\sim0.1$ & $\sim0.025$ & $\sim2\%$ \\
\hline \\[-0.3cm]
%\hline \\[-0.3cm]
\end{tabular}
 \end{center}
 \end{table*}

\subsection{Calibration errors}
In order to study the propagation of calibration errors, we took standard deviations of the simulated noise equal to the calibration errors presented in Sect.\,\ref{HFI data}. For the two HFI bands at 545 and 857\,GHz, we assumed that the calibration errors are fully correlated between the two bands, so in the simulation we took the same realization of the simulated Gaussian noise. In contrast, the calibration error at 143\,GHz is not correlated with the calibration error of the two high frequency bands. 

The fitted values of $T$ and $\beta$ computed for the 1000 synthetic spectra are presented in Fig.\,\ref{fig_simu_beta_T_opacity} (black symbols). We observe the classical anti-correlation intrinsic to the noise identified by several authors \citep[e.g., ][]{2009ApJ...696.2234S}. The dust optical depth at 250\microns\ $\tau_{250}$ is also anti-correlated with $T$ %(Fig.\,\ref{fig_simu_beta_T_opacity}) 
owing to the temperature dependance of the Planck function. Finally $\tau_{250}$ is correlated with $\beta$, 
%(Fig.\,\ref{fig_simu_beta_T_opacity})
as expected, since both $T$--$\beta$ and $T$--$\tau_{250}$
%[CRL---check this.  It seemed that a minus sign could not have been intended, but please make sure.]
 are anti-correlated. 
 
 To first order, we can consider that for each band the calibration error is constant over the maps, so the errors on $T, \beta$, and $\tau_{250}$ (Fig.\,\ref{fig_simu_beta_T_opacity} and Table\,A.1) systematically affect (in the same direction for all pixels) the three parameters derived from the fits. Obviously, these calculations are preliminary, since other systematic effects are neglected, but they give for this early analysis an estimate for the systematic errors on the three parameters $T, \beta$, and $\tau_{250}$, which are derived from the data.  

\subsection{Statistical errors and CIB anisotropies}
The statistical noise on the data is considered to be un-correlated both spatially and spectrally. We take standard deviations of the simulated noise equal to the statistical noise used for the pixel-per-pixel fitting of the data at 100\,$\mu$m, 857, 545, and 143\,GHz presented in Sect.\,\ref{Observations}. We also take into 
account the noise caused by the CIBA, using the standard deviations measured by \Planck/HFI \citep{planck2011-6.6} and \textit{IRAS} at 100\,$\mu$m \citep{Penin2011}. %In this early analysis, we consider that the CIBA are fully correlated spectrally. 
 
\begin{figure*}
\centering
\includegraphics[angle=90,width=\textwidth]{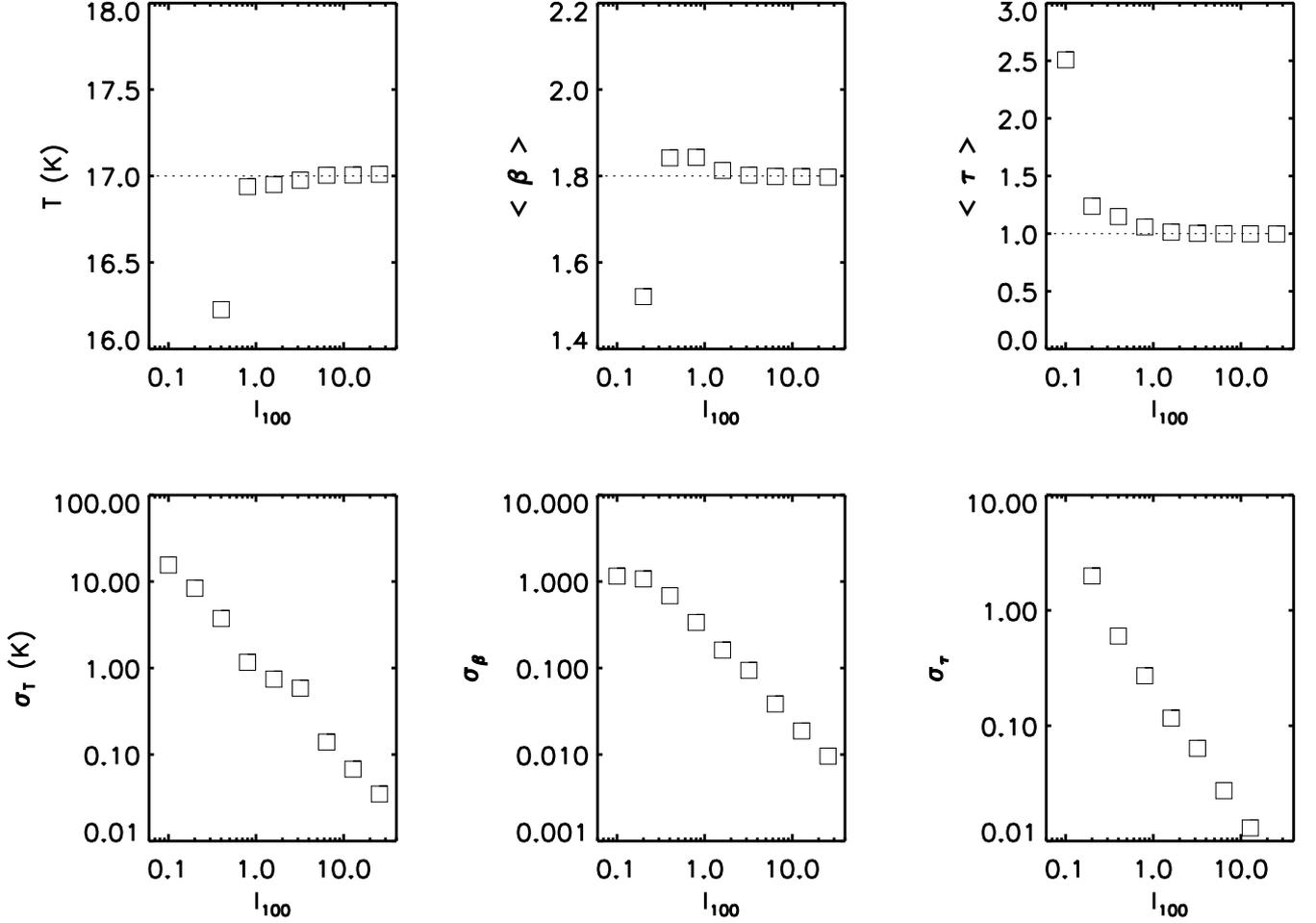}
\caption{Effects of the statistical noise and CIB anisotropies. Mean values and standard deviations of the fitted values of $T$, $\beta$ and $\tau_{250}$ (relative values) are shown as a function of the 100\,$\mu$m brightness, with $T=17\,$K, $\beta=1.8$ and $\tau_{250}=1$.}
\label{fig_simu_beta_T_vs_100}
\end{figure*}

 The correlation diagrams of Fig.\,\ref{fig_simu_beta_T_opacity} and the standard deviations given in Table\,A.1 
 %\ref{simulation_RMS_values} 
 show that the effects of the statistical noise and CIBA have a smaller amplitude than the effects of the systematic errors, but they depend on the absolute brightness, as illustrated in Fig.\,\ref{fig_simu_beta_T_vs_100}. It is interesting to note that above $I\,(100\,\mu$m$)= 1\,$MJy\,sr$^{-1}$, the fitted values of $T$, $\beta$, and $\tau_{250}$ are not biased, since $\langle T\rangle\simeq17\,$K, $\langle\beta\rangle\simeq1.8$, and $\langle\tau_{250}\rangle=1$ (relative value). On the contrary, below $I\,(100\,\mu$m$)= 1\,$MJy\,sr$^{-1}$, we see that $\langle T\rangle$ decreases, while $\langle\beta\rangle$ and $\langle\tau_{250}\rangle$ increase. Therefore, we consider that pixels with $I\,(100\,\mu$m$)< 1\,$MJy\,sr$^{-1}$ cannot be used for any quantitive analysis of the fitted parameters. 

\end{document}

%% file: Planck.tex
\def\setsymbol#1#2{\expandafter\def\csname #1\endcsname{#2}}
\def\getsymbol#1{\csname #1\endcsname}

\def\Planck{{\it Planck\/}}

\def\allearlypapers{\nocite{planck2011-1.1, planck2011-1.3, planck2011-1.4, planck2011-1.5, planck2011-1.6, planck2011-1.7, planck2011-1.10, planck2011-1.10sup, planck2011-5.1a, planck2011-5.1b, planck2011-5.2a, planck2011-5.2b, planck2011-5.2c, planck2011-6.1, planck2011-6.2, planck2011-6.3a, planck2011-6.3b, planck2011-6.4a, planck2011-6.4b, planck2011-6.4c, planck2011-6.6, planck2011-7.0, planck2011-7.2, planck2011-7.3, planck2011-7.7a, planck2011-7.7b, planck2011-7.12, planck2011-7.13}}

\newbox\tablebox    \newdimen\tablewidth
\def\leaderfil{\leaders\hbox to 5pt{\hss.\hss}\hfil}
\def\tablenote#1 #2\par{\begingroup \parindent=0.8em
    \abovedisplayshortskip=0pt\belowdisplayshortskip=0pt
    \noindent
    $$\hss\vbox{\hsize\tablewidth \hangindent=\parindent \hangafter=1 \noindent
    \hbox to \parindent{\sup{\rm #1}\hss}\strut#2\strut\par}\hss$$
    \endgroup}

\def\L2{\ifmmode L_2\else $L_2$\fi}

\def\DeltaT{\ifmmode \Delta T\else $\Delta T$\fi}
\def\deltat{\ifmmode \Delta t\else $\Delta t$\fi}
\def\fknee{\ifmmode f_{\rm knee}\else $f_{\rm knee}$\fi}
\def\Fmax{\ifmmode F_{\rm max}\else $F_{\rm max}$\fi}
\def\solar{\ifmmode{\rm M}_{\mathord\odot}\else${\rm M}_{\mathord\odot}$\fi}

\def\inv{\ifmmode^{-1}\else$^{-1}$\fi}
\def\mo{\ifmmode^{-1}\else$^{-1}$\fi}
\def\sup#1{\ifmmode ^{\rm #1}\else $^{\rm #1}$\fi}
\def\expo#1{\ifmmode \times 10^{#1}\else $\times 10^{#1}$\fi}
\def\,{\thinspace}
\def\lsim{\mathrel{\raise .4ex\hbox{\rlap{$<$}\lower 1.2ex\hbox{$\sim$}}}}
\def\gsim{\mathrel{\raise .4ex\hbox{\rlap{$>$}\lower 1.2ex\hbox{$\sim$}}}}

\def\simprop{\mathrel{\raise .4ex\hbox{\rlap{$\propto$}\lower 1.2ex\hbox{$\sim$}}}}
\def\deg{\ifmmode^\circ\else$^\circ$\fi}
\def\pdeg{\ifmmode $\setbox0=\hbox{$^{\circ}$}\rlap{\hskip.11\wd0 .}$^{\circ}
          \else \setbox0=\hbox{$^{\circ}$}\rlap{\hskip.11\wd0 .}$^{\circ}$\fi}
\def\arcs{\ifmmode {^{\scriptscriptstyle\prime\prime}}
          \else $^{\scriptscriptstyle\prime\prime}$\fi}
\def\arcm{\ifmmode {^{\scriptscriptstyle\prime}}
          \else $^{\scriptscriptstyle\prime}$\fi}
\newdimen\sa  \newdimen\sb
\def\parcs{\sa=.07em \sb=.03em
     \ifmmode \hbox{\rlap{.}}^{\scriptscriptstyle\prime\kern -\sb\prime}\hbox{\kern -\sa}
     \else \rlap{.}$^{\scriptscriptstyle\prime\kern -\sb\prime}$\kern -\sa\fi}
\def\parcm{\sa=.08em \sb=.03em
     \ifmmode \hbox{\rlap{.}\kern\sa}^{\scriptscriptstyle\prime}\hbox{\kern-\sb}
     \else \rlap{.}\kern\sa$^{\scriptscriptstyle\prime}$\kern-\sb\fi}
\def\ra[#1 #2 #3.#4]{#1\sup{h}#2\sup{m}#3\sup{s}\llap.#4}
\def\dec[#1 #2 #3.#4]{#1\deg#2\arcm#3\arcs\llap.#4}
\def\deco[#1 #2 #3]{#1\deg#2\arcm#3\arcs}
\def\rra[#1 #2]{#1\sup{h}#2\sup{m}}

\def\dots{\relax\ifmmode \ldots\else $\ldots$\fi}
\def\WHzsr{\ifmmode $W\,Hz\mo\,sr\mo$\else W\,Hz\mo\,sr\mo\fi}
\def\mHz{\ifmmode $\,mHz$\else \,mHz\fi}
\def\GHz{\ifmmode $\,GHz$\else \,GHz\fi}
\def\mKs{\ifmmode $\,mK\,s$^{1/2}\else \,mK\,s$^{1/2}$\fi}
\def\muKs{\ifmmode \,\mu$K\,s$^{1/2}\else \,$\mu$K\,s$^{1/2}$\fi}
\def\muKRJs{\ifmmode \,\mu$K$_{\rm RJ}$\,s$^{1/2}\else \,$\mu$K$_{\rm RJ}$\,s$^{1/2}$\fi}
\def\muKHz{\ifmmode \,\mu$K\,Hz$^{-1/2}\else \,$\mu$K\,Hz$^{-1/2}$\fi}
\def\MJysr{\ifmmode \,$MJy\,sr\mo$\else \,MJy\,sr\mo\fi}
\def\MJysrmK{\ifmmode \,$MJy\,sr\mo$\,mK$_{\rm CMB}\mo\else \,MJy\,sr\mo\,mK$_{\rm CMB}\mo$\fi}
\def\microns{\ifmmode \,\mu$m$\else \,$\mu$m\fi}
\def\micron{\microns}
\def\muK{\ifmmode \,\mu$K$\else \,$\mu$\hbox{K}\fi}
\def\microK{\ifmmode \,\mu$K$\else \,$\mu$\hbox{K}\fi}
\def\muW{\ifmmode \,\mu$W$\else \,$\mu$\hbox{W}\fi}
\def\kms{\ifmmode $\,km\,s$^{-1}\else \,km\,s$^{-1}$\fi}
\def\kmsMpc{\ifmmode $\,\kms\,Mpc\mo$\else \,\kms\,Mpc\mo\fi}
\setsymbol{LFI:center:frequency:70GHz:units}{70.3\,GHz}
\setsymbol{LFI:center:frequency:44GHz:units}{44.1\,GHz}
\setsymbol{LFI:center:frequency:30GHz:units}{28.5\,GHz}

\setsymbol{LFI:center:frequency:70GHz}{70.3}
\setsymbol{LFI:center:frequency:44GHz}{44.1}
\setsymbol{LFI:center:frequency:30GHz}{28.5}

\setsymbol{LFI:center:frequency:LFI18:Rad:M:units}{71.7\GHz}
\setsymbol{LFI:center:frequency:LFI19:Rad:M:units}{67.5\GHz}
\setsymbol{LFI:center:frequency:LFI20:Rad:M:units}{69.2\GHz}
\setsymbol{LFI:center:frequency:LFI21:Rad:M:units}{70.4\GHz}
\setsymbol{LFI:center:frequency:LFI22:Rad:M:units}{71.5\GHz}
\setsymbol{LFI:center:frequency:LFI23:Rad:M:units}{70.8\GHz}
\setsymbol{LFI:center:frequency:LFI24:Rad:M:units}{44.4\GHz}
\setsymbol{LFI:center:frequency:LFI25:Rad:M:units}{44.0\GHz}
\setsymbol{LFI:center:frequency:LFI26:Rad:M:units}{43.9\GHz}
\setsymbol{LFI:center:frequency:LFI27:Rad:M:units}{28.3\GHz}
\setsymbol{LFI:center:frequency:LFI28:Rad:M:units}{28.8\GHz}
\setsymbol{LFI:center:frequency:LFI18:Rad:S:units}{70.1\GHz}
\setsymbol{LFI:center:frequency:LFI19:Rad:S:units}{69.6\GHz}
\setsymbol{LFI:center:frequency:LFI20:Rad:S:units}{69.5\GHz}
\setsymbol{LFI:center:frequency:LFI21:Rad:S:units}{69.5\GHz}
\setsymbol{LFI:center:frequency:LFI22:Rad:S:units}{72.8\GHz}
\setsymbol{LFI:center:frequency:LFI23:Rad:S:units}{71.3\GHz}
\setsymbol{LFI:center:frequency:LFI24:Rad:S:units}{44.1\GHz}
\setsymbol{LFI:center:frequency:LFI25:Rad:S:units}{44.1\GHz}
\setsymbol{LFI:center:frequency:LFI26:Rad:S:units}{44.1\GHz}
\setsymbol{LFI:center:frequency:LFI27:Rad:S:units}{28.5\GHz}
\setsymbol{LFI:center:frequency:LFI28:Rad:S:units}{28.2\GHz}

\setsymbol{LFI:center:frequency:LFI18:Rad:M}{71.7}
\setsymbol{LFI:center:frequency:LFI19:Rad:M}{67.5}
\setsymbol{LFI:center:frequency:LFI20:Rad:M}{69.2}
\setsymbol{LFI:center:frequency:LFI21:Rad:M}{70.4}
\setsymbol{LFI:center:frequency:LFI22:Rad:M}{71.5}
\setsymbol{LFI:center:frequency:LFI23:Rad:M}{70.8}
\setsymbol{LFI:center:frequency:LFI24:Rad:M}{44.4}
\setsymbol{LFI:center:frequency:LFI25:Rad:M}{44.0}
\setsymbol{LFI:center:frequency:LFI26:Rad:M}{43.9}
\setsymbol{LFI:center:frequency:LFI27:Rad:M}{28.3}
\setsymbol{LFI:center:frequency:LFI28:Rad:M}{28.8}
\setsymbol{LFI:center:frequency:LFI18:Rad:S}{70.1}
\setsymbol{LFI:center:frequency:LFI19:Rad:S}{69.6}
\setsymbol{LFI:center:frequency:LFI20:Rad:S}{69.5}
\setsymbol{LFI:center:frequency:LFI21:Rad:S}{69.5}
\setsymbol{LFI:center:frequency:LFI22:Rad:S}{72.8}
\setsymbol{LFI:center:frequency:LFI23:Rad:S}{71.3}
\setsymbol{LFI:center:frequency:LFI24:Rad:S}{44.1}
\setsymbol{LFI:center:frequency:LFI25:Rad:S}{44.1}
\setsymbol{LFI:center:frequency:LFI26:Rad:S}{44.1}
\setsymbol{LFI:center:frequency:LFI27:Rad:S}{28.5}
\setsymbol{LFI:center:frequency:LFI28:Rad:S}{28.2}

\setsymbol{LFI:white:noise:sensitivity:70GHz:units}{152.6\muKs}
\setsymbol{LFI:white:noise:sensitivity:44GHz:units}{173.1\muKs}
\setsymbol{LFI:white:noise:sensitivity:30GHz:units}{146.8\muKs}

\setsymbol{LFI:white:noise:sensitivity:70GHz}{152.6}
\setsymbol{LFI:white:noise:sensitivity:44GHz}{173.1}
\setsymbol{LFI:white:noise:sensitivity:30GHz}{146.8}

\setsymbol{LFI:white:noise:sensitivity:LFI18:Rad:M:units}{512.0\muKs}
\setsymbol{LFI:white:noise:sensitivity:LFI19:Rad:M:units}{581.4\muKs}
\setsymbol{LFI:white:noise:sensitivity:LFI20:Rad:M:units}{590.8\muKs}
\setsymbol{LFI:white:noise:sensitivity:LFI21:Rad:M:units}{455.2\muKs}
\setsymbol{LFI:white:noise:sensitivity:LFI22:Rad:M:units}{492.0\muKs}
\setsymbol{LFI:white:noise:sensitivity:LFI23:Rad:M:units}{507.7\muKs}
\setsymbol{LFI:white:noise:sensitivity:LFI24:Rad:M:units}{462.2\muKs}
\setsymbol{LFI:white:noise:sensitivity:LFI25:Rad:M:units}{413.6\muKs}
\setsymbol{LFI:white:noise:sensitivity:LFI26:Rad:M:units}{478.6\muKs}
\setsymbol{LFI:white:noise:sensitivity:LFI27:Rad:M:units}{277.7\muKs}
\setsymbol{LFI:white:noise:sensitivity:LFI28:Rad:M:units}{312.3\muKs}
\setsymbol{LFI:white:noise:sensitivity:LFI18:Rad:S:units}{465.7\muKs}
\setsymbol{LFI:white:noise:sensitivity:LFI19:Rad:S:units}{555.6\muKs}
\setsymbol{LFI:white:noise:sensitivity:LFI20:Rad:S:units}{623.2\muKs}
\setsymbol{LFI:white:noise:sensitivity:LFI21:Rad:S:units}{564.1\muKs}
\setsymbol{LFI:white:noise:sensitivity:LFI22:Rad:S:units}{534.4\muKs}
\setsymbol{LFI:white:noise:sensitivity:LFI23:Rad:S:units}{542.4\muKs}
\setsymbol{LFI:white:noise:sensitivity:LFI24:Rad:S:units}{399.2\muKs}
\setsymbol{LFI:white:noise:sensitivity:LFI25:Rad:S:units}{392.6\muKs}
\setsymbol{LFI:white:noise:sensitivity:LFI26:Rad:S:units}{418.6\muKs}
\setsymbol{LFI:white:noise:sensitivity:LFI27:Rad:S:units}{302.9\muKs}
\setsymbol{LFI:white:noise:sensitivity:LFI28:Rad:S:units}{285.3\muKs}

\setsymbol{LFI:white:noise:sensitivity:LFI18:Rad:M}{512.0}
\setsymbol{LFI:white:noise:sensitivity:LFI19:Rad:M}{581.4}
\setsymbol{LFI:white:noise:sensitivity:LFI20:Rad:M}{590.8}
\setsymbol{LFI:white:noise:sensitivity:LFI21:Rad:M}{455.2}
\setsymbol{LFI:white:noise:sensitivity:LFI22:Rad:M}{492.0}
\setsymbol{LFI:white:noise:sensitivity:LFI23:Rad:M}{507.7}
\setsymbol{LFI:white:noise:sensitivity:LFI24:Rad:M}{462.2}
\setsymbol{LFI:white:noise:sensitivity:LFI25:Rad:M}{413.6}
\setsymbol{LFI:white:noise:sensitivity:LFI26:Rad:M}{478.6}
\setsymbol{LFI:white:noise:sensitivity:LFI27:Rad:M}{277.7}
\setsymbol{LFI:white:noise:sensitivity:LFI28:Rad:M}{312.3}
\setsymbol{LFI:white:noise:sensitivity:LFI18:Rad:S}{465.7}
\setsymbol{LFI:white:noise:sensitivity:LFI19:Rad:S}{555.6}
\setsymbol{LFI:white:noise:sensitivity:LFI20:Rad:S}{623.2}
\setsymbol{LFI:white:noise:sensitivity:LFI21:Rad:S}{564.1}
\setsymbol{LFI:white:noise:sensitivity:LFI22:Rad:S}{534.4}
\setsymbol{LFI:white:noise:sensitivity:LFI23:Rad:S}{542.4}
\setsymbol{LFI:white:noise:sensitivity:LFI24:Rad:S}{399.2}
\setsymbol{LFI:white:noise:sensitivity:LFI25:Rad:S}{392.6}
\setsymbol{LFI:white:noise:sensitivity:LFI26:Rad:S}{418.6}
\setsymbol{LFI:white:noise:sensitivity:LFI27:Rad:S}{302.9}
\setsymbol{LFI:white:noise:sensitivity:LFI28:Rad:S}{285.3}

\setsymbol{LFI:knee:frequency:70GHz:units}{29.5\mHz}
\setsymbol{LFI:knee:frequency:44GHz:units}{56.2\mHz}
\setsymbol{LFI:knee:frequency:30GHz:units}{113.7\mHz}

\setsymbol{LFI:knee:frequency:70GHz}{29.5}
\setsymbol{LFI:knee:frequency:44GHz}{56.2}
\setsymbol{LFI:knee:frequency:30GHz}{113.7}

\setsymbol{LFI:knee:frequency:LFI18:Rad:M:units}{16.3\mHz}
\setsymbol{LFI:knee:frequency:LFI19:Rad:M:units}{15.1\mHz}
\setsymbol{LFI:knee:frequency:LFI20:Rad:M:units}{18.7\mHz}
\setsymbol{LFI:knee:frequency:LFI21:Rad:M:units}{37.2\mHz}
\setsymbol{LFI:knee:frequency:LFI22:Rad:M:units}{12.7\mHz}
\setsymbol{LFI:knee:frequency:LFI23:Rad:M:units}{34.6\mHz}
\setsymbol{LFI:knee:frequency:LFI24:Rad:M:units}{46.2\mHz}
\setsymbol{LFI:knee:frequency:LFI25:Rad:M:units}{24.9\mHz}
\setsymbol{LFI:knee:frequency:LFI26:Rad:M:units}{67.6\mHz}
\setsymbol{LFI:knee:frequency:LFI27:Rad:M:units}{187.4\mHz}
\setsymbol{LFI:knee:frequency:LFI28:Rad:M:units}{122.2\mHz}
\setsymbol{LFI:knee:frequency:LFI18:Rad:S:units}{17.7\mHz}
\setsymbol{LFI:knee:frequency:LFI19:Rad:S:units}{22.0\mHz}
\setsymbol{LFI:knee:frequency:LFI20:Rad:S:units}{8.7\mHz}
\setsymbol{LFI:knee:frequency:LFI21:Rad:S:units}{25.9\mHz}
\setsymbol{LFI:knee:frequency:LFI22:Rad:S:units}{15.8\mHz}
\setsymbol{LFI:knee:frequency:LFI23:Rad:S:units}{129.8\mHz}
\setsymbol{LFI:knee:frequency:LFI24:Rad:S:units}{100.9\mHz}
\setsymbol{LFI:knee:frequency:LFI25:Rad:S:units}{38.9\mHz}
\setsymbol{LFI:knee:frequency:LFI26:Rad:S:units}{58.9\mHz}
\setsymbol{LFI:knee:frequency:LFI27:Rad:S:units}{104.4\mHz}
\setsymbol{LFI:knee:frequency:LFI28:Rad:S:units}{40.7\mHz}

\setsymbol{LFI:knee:frequency:LFI18:Rad:M}{16.3}
\setsymbol{LFI:knee:frequency:LFI19:Rad:M}{15.1}
\setsymbol{LFI:knee:frequency:LFI20:Rad:M}{18.7}
\setsymbol{LFI:knee:frequency:LFI21:Rad:M}{37.2}
\setsymbol{LFI:knee:frequency:LFI22:Rad:M}{12.7}
\setsymbol{LFI:knee:frequency:LFI23:Rad:M}{34.6}
\setsymbol{LFI:knee:frequency:LFI24:Rad:M}{46.2}
\setsymbol{LFI:knee:frequency:LFI25:Rad:M}{24.9}
\setsymbol{LFI:knee:frequency:LFI26:Rad:M}{67.6}
\setsymbol{LFI:knee:frequency:LFI27:Rad:M}{187.4}
\setsymbol{LFI:knee:frequency:LFI28:Rad:M}{122.2}
\setsymbol{LFI:knee:frequency:LFI18:Rad:S}{17.7}
\setsymbol{LFI:knee:frequency:LFI19:Rad:S}{22.0}
\setsymbol{LFI:knee:frequency:LFI20:Rad:S}{8.7}
\setsymbol{LFI:knee:frequency:LFI21:Rad:S}{25.9}
\setsymbol{LFI:knee:frequency:LFI22:Rad:S}{15.8}
\setsymbol{LFI:knee:frequency:LFI23:Rad:S}{129.8}
\setsymbol{LFI:knee:frequency:LFI24:Rad:S}{100.9}
\setsymbol{LFI:knee:frequency:LFI25:Rad:S}{38.9}
\setsymbol{LFI:knee:frequency:LFI26:Rad:S}{58.9}
\setsymbol{LFI:knee:frequency:LFI27:Rad:S}{104.4}
\setsymbol{LFI:knee:frequency:LFI28:Rad:S}{40.7}

\setsymbol{LFI:slope:70GHz:units}{$-1.03$\mHz}
\setsymbol{LFI:slope:44GHz:units}{$-0.89$\mHz}
\setsymbol{LFI:slope:30GHz:units}{$-0.87$\mHz}

\setsymbol{LFI:slope:70GHz}{$-1.03$}
\setsymbol{LFI:slope:44GHz}{$-0.89$}
\setsymbol{LFI:slope:30GHz}{$-0.87$}

\setsymbol{LFI:slope:LFI18:Rad:M:units}{$-1.04$\mHz}
\setsymbol{LFI:slope:LFI19:Rad:M:units}{$-1.09$\mHz}
\setsymbol{LFI:slope:LFI20:Rad:M:units}{$-0.69$\mHz}
\setsymbol{LFI:slope:LFI21:Rad:M:units}{$-1.56$\mHz}
\setsymbol{LFI:slope:LFI22:Rad:M:units}{$-1.01$\mHz}
\setsymbol{LFI:slope:LFI23:Rad:M:units}{$-0.96$\mHz}
\setsymbol{LFI:slope:LFI24:Rad:M:units}{$-0.83$\mHz}
\setsymbol{LFI:slope:LFI25:Rad:M:units}{$-0.91$\mHz}
\setsymbol{LFI:slope:LFI26:Rad:M:units}{$-0.95$\mHz}
\setsymbol{LFI:slope:LFI27:Rad:M:units}{$-0.87$\mHz}
\setsymbol{LFI:slope:LFI28:Rad:M:units}{$-0.88$\mHz}
\setsymbol{LFI:slope:LFI18:Rad:S:units}{$-1.15$\mHz}
\setsymbol{LFI:slope:LFI19:Rad:S:units}{$-1.00$\mHz}
\setsymbol{LFI:slope:LFI20:Rad:S:units}{$-0.95$\mHz}
\setsymbol{LFI:slope:LFI21:Rad:S:units}{$-0.92$\mHz}
\setsymbol{LFI:slope:LFI22:Rad:S:units}{$-1.01$\mHz}
\setsymbol{LFI:slope:LFI23:Rad:S:units}{$-0.95$\mHz}
\setsymbol{LFI:slope:LFI24:Rad:S:units}{$-0.73$\mHz}
\setsymbol{LFI:slope:LFI25:Rad:S:units}{$-1.16$\mHz}
\setsymbol{LFI:slope:LFI26:Rad:S:units}{$-0.79$\mHz}
\setsymbol{LFI:slope:LFI27:Rad:S:units}{$-0.82$\mHz}
\setsymbol{LFI:slope:LFI28:Rad:S:units}{$-0.91$\mHz}

\setsymbol{LFI:slope:LFI18:Rad:M}{$-1.04$}
\setsymbol{LFI:slope:LFI19:Rad:M}{$-1.09$}
\setsymbol{LFI:slope:LFI20:Rad:M}{$-0.69$}
\setsymbol{LFI:slope:LFI21:Rad:M}{$-1.56$}
\setsymbol{LFI:slope:LFI22:Rad:M}{$-1.01$}
\setsymbol{LFI:slope:LFI23:Rad:M}{$-0.96$}
\setsymbol{LFI:slope:LFI24:Rad:M}{$-0.83$}
\setsymbol{LFI:slope:LFI25:Rad:M}{$-0.91$}
\setsymbol{LFI:slope:LFI26:Rad:M}{$-0.95$}
\setsymbol{LFI:slope:LFI27:Rad:M}{$-0.87$}
\setsymbol{LFI:slope:LFI28:Rad:M}{$-0.88$}
\setsymbol{LFI:slope:LFI18:Rad:S}{$-1.15$}
\setsymbol{LFI:slope:LFI19:Rad:S}{$-1.00$}
\setsymbol{LFI:slope:LFI20:Rad:S}{$-0.95$}
\setsymbol{LFI:slope:LFI21:Rad:S}{$-0.92$}
\setsymbol{LFI:slope:LFI22:Rad:S}{$-1.01$}
\setsymbol{LFI:slope:LFI23:Rad:S}{$-0.95$}
\setsymbol{LFI:slope:LFI24:Rad:S}{$-0.73$}
\setsymbol{LFI:slope:LFI25:Rad:S}{$-1.16$}
\setsymbol{LFI:slope:LFI26:Rad:S}{$-0.79$}
\setsymbol{LFI:slope:LFI27:Rad:S}{$-0.82$}
\setsymbol{LFI:slope:LFI28:Rad:S}{$-0.91$}

\setsymbol{LFI:FWHM:70GHz:units}{13\parcm01}
\setsymbol{LFI:FWHM:44GHz:units}{27\parcm92}
\setsymbol{LFI:FWHM:30GHz:units}{32\parcm65}

\setsymbol{LFI:FWHM:70GHz}{13.01}
\setsymbol{LFI:FWHM:44GHz}{27.92}
\setsymbol{LFI:FWHM:30GHz}{32.65}

\setsymbol{LFI:FWHM:LFI18:units}{13\parcm39}
\setsymbol{LFI:FWHM:LFI19:units}{13\parcm01}
\setsymbol{LFI:FWHM:LFI20:units}{12\parcm75}
\setsymbol{LFI:FWHM:LFI21:units}{12\parcm74}
\setsymbol{LFI:FWHM:LFI22:units}{12\parcm87}
\setsymbol{LFI:FWHM:LFI23:units}{13\parcm27}
\setsymbol{LFI:FWHM:LFI24:units}{22\parcm98}
\setsymbol{LFI:FWHM:LFI25:units}{30\parcm46}
\setsymbol{LFI:FWHM:LFI26:units}{30\parcm31}
\setsymbol{LFI:FWHM:LFI27:units}{32\parcm65}
\setsymbol{LFI:FWHM:LFI28:units}{32\parcm66}

\setsymbol{LFI:FWHM:LFI18}{13.39}
\setsymbol{LFI:FWHM:LFI19}{13.01}
\setsymbol{LFI:FWHM:LFI20}{12.75}
\setsymbol{LFI:FWHM:LFI21}{12.74}
\setsymbol{LFI:FWHM:LFI22}{12.87}
\setsymbol{LFI:FWHM:LFI23}{13.27}
\setsymbol{LFI:FWHM:LFI24}{22.98}
\setsymbol{LFI:FWHM:LFI25}{30.46}
\setsymbol{LFI:FWHM:LFI26}{30.31}
\setsymbol{LFI:FWHM:LFI27}{32.65}
\setsymbol{LFI:FWHM:LFI28}{32.66}

\setsymbol{LFI:FWHM:uncertainty:LFI18:units}{0.170\arcm}
\setsymbol{LFI:FWHM:uncertainty:LFI19:units}{0.174\arcm}
\setsymbol{LFI:FWHM:uncertainty:LFI20:units}{0.170\arcm}
\setsymbol{LFI:FWHM:uncertainty:LFI21:units}{0.156\arcm}
\setsymbol{LFI:FWHM:uncertainty:LFI22:units}{0.164\arcm}
\setsymbol{LFI:FWHM:uncertainty:LFI23:units}{0.171\arcm}
\setsymbol{LFI:FWHM:uncertainty:LFI24:units}{0.652\arcm}
\setsymbol{LFI:FWHM:uncertainty:LFI25:units}{1.075\arcm}
\setsymbol{LFI:FWHM:uncertainty:LFI26:units}{1.131\arcm}
\setsymbol{LFI:FWHM:uncertainty:LFI27:units}{1.266\arcm}
\setsymbol{LFI:FWHM:uncertainty:LFI28:units}{1.287\arcm}

\setsymbol{LFI:FWHM:uncertainty:LFI18}{0.170}
\setsymbol{LFI:FWHM:uncertainty:LFI19}{0.174}
\setsymbol{LFI:FWHM:uncertainty:LFI20}{0.170}
\setsymbol{LFI:FWHM:uncertainty:LFI21}{0.156}
\setsymbol{LFI:FWHM:uncertainty:LFI22}{0.164}
\setsymbol{LFI:FWHM:uncertainty:LFI23}{0.171}
\setsymbol{LFI:FWHM:uncertainty:LFI24}{0.652}
\setsymbol{LFI:FWHM:uncertainty:LFI25}{1.075}
\setsymbol{LFI:FWHM:uncertainty:LFI26}{1.131}
\setsymbol{LFI:FWHM:uncertainty:LFI27}{1.266}
\setsymbol{LFI:FWHM:uncertainty:LFI28}{1.287}

\setsymbol{HFI:center:frequency:100GHz:units}{100\,GHz}
\setsymbol{HFI:center:frequency:143GHz:units}{143\,GHz}
\setsymbol{HFI:center:frequency:217GHz:units}{217\,GHz}
\setsymbol{HFI:center:frequency:353GHz:units}{353\,GHz}
\setsymbol{HFI:center:frequency:545GHz:units}{550\,GHz}
\setsymbol{HFI:center:frequency:857GHz:units}{857\,GHz}

\setsymbol{HFI:center:frequency:100GHz}{100}
\setsymbol{HFI:center:frequency:143GHz}{143}
\setsymbol{HFI:center:frequency:217GHz}{217}
\setsymbol{HFI:center:frequency:353GHz}{353}
\setsymbol{HFI:center:frequency:545GHz}{550}
\setsymbol{HFI:center:frequency:857GHz}{857}

\setsymbol{HFI:Ndetectors:100GHz}{8}
\setsymbol{HFI:Ndetectors:143GHz}{11}
\setsymbol{HFI:Ndetectors:217GHz}{12}
\setsymbol{HFI:Ndetectors:353GHz}{12}
\setsymbol{HFI:Ndetectors:545GHz}{3}
\setsymbol{HFI:Ndetectors:857GHz}{4}

\setsymbol{HFI:FWHM:Maps:100GHz:units}{9\parcm88}
\setsymbol{HFI:FWHM:Maps:143GHz:units}{7\parcm18}
\setsymbol{HFI:FWHM:Maps:217GHz:units}{4\parcm87}
\setsymbol{HFI:FWHM:Maps:353GHz:units}{4\parcm65}
\setsymbol{HFI:FWHM:Maps:545GHz:units}{4\parcm72}
\setsymbol{HFI:FWHM:Maps:857GHz:units}{4\parcm39}
\setsymbol{HFI:FWHM:Maps:100GHz}{9.88}
\setsymbol{HFI:FWHM:Maps:143GHz}{7.18}
\setsymbol{HFI:FWHM:Maps:217GHz}{4.87}
\setsymbol{HFI:FWHM:Maps:353GHz}{4.65}
\setsymbol{HFI:FWHM:Maps:545GHz}{4.72}
\setsymbol{HFI:FWHM:Maps:857GHz}{4.39}

\setsymbol{HFI:beam:ellipticity:Maps:100GHz}{1.15}
\setsymbol{HFI:beam:ellipticity:Maps:143GHz}{1.01}
\setsymbol{HFI:beam:ellipticity:Maps:217GHz}{1.06}
\setsymbol{HFI:beam:ellipticity:Maps:353GHz}{1.05}
\setsymbol{HFI:beam:ellipticity:Maps:545GHz}{1.14}
\setsymbol{HFI:beam:ellipticity:Maps:857GHz}{1.19}

\setsymbol{HFI:FWHM:Mars:100GHz:units}{9\parcm37}
\setsymbol{HFI:FWHM:Mars:143GHz:units}{7\parcm04}
\setsymbol{HFI:FWHM:Mars:217GHz:units}{4\parcm68}
\setsymbol{HFI:FWHM:Mars:353GHz:units}{4\parcm43}
\setsymbol{HFI:FWHM:Mars:545GHz:units}{3\parcm80}
\setsymbol{HFI:FWHM:Mars:857GHz:units}{3\parcm67}

\setsymbol{HFI:FWHM:Mars:100GHz}{9.37}
\setsymbol{HFI:FWHM:Mars:143GHz}{7.04}
\setsymbol{HFI:FWHM:Mars:217GHz}{4.68}
\setsymbol{HFI:FWHM:Mars:353GHz}{4.43}
\setsymbol{HFI:FWHM:Mars:545GHz}{3.80}
\setsymbol{HFI:FWHM:Mars:857GHz}{3.67}

\setsymbol{HFI:beam:ellipticity:Mars:100GHz}{1.18}
\setsymbol{HFI:beam:ellipticity:Mars:143GHz}{1.03}
\setsymbol{HFI:beam:ellipticity:Mars:217GHz}{1.14}
\setsymbol{HFI:beam:ellipticity:Mars:353GHz}{1.09}
\setsymbol{HFI:beam:ellipticity:Mars:545GHz}{1.25}
\setsymbol{HFI:beam:ellipticity:Mars:857GHz}{1.03}

\setsymbol{HFI:CMB:relative:calibration:100GHz}{$\lsim 1\%$}
\setsymbol{HFI:CMB:relative:calibration:143GHz}{$\lsim 1\%$}
\setsymbol{HFI:CMB:relative:calibration:217GHz}{$\lsim 1\%$}
\setsymbol{HFI:CMB:relative:calibration:353GHz}{$\lsim 1\%$}
\setsymbol{HFI:CMB:relative:calibration:545GHz}{}
\setsymbol{HFI:CMB:relative:calibration:857GHz}{}

\setsymbol{HFI:CMB:absolute:calibration:100GHz}{$\lsim 2\%$}
\setsymbol{HFI:CMB:absolute:calibration:143GHz}{$\lsim 2\%$}
\setsymbol{HFI:CMB:absolute:calibration:217GHz}{$\lsim 2\%$}
\setsymbol{HFI:CMB:absolute:calibration:353GHz}{$\lsim 2\%$}
\setsymbol{HFI:CMB:absolute:calibration:545GHz}{}
\setsymbol{HFI:CMB:absolute:calibration:857GHz}{}

\setsymbol{HFI:FIRAS:gain:calibration:accuracy:statistical:100GHz}{}
\setsymbol{HFI:FIRAS:gain:calibration:accuracy:statistical:143GHz}{}
\setsymbol{HFI:FIRAS:gain:calibration:accuracy:statistical:217GHz}{}
\setsymbol{HFI:FIRAS:gain:calibration:accuracy:statistical:353GHz}{2.5\%}
\setsymbol{HFI:FIRAS:gain:calibration:accuracy:statistical:545GHz}{1\%}
\setsymbol{HFI:FIRAS:gain:calibration:accuracy:statistical:857GHz}{0.5\%}

\setsymbol{HFI:FIRAS:gain:calibration:accuracy:systematic:100GHz}{}
\setsymbol{HFI:FIRAS:gain:calibration:accuracy:systematic:143GHz}{}
\setsymbol{HFI:FIRAS:gain:calibration:accuracy:systematic:217GHz}{}
\setsymbol{HFI:FIRAS:gain:calibration:accuracy:systematic:353GHz}{}
\setsymbol{HFI:FIRAS:gain:calibration:accuracy:systematic:545GHz}{7\%}
\setsymbol{HFI:FIRAS:gain:calibration:accuracy:systematic:857GHz}{7\%}

\setsymbol{HFI:FIRAS:zero:point:accuracy:100GHz:units}{0.8\MJysr}
\setsymbol{HFI:FIRAS:zero:point:accuracy:143GHz:units}{}
\setsymbol{HFI:FIRAS:zero:point:accuracy:217GHz:units}{}
\setsymbol{HFI:FIRAS:zero:point:accuracy:353GHz:units}{1.4\MJysr}
\setsymbol{HFI:FIRAS:zero:point:accuracy:545GHz:units}{2.2\MJysr}
\setsymbol{HFI:FIRAS:zero:point:accuracy:857GHz:units}{1.7\MJysr}

\setsymbol{HFI:FIRAS:zero:point:accuracy:100GHz}{0.8}
\setsymbol{HFI:FIRAS:zero:point:accuracy:143GHz}{}
\setsymbol{HFI:FIRAS:zero:point:accuracy:217GHz}{}
\setsymbol{HFI:FIRAS:zero:point:accuracy:353GHz}{1.4}
\setsymbol{HFI:FIRAS:zero:point:accuracy:545GHz}{2.2}
\setsymbol{HFI:FIRAS:zero:point:accuracy:857GHz}{1.7}

\setsymbol{HFI:unit:conversion:100GHz:units}{0.2415\MJysrmK}
\setsymbol{HFI:unit:conversion:143GHz:units}{0.3694\MJysrmK}
\setsymbol{HFI:unit:conversion:217GHz:units}{0.4811\MJysrmK}
\setsymbol{HFI:unit:conversion:353GHz:units}{0.2883\MJysrmK}
\setsymbol{HFI:unit:conversion:545GHz:units}{0.05826\MJysrmK}
\setsymbol{HFI:unit:conversion:857GHz:units}{0.002238\MJysrmK}

\setsymbol{HFI:unit:conversion:100GHz}{0.2415}
\setsymbol{HFI:unit:conversion:143GHz}{0.3694}
\setsymbol{HFI:unit:conversion:217GHz}{0.4811}
\setsymbol{HFI:unit:conversion:353GHz}{0.2883}
\setsymbol{HFI:unit:conversion:545GHz}{0.05826}
\setsymbol{HFI:unit:conversion:857GHz}{0.002238}

\setsymbol{HFI:colour:correction:alpha=-2:V1.01:100GHz}{0.9893}
\setsymbol{HFI:colour:correction:alpha=-2:V1.01:143GHz}{0.9759}
\setsymbol{HFI:colour:correction:alpha=-2:V1.01:217GHz}{1.0007}
\setsymbol{HFI:colour:correction:alpha=-2:V1.01:353GHz}{1.0028}
\setsymbol{HFI:colour:correction:alpha=-2:V1.01:545GHz}{1.0019}
\setsymbol{HFI:colour:correction:alpha=-2:V1.01:857GHz}{0.9889}

\setsymbol{HFI:colour:correction:alpha=0:V1.01:100GHz}{1.0008}
\setsymbol{HFI:colour:correction:alpha=0:V1.01:143GHz}{1.0148}
\setsymbol{HFI:colour:correction:alpha=0:V1.01:217GHz}{0.9909}
\setsymbol{HFI:colour:correction:alpha=0:V1.01:353GHz}{0.9888}
\setsymbol{HFI:colour:correction:alpha=0:V1.01:545GHz}{0.9878}
\setsymbol{HFI:colour:correction:alpha=0:V1.01:857GHz}{1.0014}

%% file: Proj_Ref_7_13_authors_and_institutes.tex
%This author list corresponds to \title{Author list for Proj. Ref. 7.13: Thermal dust in nearby molecular clouds}
%Prepared by R. Leonardi (rleonardi@sciops.esa.int), ESAC/ESA, on 10MAY2011
%This version is from 16 May 2011 at 12:00 CET
%\subtitle{There are 199 co-authors in this list}
\author{\small
Planck Collaboration:
A.~Abergel\inst{46}
\and
P.~A.~R.~Ade\inst{70}
\and
N.~Aghanim\inst{46}
\and
M.~Arnaud\inst{57}
\and
M.~Ashdown\inst{55, 4}
\and
J.~Aumont\inst{46}
\and
C.~Baccigalupi\inst{68}
\and
A.~Balbi\inst{28}
\and
A.~J.~Banday\inst{74, 7, 62}
\and
R.~B.~Barreiro\inst{52}
% ...
}
\institute{\small
Aalto University Mets\"{a}hovi Radio Observatory, Mets\"{a}hovintie 114, FIN-02540 Kylm\"{a}l\"{a}, Finland\\
\and
Agenzia Spaziale Italiana Science Data Center, c/o ESRIN, via Galileo Galilei, Frascati, Italy\\
\and
Astroparticule et Cosmologie, CNRS (UMR7164), Universit\'{e} Denis Diderot Paris 7, B\^{a}timent Condorcet, 10 rue A. Domon et L\'{e}onie Duquet, Paris, France\\
\and
Astrophysics Group, Cavendish Laboratory, University of Cambridge, J J Thomson Avenue, Cambridge CB3 0HE, U.K.\\
\and
Atacama Large Millimeter/submillimeter Array, ALMA Santiago Central Offices, Alonso de Cordova 3107, Vitacura, Casilla 763 0355, Santiago, Chile\\
\and
CITA, University of Toronto, 60 St. George St., Toronto, ON M5S 3H8, Canada\\
\and
CNRS, IRAP, 9 Av. colonel Roche, BP 44346, F-31028 Toulouse cedex 4, France\\
\and
California Institute of Technology, Pasadena, California, U.S.A.\\
\and
DAMTP, University of Cambridge, Centre for Mathematical Sciences, Wilberforce Road, Cambridge CB3 0WA, U.K.\\
\and
DSM/Irfu/SPP, CEA-Saclay, F-91191 Gif-sur-Yvette Cedex, France\\
\and
DTU Space, National Space Institute, Juliane Mariesvej 30, Copenhagen, Denmark\\
\and
Departamento de F\'{\i}sica, Universidad de Oviedo, Avda. Calvo Sotelo s/n, Oviedo, Spain\\
\and
Department of Astronomy and Astrophysics, University of Toronto, 50 Saint George Street, Toronto, Ontario, Canada\\
\and
Department of Astronomy and Earth Sciences, Tokyo Gakugei University, Koganei, Tokyo 184-8501, Japan\\
\and
Department of Physics \& Astronomy, University of British Columbia, 6224 Agricultural Road, Vancouver, British Columbia, Canada\\
\and
Department of Physics, Gustaf H\"{a}llstr\"{o}min katu 2a, University of Helsinki, Helsinki, Finland\\
\and
Department of Physics, Princeton University, Princeton, New Jersey, U.S.A.\\
\and
Department of Physics, Purdue University, 525 Northwestern Avenue, West Lafayette, Indiana, U.S.A.\\
\and
Department of Physics, University of California, Berkeley, California, U.S.A.\\
\and
Department of Physics, University of California, One Shields Avenue, Davis, California, U.S.A.\\
\and
Department of Physics, University of California, Santa Barbara, California, U.S.A.\\
\and
Department of Physics, University of Illinois at Urbana-Champaign, 1110 West Green Street, Urbana, Illinois, U.S.A.\\
\and
Dipartimento di Fisica G. Galilei, Universit\`{a} degli Studi di Padova, via Marzolo 8, 35131 Padova, Italy\\
\and
Dipartimento di Fisica, Universit\`{a} La Sapienza, P. le A. Moro 2, Roma, Italy\\
\and
Dipartimento di Fisica, Universit\`{a} degli Studi di Milano, Via Celoria, 16, Milano, Italy\\
\and
Dipartimento di Fisica, Universit\`{a} degli Studi di Trieste, via A. Valerio 2, Trieste, Italy\\
\and
Dipartimento di Fisica, Universit\`{a} di Ferrara, Via Saragat 1, 44122 Ferrara, Italy\\
\and
Dipartimento di Fisica, Universit\`{a} di Roma Tor Vergata, Via della Ricerca Scientifica, 1, Roma, Italy\\
\and
Discovery Center, Niels Bohr Institute, Blegdamsvej 17, Copenhagen, Denmark\\
\and
Dpto. Astrof\'{i}sica, Universidad de La Laguna (ULL), E-38206 La Laguna, Tenerife, Spain\\
\and
European Southern Observatory, ESO Vitacura, Alonso de Cordova 3107, Vitacura, Casilla 19001, Santiago, Chile\\
\and
European Space Agency, ESAC, Planck Science Office, Camino bajo del Castillo, s/n, Urbanizaci\'{o}n Villafranca del Castillo, Villanueva de la Ca\~{n}ada, Madrid, Spain\\
\and
European Space Agency, ESTEC, Keplerlaan 1, 2201 AZ Noordwijk, The Netherlands\\
\and
Helsinki Institute of Physics, Gustaf H\"{a}llstr\"{o}min katu 2, University of Helsinki, Helsinki, Finland\\
\and
INAF - Osservatorio Astrofisico di Catania, Via S. Sofia 78, Catania, Italy\\
\and
INAF - Osservatorio Astronomico di Padova, Vicolo dell'Osservatorio 5, Padova, Italy\\
\and
INAF - Osservatorio Astronomico di Roma, via di Frascati 33, Monte Porzio Catone, Italy\\
\and
INAF - Osservatorio Astronomico di Trieste, Via G.B. Tiepolo 11, Trieste, Italy\\
\and
INAF/IASF Bologna, Via Gobetti 101, Bologna, Italy\\
\and
INAF/IASF Milano, Via E. Bassini 15, Milano, Italy\\
\and
INRIA, Laboratoire de Recherche en Informatique, Universit\'{e} Paris-Sud 11, B\^{a}timent 490, 91405 Orsay Cedex, France\\
\and
IPAG: Institut de Plan\'{e}tologie et d'Astrophysique de Grenoble, Universit\'{e} Joseph Fourier, Grenoble 1 / CNRS-INSU, UMR 5274, Grenoble, F-38041, France\\
\and
Imperial College London, Astrophysics group, Blackett Laboratory, Prince Consort Road, London, SW7 2AZ, U.K.\\
\and
Infrared Processing and Analysis Center, California Institute of Technology, Pasadena, CA 91125, U.S.A.\\
\and
Institut N\'{e}el, CNRS, Universit\'{e} Joseph Fourier Grenoble I, 25 rue des Martyrs, Grenoble, France\\
\and
Institut d'Astrophysique Spatiale, CNRS (UMR8617) Universit\'{e} Paris-Sud 11, B\^{a}timent 121, Orsay, France\\
\and
Institut d'Astrophysique de Paris, CNRS UMR7095, Universit\'{e} Pierre \& Marie Curie, 98 bis boulevard Arago, Paris, France\\
\and
Institute of Astronomy and Astrophysics, Academia Sinica, Taipei, Taiwan\\
\and
Institute of Astronomy, University of Cambridge, Madingley Road, Cambridge CB3 0HA, U.K.\\
\and
Institute of Theoretical Astrophysics, University of Oslo, Blindern, Oslo, Norway\\
\and
Instituto de Astrof\'{\i}sica de Canarias, C/V\'{\i}a L\'{a}ctea s/n, La Laguna, Tenerife, Spain\\
\and
Instituto de F\'{\i}sica de Cantabria (CSIC-Universidad de Cantabria), Avda. de los Castros s/n, Santander, Spain\\
\and
Jet Propulsion Laboratory, California Institute of Technology, 4800 Oak Grove Drive, Pasadena, California, U.S.A.\\
\and
Jodrell Bank Centre for Astrophysics, Alan Turing Building, School of Physics and Astronomy, The University of Manchester, Oxford Road, Manchester, M13 9PL, U.K.\\
\and
Kavli Institute for Cosmology Cambridge, Madingley Road, Cambridge, CB3 0HA, U.K.\\
\and
LERMA, CNRS, Observatoire de Paris, 61 Avenue de l'Observatoire, Paris, France\\
\and
Laboratoire AIM, IRFU/Service d'Astrophysique - CEA/DSM - CNRS - Universit\'{e} Paris Diderot, B\^{a}t. 709, CEA-Saclay, F-91191 Gif-sur-Yvette Cedex, France\\
\and
Laboratoire Traitement et Communication de l'Information, CNRS (UMR 5141) and T\'{e}l\'{e}com ParisTech, 46 rue Barrault F-75634 Paris Cedex 13, France\\
\and
Laboratoire de Physique Subatomique et de Cosmologie, CNRS/IN2P3, Universit\'{e} Joseph Fourier Grenoble I, Institut National Polytechnique de Grenoble, 53 rue des Martyrs, 38026 Grenoble cedex, France\\
\and
Laboratoire de l'Acc\'{e}l\'{e}rateur Lin\'{e}aire, Universit\'{e} Paris-Sud 11, CNRS/IN2P3, Orsay, France\\
\and
Lawrence Berkeley National Laboratory, Berkeley, California, U.S.A.\\
\and
Max-Planck-Institut f\"{u}r Astrophysik, Karl-Schwarzschild-Str. 1, 85741 Garching, Germany\\
\and
MilliLab, VTT Technical Research Centre of Finland, Tietotie 3, Espoo, Finland\\
\and
National University of Ireland, Department of Experimental Physics, Maynooth, Co. Kildare, Ireland\\
\and
Niels Bohr Institute, Blegdamsvej 17, Copenhagen, Denmark\\
\and
Observational Cosmology, Mail Stop 367-17, California Institute of Technology, Pasadena, CA, 91125, U.S.A.\\
\and
Optical Science Laboratory, University College London, Gower Street, London, U.K.\\
\and
SISSA, Astrophysics Sector, via Bonomea 265, 34136, Trieste, Italy\\
\and
SUPA, Institute for Astronomy, University of Edinburgh, Royal Observatory, Blackford Hill, Edinburgh EH9 3HJ, U.K.\\
\and
School of Physics and Astronomy, Cardiff University, Queens Buildings, The Parade, Cardiff, CF24 3AA, U.K.\\
\and
Space Sciences Laboratory, University of California, Berkeley, California, U.S.A.\\
\and
Spitzer Science Center, 1200 E. California Blvd., Pasadena, California, U.S.A.\\
\and
Stanford University, Dept of Physics, Varian Physics Bldg, 382 Via Pueblo Mall, Stanford, California, U.S.A.\\
\and
Universit\'{e} de Toulouse, UPS-OMP, IRAP, F-31028 Toulouse cedex 4, France\\
% ...
}